\documentclass[10pt,onecolumn]{IEEEtran}

\usepackage{amsmath}
\usepackage{amsthm}
\usepackage{amssymb}
\usepackage{graphicx}
\usepackage{subfig}
\usepackage{rotating}
\usepackage{flushend}
\usepackage{verbatim}
\usepackage{enumitem}

\newsavebox{\tempbox}

\usepackage{multirow}
\usepackage{xcolor}

\usepackage{balance}

\DeclareMathOperator*{\argmax}{arg\,max}

\newcommand{\logt}{\log_{2}}

\newcommand{\abs}[1]{\left| #1 \right|}

\newcommand{\defn}{\triangleq}

\newcommand{\g}[3]{g^{n}_{ #1  }\left( #2,#3 \right)}
\newcommand{\bg}[3]{\bar{g}^{n}_{ #1  }\left( #2,#3 \right)}

\newcommand{\mbf}[1]{\mathbf{#1}}
\newcommand{\mcf}[1]{\mathcal{#1}}
\newcommand{\idc}[1]{\mathfrak{1}\left({#1}\right) }

\newcommand{\set}[1]{\left\{ #1 \right\}}
\newcommand{\stab}{\scriptscriptstyle{(\text{stable})}}
\newcommand{\sat}{\scriptscriptstyle{(\text{saturate})}}

\DeclareMathOperator{\markov}{\setlength{\unitlength}{.5cm} \begin{picture}(1,1)  \put(0,.22){\line(1,0){1}}  \put(.5,.22){\circle{.3}}   \end{picture}}

\newtheorem{theorem}{Theorem}
\newtheorem{define}[theorem]{Definition}

\newtheorem{lemma}[theorem]{Lemma}
\newtheorem{cor}[theorem]{Corollary}

\newtheorem{remark}[theorem]{Remark}
\let\oldremark\remark
\renewcommand{\remark}{\oldremark\normalfont}

\IEEEoverridecommandlockouts

\begin{document}

\title{Inducing information stability and applications thereof to obtaining information theoretic necessary conditions directly from operational requirements}

\author{Eric Graves and Tan F. Wong% <-this % stops a space
\thanks{Eric Graves is with Army Research Lab,  Adelphi, MD
  20783, U.S.A. \texttt{ericsgra@ufl.edu} }%
\thanks{Tan F. Wong is with Department of Electrical and Computer Engineering,
 University of Florida,
 Gainesville, FL 32611, U.S.A. 
\texttt{twong@ufl.edu}}%
\thanks{This research was presented in part at the \emph{2014, 2016, and 2017 IEEE International Symposium on Information Theory}.  T. F. Wong was supported by the National Science Foundation
under Grant CCF-1320086. Eric Graves was supported by something at Army Research Lab.}% <-this % stops a space
}

\maketitle

\begin{abstract}
  This work constructs a discrete random variable that, when
  conditioned upon, ensures information stability of quasi-images.
  Using this construction, a new methodology is derived to obtain
  information theoretic necessary conditions directly from operational
  requirements.  In particular, this methodology is used to derive new
  necessary conditions for keyed authentication over discrete
  memoryless channels and to establish the capacity region of the
  wiretap channel, subject to finite leakage and finite error, under
  two different secrecy metrics.  These examples establish the
  usefulness of the proposed methodology.
\end{abstract}

%\begin{IEEEkeywords}
%\end{IEEEkeywords}

\section{Introduction}

Consider arbitrary \emph{discrete random variables} (DRVs)
$(M,\mbf{X},\mbf{Y})$, which form a Markov chain in that order, where
$\mbf{X} = (X_1,X_2,\dots,X_n)$, $\mbf{Y} = (Y_1,Y_2,\dots,Y_n)$, and
\begin{align*}
\Pr \left( \mbf{Y} = \mbf{y} \middle|\mbf{X} = \mbf{x}\right) 
&= \prod_{i=1}^n \Pr \left( Y_i = y_i \middle| X_i = x_i \right) \\
%&= \prod_{i=1}^n \Pr \left( Y = y_i \middle| X = x_i \right).
&= \prod_{i=1}^n p_{Y|X}\left(y_i \middle|x_i \right)
\end{align*}  
for some conditional distribution $p_{Y|X}$.  DRVs of this nature are
often found in the literature related to the information theory,
starting when Shannon~\cite{shannon1948mathematical} considered one
way communication over a \emph{discrete memoryless channel} (DMC)
where $M$ represents the message, $\mbf{X}$ the output of the channel
encoder, and $\mbf{Y}$ the input to the channel decoder. The purpose
of this work is to provide a DRV $U$ such that
\begin{itemize}
\item the cardinality of its alphabet, $|\mcf{U}|$, grows
  sub-exponentially with $n$,
\item $U,\mbf{X},\mbf{Y}$ form a Markov chain in that order, and
\item $U$ induces information stability
(see~\cite{dobrushin1963general,pinsker1960information}).
\end{itemize}
The last property above means that for probability that converges to
unity with $n$, a $u \in \mcf{U}$ chosen randomly according to $U$
will exhibit
% \begin{align}
% -\logt p_{\mbf{Y}|M,U} \left( \mbf{Y}  \middle| M,u \right) \rightarrow \mathbb{H}(\mbf{Y}|M,U=u) \text{ (in probability)}
% \end{align}
% and
% \begin{align}
% -\logt p_{\mbf{Y}|U} \left( \mbf{Y}|u  \right) \rightarrow \mathbb{H}(\mbf{Y}|U=u) \text{ (in probability)}
% \end{align}
convergence in probability of
\begin{align*}
  -\logt p_{\mbf{Y}|U} \left( \mbf{Y}|u \right) 
&\rightarrow
\mathbb{H}(\mbf{Y}|U=u) \\
-\logt p_{\mbf{Y}|M,U} \left( \mbf{Y} \middle| M,u \right)
&\rightarrow
\mathbb{H}(\mbf{Y}|M,U=u)
\end{align*}
where
%$p_{\mbf{Y}|M,U}(\mbf{y}|m,u) = \Pr \left( \mbf{Y} = \mbf{y}\middle| M
%  = m, U = u \right)$ 
\begin{align*}
\mathbb{H}(\mbf{Y}|U = u) 
&= - \sum_{\mbf{y}} p_{\mbf{Y}|U}(\mbf{y}|u)
\logt p_{\mbf{Y}|U}(\mbf{y}|u) \\
\mathbb{H}(\mbf{Y}|M,U=u) 
&= - \sum_{m,\mbf{y}}
p_{\mbf{Y},M|U}(\mbf{y},m|u) \logt p_{\mbf{Y}|M,U}(\mbf{y}|m,u)
\end{align*}
are conditional entropies. The construct of $U$ is similar to that of
a DRV that determines the empirical distribution, or type as defined
in~\cite[Chap.~2]{CK}, of $\mbf{X}$.

Such a property proves to be extremely useful in establishing
information-theoretic necessary conditions directly from operational
requirements. To understand the importance of such a capability,
consider Fano's inequality~\cite{fano1952class}, which states that
given DRVs $M,\hat M$ and $\epsilon \in (0,1)$, if
$\Pr(M = \hat M) < \epsilon$ then
\begin{align*}
\mathbb{H}(M|\hat M) \leq \epsilon \logt |\mcf{M}| + \mathbb{H}(B_{\epsilon}),
\end{align*}
where $B_{\epsilon}$ is a Bernoulli random variable with parameter
$\epsilon$. A typical application of Fano's inequality generally views
the DRV $M$ and $\hat M$ as respectively being a message sent at the
transmitter and an estimate of the message made at the receiver of a
communication system. Thus Fano's inequality gives us an upper bound
on $\mathbb{H}(M|\hat M)$ from the operation requirement of
maintaining a small transmission error probability, i.e.,
$\Pr(M = \hat M) < \epsilon$.
% Because any communication system requires the reconstruction and the
% original to be equal with high probability (i.e.,
% $\Pr(M = \hat M) < \epsilon$), Fano's inequality gives us upper bounds
% on $\mathbb{H}(M|\hat M)$ which apply to all operational communication
% systems.
While no one would argue against the utility of Fano's inequality, it
is clear that it can only provide a bound for one specific operational
requirement, namely a small transmission error probability. In
contrast, inducing information stability can allow for a replacement
of stochastic terms with information theoretic averages, directly in
the operational quantities. Thus, such a method is applicable to all
operational requirements that can be written as functions of
distributions of DRVs involved. %probability space.
To demonstrate this methodology we have provided three different
examples: one way communication over a discrete memoryless channel;
tighter bounds on the probability of intrusion in a generalization of
a problem introduced by Lai et al.~\cite{lai2009authentication}; and
establishing the capacity of the wire-tap channel with finite error
and leakage under two different secrecy metrics. These problems are
chosen as to present a wide-range of operational requirements for
which this new methodology can extract information theoretic necessary
conditions. Furthermore, while our first example is chosen simply to
present the reader with a well-studied problem in information theory,
the second and third examples establish new results.

The rest of the paper is organized as follows. First we conclude the
introduction by describing the notation used through the rest of the
paper in Section~\ref{sec:notation}, particular attention should be
given to the definition of a regular collection of DRVs
(Definition~\ref{def:rvreq}) which defines where the theorems can be
applied. Following this, we highlight relevant work in
Section~\ref{sec:background}. We then present our main results in
Section~\ref{sec:preview}, and applications thereof in
Section~\ref{sec:applications}.  The proofs of each main theorem is
given its own unique treatment from Sections~\ref{sec:mt}
to~\ref{sec:thm:mt_aug:2}.  This is done so that we may first present a
suite of lemmas which serve to reduce the complexity of their
proofs. Conclusions are found in section~\ref{sec:conclusion}, and
some miscellaneous proofs are found in the appendix.

\subsection{Notation}\label{sec:notation}
Constants, random variables (RVs), and sets will be denoted by lower
case, upper case and script letters respectively. Function
$\Pr(\cdot)$ returns the probability of the event in the predicate. We
will always employ the corresponding script form of a letter to denote
the support set of any DRV. That is, if $X$ is a DRV, then $\mcf{X}$
is the set of all $x$ for which $\Pr(X = x) >0$. Functions will be
lower case or upper case depending on if they are random or
not. Conditional DRVs and events will be denoted by $|$, for example
the DRV $X$ given the event $\{Y=y\}$ is written $X|\{Y=y\}$.

The set of positive integers is written as $\mathbb{N}_+$, and the set
of positive real numbers is written as $\mathbb{R}_+$. Furthermore,
$\left[i: j \right]$ denotes the set of integers starting at $i$ and
ending at $j$, inclusively. We use $\bigotimes$ to denote collections
of constants, DRVs, etc.  For instance the collection of three DRVs
$(X_1,X_2,X_3) = \bigotimes_{i=1}^3 X_i$. Throughout this paper
$\mbf{X} \defn \bigotimes_{i=1}^n X_i$. That is, $\mbf{X}$ denotes a
sequence of $n$ possibly mutually dependent DRVs, and $\mbf{x}$
denotes a sequence $n$ constants all from $\mcf{X}$.  Note that we
have omitted the dependence on $n$ for simpler notation, and will
continue to do so for the rest of the paper unless when it is
necessary to highlight the dependence.  The support set of $\mbf{X}$
is clearly a subset of $\mcf{X}^n \defn \bigotimes_{i=1}^n \mcf{X}$.
Also when $ X = \emptyset$, by convention this defines $X$ as some
unspecified constant.  From here-forth, we will only rarely need to
refer back to the individual elements in the $n$-length sequences of
$\mbf{x}$ and $\mbf{y}$. As such, the subscripts of DRVs will be
primarily used to denote a collection of multiple $n$-length DRV
sequences, such as
$\mbf{Y}_{[1:i]}\defn \bigotimes_{j=1}^{i} \mbf{Y}_j$, for some
$i \in \mathbb{N}_+$.

% We now take a brief aside to discuss notation for probability
% distributions. 
Probability distributions, being deterministic functions over their
support sets, will be denoted with lower case letters. Of particular
importance will be $p$, which will always denote a probability
distribution, and when written with the subscript of DRVs,
specifically denotes the associated probability distribution over said
random variables. For instance, $p_{X|Y}(x|y) \defn
\Pr(X=x|Y=y)$. With this notation $p_{X|Y}(X|Y)$ is itself a RV, while
$p_{X|Y}(x|y)$ is a fixed value.  When the context is clear, we may
drop the subscript entirely. Furthermore,
$p_{X}(\mcf{A}) = \sum_{x \in \mcf{A}} p_{X}(x)$ for any
$\mcf{A} \subseteq \mcf{X}$. The set of all possible conditional
distributions of the form $w(y|x)$, where $y \in \mcf{Y}$ and
$x \in \mcf{X}$, is denoted $\mcf{P}(\mcf{Y}|\mcf{X})$. For DRVs
$\mbf{Y}$ and $\mbf{X}$ if
$p_{\mbf{Y}|\mbf{X}}(\mbf{y}|\mbf{x}) = \prod_{i=1}^n p_{Y_1|X_1}
(y_i|x_i)$, we will write
$p_{\mbf{Y}|\mbf{X}}(\mbf{y}|\mbf{x}) = p_{Y|X}^n (\mbf{y}|\mbf{x})$
or when clear
$p_{\mbf{Y}|\mbf{X}}(\mbf{y}|\mbf{x}) = p^n (\mbf{y}|\mbf{x})$. The
empirical conditional distribution of $\mbf{y}|\mbf{x}$ is defined as
$p_{\mbf{y}|\mbf{x}}(a|b) \defn \frac{\abs{i \in [1:n]:(y_i,x_i) =
    (a,b) } }{\abs{i \in [1:n]: x_i = b} }$ for
$(a,b) \in \mcf{Y}\times \mcf{X}$. The set of all valid empirical
distributions for an $n$-length sequence will be denoted
$\mcf{P}_n$. For empirical conditional distributions we shall use
$\mcf{P}_n(\mcf{Y}|p)$ where $p \in \mcf{P}_n(\mcf{X})$, to denote the
set of conditional empirical distributions $w$ for which $w(y|x) p(x)$
is a valid distribution in $\mcf{P}(\mcf{Y},\mcf{X})$.

Many of the results to be presented in later sections involves DRVs
that satisfy specific sets of relationship and/or properties. For
relationships between DRVs in particular, we will use the following
two operators. First if $X \markov Y \markov Z$, then DRVs $X,Y,Z$
form a Markov chain in that order. In other words
$p_{X,Y,Z}(x,y,z) = p_{X}(x) p_{Y|X}(y|x) p_{Z|Y}(z|y)$ for all
$(x,y,z) \in \mcf{X}\times \mcf{Y} \times \mcf{Z}$. On the other hand,
if $X \gg Y$, then $Y$ can be written as a deterministic function of
$X$. For any DRVs $X,Y,Z$, if $X \gg Y$ then $Y \markov X \markov Z$.
To simplify the statements of our results, we will adopt the standard
set notation when describing DRVs satisfying a specific set of
properties. For instance, the DRVs $U, X, Y$ that satisfy the
conditions that $\abs{\mcf{U}} \leq n$ and that
$U \markov X \markov Y $ will be denoted by
$(U, X, Y) : \{ \abs{\mcf{U}} \leq n,~U \markov X \markov Y \}$.

Information-theoretic quantities which are averages over probability
distributions of DRVs will be denoted by blackboard bold letters.
%(this notation has been used once before for positive integers and positive real numbers). 
In specific, the following quantities will receive heavy use: \\
For DRVs $U,X,Y,Z$, and probability distributions
$w, \hat w, \tilde w \in \mcf{P}(\mcf{Y}|\mcf{X})$ and
$p \in \mcf{P}(\mcf{X})$,
\begin{align*}
\mathbb{H}_{u}(X|Z) &= - \sum_{(x,z) \in \mcf{X}  \times \mcf{Z}}
                      p_{X,Z|U}(x,z|u) \logt p_{X|Z,U}(x|z,u) \\
\mathbb{I}_{u}(X;Y|Z) &= \mathbb{H}_{u}(X|Z) - \mathbb{H}_{u}(X|Y,Z) \\
\mathbb{D}(w ||\tilde w | p ) &= \sum_{x,y} w(y|x) p(x) \logt \frac{w(y|x)}{\tilde w(y|x)}\\
\mathbb{D}_{\hat w} (w||\tilde w| p ) &=\sum_{y,x} \hat w(y|x) p(x) \logt \frac{ w(y|x) }{\tilde w(y|x)} \\
&= \mathbb{D}(\hat w || w|p) - \mathbb{D}(\hat w ||\tilde w|p).
\end{align*}
It should be noted that while $\mathbb{H}(X|Z,U)$ is a constant,
$\mathbb{H}_{U}(X|Z)$ is a RV. More specifically $\mathbb{H}(X|Z,U)$
is the expected value of $\mathbb{H}_{U}(X|Z)$ over $U$.  Moreover we
will employ the concept of entropy spectrum~\cite{han2003} in the
development of some of our results. More specifically, we will mostly
consider the \emph{entropy spectrum frequency} of $\mbf{y}|\mbf{x}$,
which is defined as
\begin{equation*}
h_{\mbf{Y}|\mbf{X}}(\mbf{y}|\mbf{x}) \defn - \logt
p_{\mbf{Y}|\mbf{X}}(\mbf{y}|\mbf{x} ).
\end{equation*} 
The conditioning notation will be omitted in the special cases where
$\mbf{X} = \emptyset$. Furthermore, the subscript may be omitted when
the context is clear. Finally, we note that the exact bounds obtained
in this paper quickly become unwieldy. This is unfortunate because
this detracts from the elegance of the stated results. As a
compromise, we introduce the following order terminology which is
similar in spirit to Bachmann-Landau notation, but has a formal
definition which has to be context sensitive.
\begin{define}
For any $\epsilon \in \mathbb{R}_+$, we say $f(\epsilon) =
O(g(\epsilon))$ if there exists a constant $c \in  \mathbb{R}_+$ (that
is possibly a function of the cardinalities of the alphabets involved) such that 
\begin{align*}
|f(\epsilon)| \leq c |g(\epsilon)|.
\end{align*} 
\end{define}
% This differs from the standard Bachmann--Landau notation in that
% there is not necessarily a value going to infinity.
Throughout the paper our results will be expressed in terms of
$O(g(\epsilon))$, for some $g : \epsilon \rightarrow \mathbb{R}$, with
the value of acceptable $\epsilon$ being itself a function of
$n$. 
% Because of this, all order terms in this paper can be verified in
% the following way. Given $\epsilon \in (f_{-}(n),f_{+}(n)$,
% $f_-(n) : \mathbb{N} \rightarrow \mathbb{R}$,
% $f_+ (n) : \mathbb{N} \rightarrow \mathbb{R}$, if
% \begin{align}
% \max_{n\in \mathbb{N}}  \max_{\epsilon \in (f_{-}(n),f_{+}(n)))} \frac{ f(\epsilon) }{g(\epsilon)} < c,
% \end{align}
% for some $c \in \mathbb{R}$, then $f(\epsilon) = O(g(\epsilon))$. 
The exact calculations of the order terms are cumbersome and trivial,
and we will skip most of such calculations except a few particularly
important ones.

%\subsection{When the theorem} \label{sec:concepts}
Now, we restrict the DRVs that our main theorems are applicable
to.
%Requirement~\ref{def:rvreq} if
\begin{define} \textbf{(Regular collection of DRVs)}
\label{def:rvreq}
For any arbitrary index set $\mcf{W}$ and any $l \in \mathbf{N}_+$,
DRVs $(M_{[1:l]},\mbf{X},\mbf{Y}_{\mcf{W}})$ form a \emph{regular
  collection} if
\begin{itemize}
\item $|\mcf{X}|$ and $\sup_{w \in \mcf{W}}|\mcf{Y}_{w}|$ are finite,
\item $|\mcf{X}| \geq 2$ and $|\mcf{Y}_{w}| \geq 2$ for all $w \in
  \mcf{W}$,
\item $M_{[1:l]} \markov \mbf{X} \markov \mbf{Y}_{\mcf{W}}$,
\item $\mbf{Y}_{w} | \mbf{X}$ is distributed $p^n_{Y_w|X}$, where
  $p_{Y_w|X} \in \mcf{P}(\mcf{Y}_w|\mcf{X})$ for all $w \in \mcf{W}$,
  and
\item $n \geq 27$.
\end{itemize}
\end{define}
Furthermore, to simplify notation we assume that $\mcf{Y}_w = \mcf{Y}$
for all $w \in \mcf{W}$, and when
$\mcf{W}\subset \mcf{P}(\mcf{Y}|\mcf{X})$ we will assume that
\begin{align*}
p_{\mbf{Y}_{w}|\mbf{X}}(\mbf{y}|\mbf{x}) = w^n(\mbf{y}|\mbf{x}). 
\end{align*}
Note neither of these assumptions are in the least restrictive given
the first requirement of the definition.

%Such DRVs occur frequently the study of information theory, as they
%form the basis for \emph{discrete memoryless channels} (DMC) and rate
%distortion problems, when the distortion function is applied symbol by
%symbol. For visualization, it is generally helpful to consider the
%$M_{[1:l]}$ a collection of $l$ different messages that are to be
%transmitted over a collection of $|\mcf{W}|$ different DMC $p_{Y_w|X}$.

\section{Background}\label{sec:background}

\subsection{Images and Quasi-images}\label{sec:background_image}

The manipulation of images and quasi-images will play an important
role in establishing our theorems. Let us define these concepts. For
all discussions and results in this section, it is assumed that
$(\emptyset, \mbf{X}, \mbf{Y})$ is a regular collection of DRVs.
\begin{define}
  \textbf{(\cite[Ch.~15]{CK})} Let
  $p_{Y|X} \in \mcf{P}(\mcf{Y}|\mcf{X})$. For any $\eta \in (0,1)$, a
  set $\mcf{B} \subseteq \mcf{Y}^n$ is called an \emph{$\eta$-image}
  of $\mcf{A}\subseteq \mcf{X}^n$ (generated) by $p_{Y|X}$ if
\begin{align*}
p_{Y|X}^n(\mcf{B}| \mbf{x}) \geq \eta, ~\forall \mbf{x} \in \mcf{A}.
\end{align*} 
Furthermore $g^n_{Y|X}(\mcf{A},\eta)$ denotes the minimum cardinality
(size)
of $\eta$-images of $\mcf{A}$ by
$p_{Y|X}\in \mcf{P}(\mcf{Y}|\mcf{X})$. That is,
\begin{align*}
g^n_{Y|X}(\mcf{A},\eta) = \min_{\mcf{B}\subseteq \mcf{Y}^n :\mcf{B}
  \text{ is an $\eta$-image of $\mcf{A}$ by $p_{Y|X}$}} |\mcf{B}|.
\end{align*}
\end{define}
\begin{define}
  \textbf{(\cite[Problem~15.13]{CK})} Let
  $p_{Y|X} \in \mcf{P}(\mcf{Y}|\mcf{X})$. For any $\eta \in (0,1)$, a
  set $\mcf{B} \subseteq \mcf{Y}^n$ is called an \emph{$\eta$-quasi
    image} of $\mbf{X}$ by $p_{Y|X}$ if
\begin{align*}
\sum_{\mbf{x}} p_{Y|X}^n(\mcf{B}| \mbf{x})p_{\mbf{X}}(\mbf{x}) \geq \eta.
\end{align*} 
Furthermore $\bar g^n_{Y|X}(\mbf{X},\eta)$ denotes the minimum
cardinality (size) of $\eta$-images of $\mbf{X}$ by
$p_{Y|X}\in \mcf{P}(\mcf{Y}|\mcf{X})$.
\end{define}

Image sizes were originally introduced in G{\'a}cs and
K{\"o}rner~\cite{gacs73} and Ahlswede et al.~\cite{Ahlswede76image},
and found use in proving strong converses due to the blowing up lemma,
which the authors of~\cite{gacs73} and ~\cite{Ahlswede76image} credit
to Margulis~\cite{Margulis74}. In our paper's context, the blowing up
lemma will play an important role because of how it relates image
sizes. Before pointing out the lemmas which will find use in this
paper, we refer readers to
~\cite[Chap.~5]{CK},~\cite{marton1986simple}
and~\cite[Chapter~3]{Raginsky12} for an information theoretic context
of the blowing up lemma.
\begin{lemma}
\label{lem:6.6}
\textbf{(\cite[Lemma~6.6]{CK})}
Given $\mcf{X}$, $\mcf{Y}$, $\alpha \in (0,1)$, and $\beta\in (0,1-\alpha]$, there exists $\tau_n : \mathbb{R}_+ \times \mathbb{R}_+ \rightarrow \mathbb{R}_+$, where $\lim_{n \rightarrow \infty} \tau_n(\alpha,\beta) = 0$ such that
\begin{align*}
0 \leq \frac{1}{n} \logt \frac{\g{Y|X}{\mcf{A}}{1-\beta}}{ \g{Y|X}{\mcf{A}}{\alpha}}
\leq  \tau_n(\alpha,\beta)
\end{align*}
for every $\mcf{A} \subseteq \mcf{X}^n$, and every distribution $p_{Y|X} \in \mcf{P}(\mcf{Y}|\mcf{X})$.
\end{lemma}
While it is possible to derive the theorems in
Section~\ref{sec:preview} directly from Lemma~\ref{lem:6.6}, we take
the further step now of providing an upper bound on
$\tau_n(\alpha,\beta)$. This can be done from combining a lemma which
discusses the change in probability given a blow up (see Liu et
al.~\cite{liu2017information} or\footnote{The lemma from Raginsky and
  Sason provides the same order for $\tau_n$ as can be obtained
  via~\cite[Lemma~5.3,5.4]{CK}, but is a little sharper, and much
  simpler to present.} Raginsky and
Sason~\cite[Lemma~3.6.2]{Raginsky12}), with an upper bound on the
increase in the image size due to the blow up (see Ahslwede et
al.~\cite[Lemma~3]{Ahlswede76image} or Csisz{\'a} and
K{\"o}rner~\cite[Lemma~5.1]{CK}).
\begin{lemma}
\label{lem:tau}
%\textbf{(\cite[Lemma~3]{Ahlswede76image}+\cite[Lemma~3.6.2]{Raginsky12})}
For any $\alpha \in (0,1)$ and $\beta\in (0,1-\alpha]$, we have
\begin{align*}
\tau_n(\alpha,\beta) &\leq  \mathbb{H}\left(B\right)  + \frac{\sqrt{-\ln \beta} + \sqrt{-\ln \alpha}  }{\sqrt{2n} }  \logt|\mcf{Y}|
\end{align*}
where $B$ is a Bernoulli DRV with parameter $\frac{\sqrt{-\ln \beta} + \sqrt{-\ln \alpha}  }{\sqrt{2n}}$. 
\end{lemma}
\begin{remark}
  The value of $\tau_n$ will play a pivotal role in the bounds to
  come. In fact a tighter bound on the value of $\tau_n$ would
  directly lead to tighter bounds for multiple theorems in this
  paper. Because of this, we feel it necessary to bring forth recent
  work by Liu et al.~\cite{liu2017information,liubeyond} who endeavor
  to provide an alternative to the blowing-up lemma which offers
  tighter bounds for certain information theoretic problems. By using
  functional inequalities and the reverse hypercontractivity of
  particular Markov semigroups instead of the blowing up lemma, they
  have been able to obtain order tight bounds on the hypothesis
  testing problem. While hypothesis testing does not directly extend
  to determining minimum image and quasi-image sizes, it is clear that
  two problems are closely related. In specific the geometrical
  interpretations of their work may lead to further insight which
  allow for an improvement in the $\tau_n$ term.
\end{remark}

In terms of applications Ahlswede~\cite{Ahlswede76} used the blowing
up lemma to prove a local strong converse for maximal error codes over
a two-terminal DMC, showing that all bad codes have a good subcode of
almost the same rate.
% For any set of messages with decoding error $\epsilon$, he showed
% that by randomly removing a portion of the messages for the message
% set and then blowing up the decoding sets you could get a large
% decrease in probability of error without a large decrease in the
% normalized $\log$ of the number of messages.
Using the same lemma, K{\"o}rner and Marton~\cite{KM77} developed a
general framework for determining the achievable rates of a number of
source and channel networks. On the other hand, many of the strong
converses for some of the most fundamental multi-terminal DMCs studied
in literature were proven using image size characterization
techniques. K{\"o}rner and Martin~\cite{Korner77} employed such a
technique to prove the strong converse of a discrete memoryless
asymmetric broadcast channel. Later Dueck~\cite{dueck1981strong} used
these methods, combined with an ingenious ``wringing technique'' to
prove the strong converse of the discrete memoryless multiple access
channel with independent messages.

\subsection{Other works of interest}

Here we wish to briefly highlight a few of the methods by which
information theoretic necessary conditions are generally obtained,
first and foremost being Fano's
inequality~\cite{fano1952class}. Fano's inequality and generalizations
(for instance, Han and Verd{\'u}~\cite{han1994generalizing}), directly
provide information theoretic necessary conditions from probability of
error requirements. One significant problem is that it requires that
the error probability go to zero with $n$ in order to obtain tight
bounds in certain scenarios. One such scenario is establishing bounds
on the number of message that can be reliably distinguished in one-way
communication over a DMC, which we discuss in more detail in
Section~\ref{sec:applications}. While, as first claimed by Shannon and
proven by Wolfowitz~\cite{wolfowitz1957}, this value does not change
when allowing a finite error probability, the bound obtained from
Fano's inequality does increase with the error term.

Actually this allowed Wolfowitz to introduce the concept of a capacity
dependent upon error, usually denoted by $c(\epsilon)$. Because of
this there exists a demarcation between converses which are primarily
independent of the error rate, and those which are tight only if the
probability of error vanishes. Following the terminology of
Csisz{\'a}r and K{\"o}rner~\cite[Pg.~93]{CK}, a converse result
showing $c(\epsilon) = \lim_{\epsilon' \rightarrow 0} c(\epsilon')$
for all $\epsilon \in (0,1)$ is called a \emph{strong
  converse}. Verd{\'u} and Han~\cite{verdu94} showed the stronger
assertions that this is true for all finite $n$, and that all rates
larger must have error probability approaching unity hold for all
two-terminal DMCs. More recently techniques such as the meta-converse
by Polyanskiy et al.~\cite{polyanskiy2010channel} have been able to
establish tight necessary condition as function of error probability
up to the second order. The meta-converse leverages the idea that any
decoder can be considered as a binary hypothesis test between the
correct codeword set and the incorrect codeword set. Bounding the
decoding error by the best binary hypothesis test, new bounds, which
are relatively tight even for small values of $n$, can be established.

Thus for the single operational requirement of transmission error
probability, multiple different methodologies have been derived in
order to obtain increasingly strong results. While each of these
methodologies can be applied to different channels, they all still
require the probability-of-error operational requirement as a starting
point. The only general methodology that transcends this limitation
are those related to the information spectrum as first defined by
Verd{\'u} and Han~\cite{verdu94}. For an in depth treatment of
information spectrum methods, we point the reader to Han's
book~\cite{han2003}. The information-spectrum methods, in general,
link operational quantities directly to information/entropy spectrum
frequencies. Hence solving extremal problems of the information
spectrum in turn determines the fundamental limits of these
operational quantities. These methods are incredibly strong and
universally applicable, but generally can not easily relate back to
the more traditional information theoretic quantities like entropy and
mutual information. Our work takes this further step, but at the cost
of having to restrict our attention to DRVs that form a regular
collection. %satisfy Requirement~\ref{def:rvreq}.
Since such DRVs are of most common use, we feel this trade-off is one
worth pursuing, because many operational requirements, in addition to
the transmission error probability, are of recent interest.

\section{Main results} \label{sec:preview} 

Given a regular collection $(M_{[1:l]},\mbf{X},\mbf{Y}_{\mcf{W}})$,
our primary goal is to ``stabilize'' $\mbf{Y}_{w}$, when conditioned
on $M_{j}$, where $j\in [1:l]$, in the sense that the entropy spectrum
of $\mbf{Y}_w|M_j$ is concentrated around a single frequency. More
precisely, we want
\begin{align*}
\Pr \left( | h_{\mbf{Y}_w|M_j}(\mbf{Y}_w|M_j) - c  |  > n \epsilon_n \right) <  \delta_n 
\end{align*}
for some $c \in \mathbb{R}_+$, and $\epsilon_n$ and $\delta_n$ that vanish
with increasing $n$. It is easy to see that a statement such as the
above is not true in general. But, as we will show it can be achieved by
introducing a particular stabilizing DRV. This stabilization will allow for
the direct exchange of probabilities and entropy terms, thanks to the
following lemma.
\begin{lemma}
\label{lem:daco entropy}
Given DRVs $(Y,U)$, $\mu \in \left(0,\frac{1}{2} \right)$ and
$\epsilon \in \mathbb{R}_+$, if
\begin{align*}
\Pr \left( \abs{h(Y|U)-c} >  \epsilon \right) <  \mu , 
\end{align*}
for some $c \in \mathbb{R}_+$, then
\begin{align*}
  \abs{\mathbb{H}(Y|U) - c } < \epsilon + \mu
  \logt\frac{|\mcf{Y}|}{\mu^2}.
\end{align*}
\end{lemma}
\begin{cor}
\label{cor:daco entropy}
Furthermore
\begin{align*}
  \Pr\left( |h(Y|U) -\mathbb{H}(Y|U)| 
  > 2\epsilon + \mu \logt\frac{|\mcf{Y}|}{\mu^2}  \right) < \mu.  
\end{align*}
\end{cor}
The lemma's proof can be found in Appendix~\ref{app:daco
  entropy}. Thus stabilizing $\mbf{Y}_w|M_j$ has the added benefit
that $p_{\mbf{Y}_w|M_j}(\mbf{y}_w|m_j)$ converges to
$2^{-\mathbb{H}(\mbf{Y}_w|M_j)}$ in probability for large $n$. From
this exchange, we can easily create new necessary conditions for
different information theoretic problems, as we demonstrate in
Section~\ref{sec:applications}.

In order to construct the information stabilizing random variable,
first for a given regular collection $(\emptyset,\mbf{X},\mbf{Y})$, we
find a subset $\mcf{A}^\dagger \subseteq \mcf{X}^n$ for which the
quasi-image of $\mbf{X}\,|\set{\mbf{X} \in \mcf{A}^\dagger}$ by a
specific $p_{Y|X} \in \mcf{P}(\mcf{Y}|\mcf{X})$ is stable.
\begin{theorem}
\label{lem:daco}
Given any regular collection $(\emptyset,\mbf{X},\mbf{Y})$ and any
$\alpha \in \left( \frac{\logt n}{n} , \frac{1}{8 \ln 2} \right)$,
there exists both a set $\mcf{A}^\dagger \subseteq \mcf{X}^n$, where
\begin{align}
p_{\mbf{X}}( \mcf{A}^\dagger) \geq \frac{1}{n}\logt \frac{n}{8},\label{eq:daco:property1}
\end{align}
and positive real numbers $\delta = O(-\sqrt{\alpha} \logt \alpha)$
and $r < 2n \logt|\mcf{Y}|$ such that
\begin{align}
\Pr \left(  \left| h(\mbf{Y}|U)  - r\right|  >  n\delta + h(U)
  \middle|  U=u \right)  <  3 \cdot 2^{-n\alpha}
  \label{eq:daco:property2}
\end{align}
for all $U: \{U \markov \mbf{X} \markov \mbf{Y}\}$ and
$u \in \mcf{U} : \{ \Pr \left( \mbf{X} \in \mcf{A}^\dagger\middle| U =
  u \right) = 1\}$.
\end{theorem}
The proof of Theorem~\ref{lem:daco} can be found in
Section~\ref{sec:daco}. Next we repeatedly use
Theorem~\ref{lem:daco} to continually carve out different sets which
induce stability. Thus directly building upon Theorem~\ref{lem:daco}
we construct the following theorem.
\begin{theorem}
\label{thm:mt} 
\textbf{(Information stabilizing partitions)} 
For any regular collection $(M_{[1:l]}, \mbf{X}, \mbf{Y}_{[1:k]})$ and
real number
$\alpha \in \left(\frac{\logt n}{n}, \frac{1}{8 \ln 2}\right)$, we
have
\begin{itemize}[leftmargin=*]
\item a DRV
$V: \left\{\begin{array}{c}\abs{\mcf{V}} \leq ( 2n^3 \logt |\mcf{Y}|  )^{lk}\\ V \ll
(\mbf{X},M_{[1:l]}) \end{array}\right\}$, \vspace{3pt}
\item a positive real number
$\delta = O(-\sqrt{\alpha} \logt \alpha)$, and
\item for each DRV
$U: \left\{ \begin{array}{c} U \gg V \\ (U,M_{[1:l]}) \markov \mbf{X} \markov \mbf{Y}_{[1:k]} \end{array} \right\},$
$i \in [1:k]$, and $j \in [1:l]$, there exists a set
$\mcf{U}_{i|j} \subseteq \mcf{U}$ such that
\begin{align}
p_U \left(\mcf{U}_{i|j} \right) &\geq 1 -  2^{-\frac{n}{2}  \logt \frac{n}{8} }, \label{eq:thm:mt:1}
\end{align}
and
\begin{align}
\hspace{-10pt}\Pr\left( | h(\mbf{Y}_i|M_{j}, U) - \mathbb{H}_{U}(\mbf{Y}_i|M_j) |  >  n\delta   + 3 h(U) \middle| U=u \right)  &\notag \\
&\hspace{-100pt}<  4 \cdot 2^{-n\alpha} , \label{eq:thm:mt:2} 
\end{align}
for all $u \in \mcf{U}_{i|j}$.
\end{itemize}
\end{theorem}

%\begin{remark}
%An immediate consequence of this theorem is that
%\begin{align}
%\Pr \left( | h(\mbf{Y}_i|M_{j}, U) - \mathbb{H}_{U}(\mbf{Y}_i|M_j) | > n \delta + 7 h(U) \right) &\notag \\
%&\hspace{-130pt} < 5\cdot 2^{-n\alpha} + 2^{-(n/6)  \logt n } .
%\end{align}
%\end{remark}
The proof can be found in Section~\ref{sec:mt}.  In and of itself,
Theorem~\ref{thm:mt} allows us a new methodology of providing
information theoretic necessary conditions for certain
problems. Still, the applicability of this methodology can be improved
by also stabilizing $M_{[1:l]}$.
\begin{theorem}
\label{lem:mt_aug:1}
For any DRVs $M_{[1:l]}$, positive integer $\psi$, and positive real
numbers $\rho \in [1,\infty)$, we have:
\begin{itemize}[leftmargin=*]
\item a DRV
  $Q: \left\{ \begin{array}{c}|\mcf{Q}| \leq (\psi+1)^l\\ Q \ll
      M_{[1:l]}\end{array} \right\},$ \vspace{1pt}
\item a real number $\beta = O(\rho + 2^{-\rho}\psi)$, and
\item for each DRV $U : \{ U \gg Q\}$ and $j\in [1:l]$, there exists
  sets $\mcf{U}_{j,\stab} \subseteq \mcf{U}$ and
  $\mcf{U}_{j,\sat} \subseteq \mcf{U}$ such that
\begin{align}
p_U\left(\mcf{U}_{j,\stab}  \cup \mcf{U}_{j,\sat}  \right) &\geq 1-
                                                             2^{-\rho}
                                                             ,  \label{eq:thm:mt_aug:1}
\end{align}
\begin{align}
 \Pr\left( |h(M_j|U) - \mathbb{H}_{U}(M_j)| >  \beta + 3\logt|\mcf{U}|   \middle| U =u \right)  &< 2^{-\rho} \label{eq:thm:mt_aug:2stab} 
\end{align}
for all $u \in \mcf{U}_{j,\stab}$, and 
\begin{align}
\Pr\left( h(M_j|U) < \psi - \beta - 3\logt|\mcf{U}|    \middle| U = u \right) & < 2^{-\rho}  \label{eq:thm:mt_aug:2sat} 
\end{align}
for all $u \in \mcf{U}_{j,\sat}$.
\end{itemize}
Furthermore, if $M_{j}$ is uniform over $\mcf{M}_j$, then 
\begin{align*}
p_U\left(\mcf{U}_{j,\stab} \right) \geq 1- 2^{-\rho}
\end{align*} 
and 
\begin{align}
 \Pr\left( \left|h(M_j|U) - \logt|\mcf{M}_j| \right| > \beta + 3
  \logt|\mcf{U}|  \middle| U = u \right)  &< 2^{-\rho} 
\label{eq:thm:mt_aug:2stabprime}
\end{align}
for all $j \in [1:l]$ and $u \in \mcf{U}_{j,\stab}$.
\end{theorem}
Notice that providing stability to $\mbf{Y}_i|M_j$, $\mbf{Y}_i$ and
$M_j$, also would then provide stability to $(\mbf{Y}_i,M_j)$ and
$M_j|\mbf{Y}_i$. Providing stability to $M_j|\mbf{Y}_i$ may be
instantly recognizable to the reader as stabilizing a message given an
observation.

The need of our second augmentation theorem arises from the fact that Theorem~\ref{thm:mt} cannot in and of itself simultaneously provide stable quasi images for all product distributions in $\mcf{P}(\mcf{Y}|\mcf{X})$. Indeed, the reason for this being that there are an infinite number of such distributions, with even the number of conditional empirical distributions possible for $n$ symbols growing polynomial with $n$. In turn then, the support set of $Q$ would have to grow exponentially with $n$, which is something which we are trying to avoid as it would make our results trivial. The following augmentation theorem rectifies this problem by providing a set $\mcf{\tilde P} \subset \mcf{P}(\mcf{Y}|\mcf{X})$ which if stabilized then guarantee stability for all $\mcf{P}(\mcf{Y}|\mcf{X})$. 

First, a quick point of emphasis. For the upcoming theorem, we begin
to adopt the notation outlined previously where
$\mbf{Y}_{\mcf{P}(\mcf{Y}|\mcf{X})} \defn \bigotimes_{ w \in
  \mcf{P}(\mcf{Y}|\mcf{X})} \mbf{Y}_{w}$ and $\mbf{Y}_{w}|\mbf{X}$ is
distributed $w^n(\mbf{y}|\mbf{x})$ for
$w \in \mcf{P}(\mcf{Y}|\mcf{X})$.
\begin{theorem}
\label{lem:mt_aug:2}
For any real number
$\epsilon \in \left( 
  \frac{4 |\mcf{X}||\mcf{Y}|}{n} \logt n , 1 \right)$,
there exists a subset
\begin{align*}
\mcf{\tilde P} \subseteq \mcf{P}(\mcf{Y}|\mcf{X}) : \left\{
  |\mcf{\tilde P} | \leq \left(|\mcf{Y}| \left(1 +  \left\lfloor
  \frac{4|\mcf{Y}|^2}{\epsilon} \right\rfloor \right) \right)
  ^{|\mcf{X}||\mcf{Y}|} \right\}
\end{align*}
with the following property: \\
Given a regular collection
$(M, \mbf{X}, \mbf{Y}_{\mcf{P}(\mcf{Y}|\mcf{X})})$, for each
$w \in \mcf{P}(\mcf{Y}|\mcf{X})$ there exists a
$\tilde w_{w} \in \mcf{\tilde P}$ such that if
\begin{align*}
\Pr \left( |h(\mbf{Y}_{\tilde w_w}|M) - \mathbb{H}(\mbf{Y}_{\tilde w_{w}}|M) | > n\delta   \right) &< 2^{-n\alpha}
\end{align*}
for some $\delta,\alpha \in \mathbb{R}_+$, then
\begin{align*}
\Pr \left( |h(\mbf{Y}_{ w}|M) - \mathbb{H}(\mbf{Y}_{w}|M) | > n \tilde \delta  \right) &< 2^{-n\epsilon} + 2^{-n(\alpha-\epsilon)} 
\end{align*}
where
\begin{align*}
\tilde \delta &= (2+ 2^{-n\epsilon} + 2^{-n(\alpha-\epsilon)}) ( \delta + \epsilon) \notag \\
&+ (2^{-n\epsilon} \!+\! 2^{-n(\alpha-\epsilon)}) \left[
  \logt|\mcf{Y}| \!- \!\frac{2}{n}\logt( 2^{-n\epsilon}
  \!+\!2^{-n(\alpha - \epsilon)}) \right].
\end{align*} 
\end{theorem}

At this point Theorems~\ref{thm:mt},~\ref{lem:mt_aug:1}, and~\ref{lem:mt_aug:2} represent the main breadth of our contribution. But, it is clear that these Theorems are somewhat unwieldy. To simplify this procedure we will essentially combine Theorems~\ref{thm:mt},~\ref{lem:mt_aug:1} and~\ref{lem:mt_aug:2} into a single corollary which simultaneously stabilizes $\mbf{Y}_{w} |M_{i}$ and $M_{i}$ for all $w \in \mcf{P}(\mcf{Y}|\mcf{X})$ and $i \in [1:l]$. Because of the tension between the accuracy of the stabilization, and the support set of the stabilizing random variable, we will construct the following corollary to only stabilize $m_i$ such that $- \logt p_{M_i}(m_i) < n^2$, with the remaining $m_i$ being contained in their own set. Similar corollaries can be obtained with less accuracy on the stabilization, but with a much larger range of stabilized values (e.g., stabilize all $m_i$ such that $-\logt p_{M_i}(m_i) < 2^{n}$). While such a trade off would be useful for scenarios such as ID coding, they would not be appropriate for the examples presented here.

In order to simplify analysis we introduce the following definition. 
\begin{define}
  For any regular collection
  $(M_{[1:l]},\mbf{X},\mbf{Y}_{\mcf{P}(\mcf{Y}|\mcf{X})})$ and DRV
  $U:\{(U,M_{[1:l]})\markov \mbf{X} \markov
  \mbf{Y}_{\mcf{P}(\mcf{Y}|\mcf{X})}\}$, the $\nu$-stable %characterized
  sets for $u \in \mcf{U}$, $w \in \mcf{P}(\mcf{Y}|\mcf{X})$, and
  $M_j$ are
\begin{align}
&\mcf{D}_{\stab,(M_j)}(u,w;\nu)  \notag\\
&\defn \set{  \begin{array}{r l} 
    (\mbf{y}_{w}, m_{[1:l]}) \in \mcf{Y}^n \times \mcf{M}_{[1:l]}: \hspace{30pt}&\\ 
     |h(\mbf{y}_{w}|m_j,u) - \mathbb{H}_u(\mbf{Y}_{w}|M_j)| &\leq  n\nu  \\ 
     |h(m_j|u) - \mathbb{H}_u(M_j)| &\leq n\nu  \\
     |h(m_j|u) - h(m_j)| &\leq n\nu  \end{array} } ,
\end{align}
while the the $\nu$-saturated %characterized 
sets for $u \in \mcf{U}$, $w \in \mcf{P}(\mcf{Y}|\mcf{X})$, and $M_j$ are
\begin{align}
&\mcf{D}_{\sat,(M_j)}(u,w;\nu)  \notag\\
&\defn \set{\begin{array}{r l} 
    (\mbf{y}_{w}, m_{[1:l]}) \in \mcf{Y}^n \times \mcf{M}_{[1:l]}: \hspace{30pt}&\\  
     |h(\mbf{y}_{w}|m_j,u) - \mathbb{H}_u(\mbf{Y}_{w}|M_j) |&\leq  n\nu  \\ 
       h(m_j|u) - n^2 &\geq -n \nu    \\
     |h(m_j|u) - h(m_j)| &\leq n\nu  \end{array} }.
\end{align}
If $M_j$ is uniform over $\mcf{M}_j$, replace $\mathbb{H}_u(M_j)$ in
the above with $\logt |\mcf{M}_j|$.
\end{define}
Now if
$(\mbf{y}_w, m_{[1:l]}) \in \mcf{D}_{\stab,(M_j)}(u,w;\delta) \cup
\mcf{D}_{\sat,(M_j)}(u,w;\delta) $ then
$p(\mbf{y}_w|m_j,u) \approx 2^{-\mathbb{H}_u(\mbf{Y}_w|M_jx) \pm n
  \delta}$. In addition, if $h(m_j|u) < n^2 - 2 n\delta$, then
$p(m_j|u) \approx p(m_j) \approx 2^{-\mathbb{H}_u(M_j) \pm n
  \delta}$. In that sense $\mcf{D}_{\stab,(M_j)}(u,w;\delta)$ and
$\mcf{D}_{\sat,(M_j)}(u,w;\delta)$ consists of the probability terms
which are well described by information theoretic
quantities. Combining
Theorems~\ref{thm:mt},~\ref{lem:mt_aug:1},~\ref{lem:mt_aug:2} and
Lemma~\ref{lem:only_lemma} allow us to establish the following result.
\begin{cor}
\label{cor:mt}
Let $\varepsilon_n \defn n^{-\frac{1}{|\mcf{X}||\mcf{Y}|+1}}$. For any
regular collection
$(M_{[1:l]},\mbf{X},\mbf{Y}_{\mcf{P}(\mcf{Y}|\mcf{X})})$ and any DRV
$T :\{ (T,M_{[1:l]}) \markov \mbf{X} \markov
\mbf{Y}_{\mcf{P}(\mcf{Y}|\mcf{X})}\}$, there exists:
\begin{itemize}[leftmargin=*]
\item a DRV $U :\left\{ \begin{array}{c} U \gg T \\ \logt|\mcf{U}| =
                          O( \logt|\mcf{T}| -  l\, n\, \varepsilon_n \logt \varepsilon_n ) \\ (U,M_{[1:l]}) \markov \mbf{X} \markov \mbf{Y}_{\mcf{P}(\mcf{Y}|\mcf{X})} \end{array} \right\}$,
\item positive real number $\nu_n = O\left(n^{-1} \logt|\mcf{T}| -l \sqrt{\varepsilon_n} \logt \varepsilon_n\right)$, and 
\item set $\mcf{\tilde U} \subseteq \mcf{U}$ such that 
\begin{align}
p_{U}(\mcf{\tilde U}) \geq 1 - O( l  2^{-n\varepsilon_n})
\end{align}
and for each $j \in [1:l]$ and $u  \in \mcf{\tilde U}$ either 
\begin{align}
 \inf_{w \in \mcf{P}(\mcf{Y}|\mcf{X})} \hspace{-8pt}  \Pr \left( (\mbf{Y}_w, M_{[1:l]}) \in \mcf{D}_{\stab,(M_j)}(U,w;\nu_n) \middle| U =u \right) &\notag \\
&\hspace{-120pt} \geq 1 - 8 \cdot 2^{-n\varepsilon_n}, \label{eq:cor:d1}
\end{align}
or
\begin{align}
 \inf_{w \in \mcf{P}(\mcf{Y}|\mcf{X})} \hspace{-8pt} \Pr \left( (\mbf{Y}_w, M_{[1:l]}) \in \mcf{D}_{\sat,(M_j)}(U,w;\nu_n) \middle| U =  u \right) &\notag \\
&\hspace{-120pt} \geq 1 - 8 \cdot 2^{-n\varepsilon_n}. \label{eq:cor:d2}
\end{align}
Furthermore if $M_j$ is uniform over $\mcf{M}_j$, then~\eqref{eq:cor:d1} holds. 
\end{itemize}
\end{cor}
The proof of which is in Appendix~\ref{app:p:cor:mt}. Note the error term is primarily due to the result holding simultaneously for all distributions in $\mcf{P}(\mcf{Y}|\mcf{X})$. If this term is of importance in a potential application, and if only a finite and fixed number of quasi-images need to be stabilized, then the order term can be improved by simply combining Theorem~\ref{thm:mt} and~\ref{lem:mt_aug:1}.

\section{Applications} \label{sec:applications}

In this section we will highlight a new methodology by which to obtain information theoretic necessary conditions. First we will apply this new methodology to a classical problem to highlight how it works, and how it differs from conventional approaches. In doing so we will provide extra commentary at each step in order that we make general application of the methodology plain.  Next, we apply this methodology to establish new results for channels which require authentication, and wire-tap channels. These examples were chosen in order to demonstrate how this new methodology obtains information theoretic necessary conditions from a wide range of operational requirements.

\subsection{One way communications over a DMC} 

Here we consider a classical problem in information theory, channel coding over a DMC $p_{Y|X}$. In this model a source wants to send a message $M$, which will be chosen at random according to some arbitrary distribution over $\mcf{M}$, to the destination. Connecting the source and destination is a DMC characterized by the conditional probability distribution $p_{Y|X} \in \mcf{P}(\mcf{Y}|\mcf{X})$. To facilitate communications, the source and destination ahead of time agree upon a ``code'' consisting of an encoder $F : \mcf{M} \rightarrow \mcf{X}^n$ and a decoder $\Phi : \mcf{Y}^n \rightarrow \mcf{M}$, both of which may be stochastic. 

For the code to be considered operational it must satisfy the
following error probability criterion for some pre-arranged
$\delta \in (0,1)$:
\begin{align}
%\Pr \left( \Phi(\mbf{Y}) \neq  M \middle| \mbf{X} = F(M) \right) < \delta, \label{eq:app:1:nec}
\Pr \left( \Phi(\mbf{Y}) \neq  M \right) < \delta. \label{eq:app:1:nec}
\end{align}
We note that the distribution of $\mbf{Y}$ is induced by $p^n_{Y|X}$
with $\mbf{X} = F(M)$.  Since $((M,\emptyset),\mbf{X},\mbf{Y})$ form a
regular collection of DRVs, we can apply Corollary~\ref{cor:mt}. Doing
so, will allow for us to directly transform~\eqref{eq:app:1:nec} into
a set of information theoretic necessary conditions. Before a
demonstration of this, we shall describe how one would apply Fano's
inequality to attempt to achieve the same task, and what the
shortcomings of doing so are.

\subsubsection{Fano's inequality}
Without a uniform distribution over $M$, Fano's inequality can only (essentially) provide 
\begin{align}
\mathbb{H}(M) < \mathbb{I}(\Phi(\mbf{Y}) ; M)  + \Pr \left( \Phi(\mbf{Y}) \neq  M \right) \logt|\mcf{M}|. \label{eq:app:1:fi1}
\end{align}
Now, if it were the case that $M$ was uniform over $\mcf{M}$, then~\eqref{eq:app:1:fi1} reduces to 
\begin{align}
\logt|\mcf{M}|  < \frac{1}{1-\Pr \left( \Phi(\mbf{Y}) \neq  M \right)} \mathbb{I}(\Phi(\mbf{Y}) ; M), \label{eq:app:1:fi2}
\end{align}
and if we were further to assume that $\Pr \left( \Phi(\mbf{Y}) \neq  M \right) \leq \delta_n$, for some $\delta_n \rightarrow 0$, then asymptotically we could say
\begin{align}
\logt |\mcf{M}| \leq \mathbb{I}(\Phi(\mbf{Y});M) + O(n\delta_n).  \label{eq:app:1:fi3}
\end{align}
These bounds can be further simplified using the data processing inequality, and single-letterization techniques to show
\begin{align}
\mathbb{I}(\Phi(\mbf{Y});M) \leq \mathbb{I}(\mbf{Y};\mbf{X}) \leq n \max_{p(x)} \mathbb{I}(Y;X), \label{eq:app:1:sl}
\end{align} 
which when substituted back into~\eqref{eq:app:1:fi3} yields
\begin{align}
\lim_{n \rightarrow \infty} n^{-1}\logt |\mcf{M}| \leq \max_{p(x)} \mathbb{I}(Y;X).
\end{align}
But notice the assumptions that had to be made to obtain this result. First we had to assume $M$ was uniform, and second we had to assume that the probability of error decay to zero with $n$. The second assumption has already been the subject of much study, leading to the eventual demarcation between the strong and weak converse. With this in mind we instead consider what happens when you void the first assumption. In fact, repeating these steps with a non-uniform $M$, gives
\begin{align}
\lim_{n \rightarrow \infty} n^{-1}H(M) \leq \max_{p(x)} \mathbb{I}(Y;X). \label{eq:app:1:finec}
\end{align}
Notice that Equation~\eqref{eq:app:1:finec} \emph{looks} like a sufficient condition as well, and actually is if $M$ is information stable. But, this condition is actually not sufficient for general $M$. Consider the following example to convince yourself of this fact. Let $M$, $\mcf{M} = [0: 2^{2 n \max_{p(x)} \mathbb{I}(Y;X) }]$, have the following distribution
\begin{align}
p(m) &= \begin{cases}
\frac{3}{4} &\text{ if } m = 0 \\
\frac{1}{4} 2^{-2 n \max_{p(x)} \mathbb{I}(Y;X)} &\text{ if } m \in \mcf{M}\setminus\set{0}
\end{cases}.
\end{align}
This $M$, on average, can not be reliably transmitted over the channel. To see this consider a case where any potential decoder is given the side information that determines whether $M =0$ or $M \neq 0$. When the decoder is informed that $M=0$, then clearly the probability of error of the decoder can be eliminated. On the other hand when $M\neq 0$, then the number of potential messages greatly exceeds capacity and as a result the probability of error must be close to $1$. This later fact being a by-product of the strong converse for the DMC. Thus, even with this side information, the best possible decoder could only obtain a minimum probability of error of just below $1/4$. At the same time though, it is easy to calculate 
\begin{align}
\mathbb{H}(M) = \frac{3}{4} \logt \frac{4}{3} + \frac{1}{2} + \frac{n}{2} \max_{p(x)} \mathbb{I}(Y;X),
\end{align}
which is less than $n \max_{p(x)} \mathbb{I}(Y;X)$ for large enough $n$ as long as $\max_{p(x)} \mathbb{I}(Y;X)>0$.  As a consequence Equation~\eqref{eq:app:1:finec} can not also provide a matching sufficient condition, or in other words, Equation~\eqref{eq:app:1:finec} only provides a loose necessary condition.

\subsubsection{Information stable partitions} \label{sec:app:1.1}
Now we move onto our methodology, which even without the assumption that $M$ is information stable, nor that $\Pr \left( \Phi(\mbf{Y}) \neq  M \right)  \rightarrow 0$ as a function of $n$, yields 
\begin{align}
\Pr \left( n^{-1} h(M) >  \max_{p(x)} \mathbb{I}(Y;X) + \zeta_n \right) < \delta + 2^{-n\zeta_n}, \label{eq:app:1:itnec}
\end{align}
for some $\zeta_n: \zeta_n \rightarrow 0$, as necessary to ensure~\eqref{eq:app:1:nec}. First shown by Han~\cite[Theorem~3.8.5~\&~3.8.6]{han2003}, Equation~\eqref{eq:app:1:itnec} is not only necessary, but also has a matching\footnote{With the usual asymmetry in the sign of the negligible terms.} sufficient condition. We briefly discuss the sufficiency before completing the example. Observe that there can only be $2^{n \max_{p(x)} \mathbb{I}(Y;X)}$ values of $m \in \mcf{M}$ such that $h(m) \leq  n \max_{p(x)} \mathbb{I}(Y;X)$. We will refer to this set of messages as the transmissible set. It would be simple to construct a reliable channel code for the transmissible set, while mapping all messages not from the transmissible set to some fixed codeword. As a result, if $M$ is chosen from the transmissible set, the probability of error would be near $0$, and otherwise $1$. Hence there exists a coding scheme for which the error probability converges to the probability $M$ is chosen from the transmissible set. This previous statement is essentially Equation~\eqref{eq:app:1:itnec}.

Returning to establishing the necessary conditions. In general, our
methodology looks to directly replace the operational requirement's
probability terms with information theoretic quantities. Here the
operational requirement (Equation~\eqref{eq:app:1:nec}) can be written
as
\begin{align}
\Pr \left( \Phi(\mbf{Y}) = M \right) = \sum_{\mbf{y},m} p_{\Phi|\mbf{Y}}(m|\mbf{y}) p_{\mbf{Y},M}(\mbf{y},m) > 1 - \delta. \label{eq:app:1:nec'}
\end{align}
Next, because $((M,\emptyset),\mbf{X},\mbf{Y})$ constitute a regular
collection of DRVs, there exists
\begin{itemize} 
\item  a DRV $U :\left\{ \begin{array}{c}  \logt|\mcf{U}| = O(-n\varepsilon_n \logt \varepsilon_n)\\   (U,M) \markov \mbf{X} \markov \mbf{Y}_{\mcf{P}(\mcf{Y}|\mcf{X})} \end{array} \right\}$,
\item  positive real number $\nu_n = O(-\sqrt{\varepsilon_n} \logt \varepsilon_n)$, and
\item  $\mcf{\tilde U} \subseteq \mcf{U}$ such that
\begin{align}
p_{U}(\mcf{\tilde U}) \geq 1 - O(  2^{-n\varepsilon_n}) \label{eq:app:1:uprob1}
\end{align}
such that
\begin{align}
\Pr \left( (\mbf{Y}, M) \notin \mcf{D}_{\stab,\emptyset}(U,p_{Y|X};\nu_n) \middle|U = u \right) 
&\notag \\
&\hspace{-80pt}
  <  8\cdot 2^{-n\varepsilon_n}, \label{eq:app:1:d0stab}
\end{align}
and either 
\begin{align}
\Pr \left( (\mbf{Y}, M) \notin \mcf{D}_{\stab,(M)}(U,p_{Y|X};\nu_n) \middle| U = u \right) 
&\notag \\
&\hspace{-80pt} 
< 8\cdot 2^{-n\varepsilon_n},\label{eq:app:1:dmstab}
\end{align}
or
\begin{align}
\Pr \left( (\mbf{Y}, M) \notin \mcf{D}_{\sat,(M)}(U,p_{Y|X};\nu_n) \middle| U = u \right)
&\notag \\
&\hspace{-80pt}  
   < 8 \cdot 2^{-n\varepsilon_n}, \label{eq:app:1:dmsat}
\end{align}
for all $u \in \mcf{\tilde U}$,
\end{itemize}
where $\varepsilon_n = n^{\frac{-1}{|\mcf{X}||\mcf{Y}|+1}}$ by
Corollary\footnote{With $T= \emptyset$}~\ref{cor:mt}. The set
$\mcf{D}_{\sat,\emptyset}(U,p_{Y|X};\nu_n)$ is not considered because
the random variable $\emptyset$ is trivially uniform by
convention. Introducing $U$ into the LHS of~\eqref{eq:app:1:nec'} via
the law of total probability yields
\begin{align}
\Pr \left( \Phi(\mbf{Y}) = M \right) 
& =
\sum_{u} \sum_{\mbf{y},m} p_{\Phi|\mbf{Y}}(m|\mbf{y})
  p_{\mbf{Y},M,U}(\mbf{y},m,u) \notag \\
&> 1 - \delta. \label{eq:app:1:isp1}
\end{align}
Now, let $\mcf{\tilde U}_{\stab} \subseteq \mcf{U}$ be the set of $u$
such that~\eqref{eq:app:1:dmstab} holds. For each $u \in \mcf{\tilde
  U}_{\stab}$, let 
\begin{align*}
&\hspace{-10pt} \mcf{D}_{+}(u,p_{Y|X};\nu_n) = \\
&
\mcf{D}_{\stab,\emptyset}(u,p_{Y|X};\nu_n) \cap
\mcf{D}_{\stab,(M)}(u,p_{Y|X};\nu_n).
\end{align*}
We are to establish that the terms for $u \in \mcf{\tilde U}_{\stab}$
and $(\mbf{y},m) \in \mcf{D}_+(u,p_{Y|X};\nu_n)$ dominate the sum
in~\eqref{eq:app:1:isp1}, and the contributions of all other terms to
the sum become negligible as $n$ increases. More specifically, we will
to show\footnote{The lower bound in~\eqref{eq:app:1:prune:1} will not
  be used in establishing the necessary condition. It is provided
  mainly to show convergence.}
\begin{align}
&\hspace{-10pt} 
  \sum_{u \in \mcf{\tilde U}_{\stab}}
  \sum_{(\mbf{y},m) \in \mcf{D}_+(u,p_{Y|X};\nu_n) } \hspace{-20pt}
  p_{\Phi|\mbf{Y}}(m|\mbf{y})
  p_{\mbf{Y},M,U}(\mbf{y},m,u) 
  \notag \\
&\leq   \Pr \left( \Phi(\mbf{Y}) = M \right) \notag \\
% \label{eq:app:1:prune:0}\\
& \leq 
  \sum_{u \in \mcf{\tilde U}_{\stab}}
  \sum_{(\mbf{y},m) \in \mcf{D}_+(u,p_{Y|X};\nu_n) } \hspace{-20pt}
  p_{\Phi|\mbf{Y}}(m|\mbf{y})
  p_{\mbf{Y},M,U}(\mbf{y},m,u) 
\notag \\
& \hspace{20pt} + O(2^{-n\varepsilon_n})
%+ \hspace{-20pt}\sum_{\substack{(\mbf{y},m,u): \\ u \in \mcf{\tilde U}_{\stab} \\ (\mbf{y},m) \in \mcf{D}_+(u,p_{Y|X};\nu_n)} }\hspace{-20pt} p_{\Phi|\mbf{Y}}(m|\mbf{y}) p_{\mbf{Y},M,U}(\mbf{y},m,u) 
  \label{eq:app:1:prune:1}
\end{align}
holds.

For the time being, let us first assume~\eqref{eq:app:1:prune:1} does
hold. Then it suffices to consider only $u \in \mcf{\tilde U}_{\stab}$
and $(\mbf{y},m) \in \mcf{D}_+(u,p_{Y|X};\nu_n)$, for which we have
stability by Corollary~\ref{cor:mt} as discussed above. That is,
$$\begin{array}{r l l l}
2^{-\mathbb{H}_u(\mbf{Y}|M) - n \nu_n}  &\leq &p(\mbf{y}|m,u) &\leq
                                                                2^{-\mathbb{H}_u(\mbf{Y}|M) + n \nu_n} \\
2^{-\mathbb{H}_u(\mbf{Y}) - n \nu_n}  &\leq &p(\mbf{y}|u) &\leq
                                                            2^{-\mathbb{H}_u(\mbf{Y}) + n \nu_n} \\
2^{-\mathbb{H}_u(\mbf{M}) - n \nu_n}  &\leq &p(\mbf{m}|u) &\leq
                                                            2^{-\mathbb{H}_u(\mbf{M}) + n \nu_n} \\
2^{-\mathbb{H}_u(\mbf{M}) - 2n \nu_n}  &\leq &p(\mbf{m}) &\leq
                                                           2^{-\mathbb{H}_u(\mbf{M}) + 2n \nu_n},
\end{array}$$
which imply
\begin{align}
p_{\mbf{Y},M,U}(\mbf{y},m,u)% \notag\\
%&= p_{\mbf{Y},U}(\mbf{y},u) \frac{p_{\mbf{Y}|M,U}(\mbf{y}|m,u) p_{M|U}(m|u) }{p_{\mbf{Y}|U}(\mbf{y}|u)} \\
%&
\leq p_{\mbf{Y},U}(\mbf{y},u) 
2^{-|\mathbb{H}_u(M) - \mathbb{I}_u(\mbf{Y};M) - 3n \nu_n |^+} \label{eq:app:1:notbloat}
\end{align}
for all $u \in \mcf{\tilde U}_{\stab}$ and
$(\mbf{y},m) \in \mcf{D}_+(u,p_{Y|X};\nu_n)$. It
is~\eqref{eq:app:1:notbloat} that allows us to substitute the
distribution terms in the necessary condition~\eqref{eq:app:1:isp1}
with the corresponding information theoretic terms. In particular,
putting~\eqref{eq:app:1:notbloat} into~\eqref{eq:app:1:prune:1} and
combining with the necessary condition in~\eqref{eq:app:1:isp1}
yields
\begin{align}
\sum_{u \in \mcf{\tilde U}_{\stab}}  p_{U}(u) \, 2^{-|\mathbb{H}(M|u) - \mathbb{I}(\mbf{Y};M|u) - 3n \nu_n  |^+}  
&\notag \\
&\hspace{-70pt} 
\geq 1 - \delta - O( 2^{-n\varepsilon_n}) . \label{eq:app:1:prune:0'''}
\end{align}

Continuing on, \eqref{eq:app:1:prune:0'''} also directly implies
\begin{align} 
\Pr\left( \begin{array}{c}  \mathbb{H}_{U}(M)  \leq \mathbb{I}_{U}(\mbf{Y};M) +3n \nu_n  + n\varepsilon_n  \\  U \in \mcf{\tilde U}_{\stab} \end{array} \right)& \notag \\
&\hspace{-110pt} >  1-  \delta - O(2^{-n\varepsilon_n}) \label{eq:app:1:fin}
\end{align}
via Markov's inequality. 
%At this point if $M$ were uniform then we would essentially be done since $\logt|\mcf{M}|$ would replace $\mathbb{H}_{U}(M)$ in Equation~\eqref{eq:app:1:fin}. As it so happens, Equation~\eqref{eq:app:1:fin} already implies 
%\begin{align} 
%\Pr\left( \mathbb{H}_{U}(M)  > \mathbb{I}_{U}(\mbf{Y};M) +3n \nu_n  + n\varepsilon_n  \right)& <   \delta + O( 2^{-n\varepsilon_n}), \label{eq:app:1:finb}
%\end{align} 
%which ends up being a form more conducive to describing transmissibility requirements for other channel models, for example the \emph{Broadcast channel with confidential communications} as shown in~\cite{graves2018transmitting}. Nevertheless we continue on to our stated goal. Since $M$ is not uniform we can instead proceed as follows
But also note that
\begin{align}
&\Pr\left( \begin{array}{c} \mathbb{H}_{U}(M)  \leq
             \mathbb{I}_{U}(\mbf{Y};M) +3n \nu_n  + n\varepsilon_n  \\
             U \in \mcf{\tilde U}_{\stab} \end{array} \right) \notag \\
%&\leq O(2^{-n\varepsilon_n}) \notag\\
%&\hspace{5pt} + \Pr\left( \begin{array}{l r l} & \mathbb{H}_{U}(M)  &\leq \mathbb{I}_{U}(\mbf{Y};M) +3n \nu_n  + n\varepsilon_n  \\ \bigcap & U &\in \mcf{\tilde U}_{\stab} \\ \bigcap & (\mbf{Y}_w,M) &\in \mcf{D}_+(U,p_{Y|X};\nu_n) \end{array} \right) \\
&\leq O( 2^{-n\varepsilon_n}) \notag\\
&\hspace{5pt} + \Pr\left(  \begin{array}{c} h(M) \leq \mathbb{I}_{U}(\mbf{Y};M) +5n \nu_n  + n\varepsilon_n  \\ U \in \mcf{\tilde U}_{\stab} \\ (\mbf{Y},M) \in \mcf{D}_+(U,p_{Y|X};\nu_n) \end{array} \right) \label{eq:app:1:rmvU0} \\
&\leq O( 2^{-n\varepsilon_n}) + \Pr\left( h(M) \leq \mathbb{I}_{U}(\mbf{Y};M) +5n \nu_n  + n\varepsilon_n \right) \label{eq:app:1:rmvU1}
\end{align}
where~\eqref{eq:app:1:rmvU0} is because
$h(m) \geq h(m|u) - n\nu_n \geq \mathbb{H}(M|u) - 2n\nu_n$ for each
$u \in \mcf{\tilde U}_{\stab}$ and
$(\mbf{y},m) \in \mcf{D}_+(u,p_{Y|X};\nu_n)$. In conclusion then,
combining~\eqref{eq:app:1:fin}, \eqref{eq:app:1:rmvU1}, and the fact
that $\mathbb{I}_{u}(\mbf{Y};M) \leq n \max_{p(x)} \mathbb{I}(Y;X)$
yields~\eqref{eq:app:1:itnec}, since
$(U,M) \markov \mbf{X} \markov \mbf{Y}$.

It remains to establish~\eqref{eq:app:1:prune:1}.  The lower bound
in~\eqref{eq:app:1:prune:1} is trivial. The upper bound
in~\eqref{eq:app:1:prune:1} can be obtained by bounding the sum over
three different sets of terms. First,
\begin{align}
\sum_{\substack{(\mbf{y},m,u): \\ u \in \mcf{U} \setminus \mcf{\tilde U}} } \hspace{-7pt} p_{\Phi|\mbf{Y}}(m|\mbf{y}) p_{\mbf{Y},M,U}(\mbf{y},m,u) &\!\leq\! 1 - p_{U}(\mcf{\tilde U}) \! < \! O(2^{-n\varepsilon_n}),
\end{align}
directly follows from Equation~\eqref{eq:app:1:uprob1}, which bounds the sum over all terms not relating to a $u \in \mcf{\tilde U}$. Given that $u \in \mcf{\tilde U}$, we bound the sum of all terms for which $u \notin \mcf{\tilde U}\setminus \mcf{\tilde U}_{\stab}$ by leveraging that $(\mbf{y},m) \in \mcf{D}_{\sat,(M)}(u,p_{Y|X};\nu_n)$ with high probability for such $u$. In specific,   
\begin{align}
&\sum_{\substack{(\mbf{y},m,u): \\ u \in \mcf{\tilde U} \setminus \mcf{\tilde U}_{\stab} }} p_{\Phi|\mbf{Y}}(m|\mbf{y}) p_{\mbf{Y},M,U}(\mbf{y},m,u) \notag \\
&\leq  O(2^{-n\varepsilon_n}) \hspace{-1pt} + \hspace{-48pt} \sum_{\substack{(\mbf{y},m,u): \\ u \in \mcf{\tilde U} \setminus \mcf{\tilde U}_{\stab} \\ (\mbf{y},m) \in \mcf{D}_{\sat,(M)}(u,p_{Y|X};\nu_n)  }} \hspace{-43pt}  p_{\Phi|\mbf{Y}}(m|\mbf{y}) p_{\mbf{Y},M,U}(\mbf{y},m,u)  \label{eq:app:1:rmvdsat}\\ 
%&\leq O(2^{-n\varepsilon_n}) \hspace{-1pt} + \hspace{-48pt} \sum_{\substack{(\mbf{y},m,u): \\ u \in \mcf{\tilde U} \setminus \mcf{\tilde U}_{\stab} \\ (\mbf{y},m) \in \mcf{D}_{\sat,(M)}(u,p_{Y|X};\nu_n)  }} \hspace{-43pt}  p_{\Phi|\mbf{Y}}(m|\mbf{y}) p_{\mbf{Y},U|M}(\mbf{y},u|m) 2^{-n^2 + 2n \nu_n} \label{eq:app:1:rmvdsat2}\\ 
&\leq O(2^{-n\varepsilon_n}) \hspace{-1pt} + \hspace{-48pt} \sum_{\substack{(\mbf{y},m,u): \\ u \in \mcf{\tilde U} \setminus \mcf{\tilde U}_{\stab} \\ (\mbf{y},m) \in \mcf{D}_{\sat,(M)}(u,p_{Y|X};\nu_n)  }} \hspace{-43pt}  p_{\Phi|\mbf{Y}}(m|\mbf{y}) 2^{-n^2 + 2n \nu_n} %\label{eq:app:1:rmvdsat3}
%\\ 
%&\leq O(2^{-n\varepsilon_n}) \hspace{-1pt} +  2^{-n^2 + 2n \nu_n + \logt|\mcf{U}| + n \logt|\mcf{Y}|} 
\leq O(2^{-n\varepsilon_n}) \label{eq:app:1:rmvdsat4}
\end{align}
where~\eqref{eq:app:1:rmvdsat} is because Equation~\eqref{eq:app:1:dmsat} must hold for all $u \in \mcf{\tilde U}\setminus\mcf{\tilde U}_{\stab}$, while~\eqref{eq:app:1:rmvdsat4} is because $p_{\mbf{Y},M,U}(\mbf{y},m,u) \leq p_{M}(m) < 2^{-n^2 + 2n\nu_n}$ for all $(\mbf{y} ,m) \in \mcf{D}_{\sat,(M)}(u,p_{Y|X};\nu_n)$. Thus all terms other than those with $u \in \mcf{\tilde U}_{\stab}$ do not contribute to the sum, and for the remaining terms it follows that
\begin{align}
&\sum_{\substack{(\mbf{y},m,u): \\ u \in \mcf{\tilde U}_{\stab} }} p_{\Phi|\mbf{Y}}(m|\mbf{y}) p_{\mbf{Y},M,U}(\mbf{y},m,u) \notag \\
&\leq O(2^{-n\varepsilon_n})  + \hspace{-25pt} \sum_{\substack{(\mbf{y},m,u): \\ u \in \mcf{\tilde U}_{\stab} \\ (\mbf{y},m) \in \mcf{D}_{+}(u,p_{Y|X};\nu_n)  }} \hspace{-25pt}  p_{\Phi|\mbf{Y}}(m|\mbf{y}) p_{\mbf{Y},M,U}(\mbf{y},m,u)  \label{eq:app:1:rmvdstab}
\end{align}
since Equation~\eqref{eq:app:1:dmstab} holds for all $u \in \mcf{\tilde U}_{\stab}$. 

The process of deriving a necessary condition in the form
of~\eqref{eq:app:1:prune:0'''} will henceforth be referred to as
\emph{our new methodology.} In essence, our new methodology is to
introduce the variable $U$ through Corollary~\ref{cor:mt}, then prune
out the unstable terms, and finally switch the distribution terms for
information theoretic ones. Since our new methodology is relatively
straightforward, in the future these steps will be described in less
detail.

\subsection{$(\delta,l)$-capacity of the wiretap channel}

In the wiretap channel, a source wants to reliably send a message $M$ chosen uniformly from $\set{1,\dots,2^{nr}}$, to a given destination while ensuring a certain level of secrecy from an eavesdropper. The source is connected to the destination through a DMC $p_{Y|X} \in \mcf{P}(\mcf{Y}|\mcf{X})$ and to the eavesdropper through a DMC $p_{Z|X} \in \mcf{P}(\mcf{Y}|\mcf{X})$. Once again a code, consisting of an encoder $F: \mcf{M} \rightarrow \mbf{X}$, and decoder $\Phi: \mcf{Y} \rightarrow \mcf{M}$, must be designed to satisfy
\begin{align}
\Pr \left( \Phi(\mbf{Y}) = M  \right) < \delta, \label{eq:wtc:pe}
\end{align}
for some fixed $\delta \in (0,1)$, but additionally there exists some secrecy requirement parameterized by $\ell$. Since Wyner~\cite{Wyner1975wiretap} first considered\footnote{Wyner only presented the case where the channel to the eavesdropper was degraded. The general case was introduced by Csisz{\'a}r and K{\"o}rner in~\cite{csiszar1978broadcast}.} the wiretap channel model, their have been a large number of secrecy metrics proposed. Here we shall restrict our focus to the original metric, and a metric more commonly found in modern literature.

The original metric, the weak information leakage rate, is formally defined for all positive real numbers $\ell$ as
\begin{align}
\mathbb{I}(\mbf{Z};M) < n\ell. \label{eq:wtc:ws}
\end{align}
On the other hand, the more common modern metric is the variational distance metric parameterized by $\ell \in (0,1]$,
\begin{align}
\frac{1}{2} \sum_{\mbf{z},m} |p_{\mbf{Z},M}(\mbf{z},m) - p_{\mbf{Z}}(\mbf{z}) p_{M}(m)|^+ < \ell. \label{eq:wtc:vl}
\end{align}
We will say a code is a $(\delta,\ell)$-code subject to weak information leakage if it satisfies Equation~\eqref{eq:wtc:pe} and~\eqref{eq:wtc:ws}, and a $(\delta,\ell)$-code subject to bounded variational distance if it satisfies Equation~\eqref{eq:wtc:pe} and~\eqref{eq:wtc:vl}. 

Regardless of secrecy metric, our primary concern will be to establish information theoretic necessary conditions for the existence of a $(\delta,\ell)$-code. Specifically we shall establish this necessary conditions upon the value of $r$. As a first point of order, note that for any $(\delta, \ell)$-code subject to weak information leakage, where $\delta \rightarrow 0$ and $\ell \rightarrow 0$ as a function of $n$, using Fano's inequality it is possible to obtain
\begin{align}
nr &\leq \mathbb{I}(\mbf{Y};M) + n\delta_1 \label{eq:wtc:csi1}\\
n\ell &> \mathbb{I}(\mbf{Z};M)  - n \delta_2 \label{eq:wtc:csi2} ,
\end{align}
where $\delta_1,\delta_2 \rightarrow 0$ with $n$, which then leads to the two following inequalities
\begin{align}
nr &< \mathbb{I}(\mbf{Y};M)  + n\delta_1 \label{eq:wtc:csi3} \\
nr &< \mathbb{I}(\mbf{Y};M) - \mathbb{I}(\mbf{Z};M) + \ell + n \delta_1 + n\delta_2 .\label{eq:wtc:csi4}
\end{align}
Equations~\eqref{eq:wtc:csi3} and~\eqref{eq:wtc:csi4} also imply 
\begin{align}
r &< c(\ell) + n(\delta_1+\delta_2) \label{eq:wtc:csi5}
\end{align}
where
\begin{align}
&c(\ell) = c(\ell,p_{Y,Z|X}) \notag\\
&\defn \max_{p(t)p(v|t)p(x|v)} \min \left(  \mathbb{I}(Y;V) ,  \mathbb{I}(Y;V|T) - \mathbb{I}(Z;V|T) + \ell  \right),
\end{align}
$|\mcf{T}|<|\mcf{X}|+3$ and $|\mcf{V}|<|\mcf{X}|^2 + 4|\mcf{X}| + 3$. Although the derivation of Equation~\eqref{eq:wtc:csi5} from~\eqref{eq:wtc:csi3} and~\eqref{eq:wtc:csi4} is outside the scope of this paper, we point those interested to~\cite[Section~V]{csiszar1978broadcast} or~\cite[Lemma~17.12]{CK} for the key identity in reducing Equation~\eqref{eq:wtc:csi4}, and to~\cite[Lemma~15.4]{CK} or~\cite[Appendix~C]{GNIT} for establishing the cardinality bounds. For our purposes, we will directly assume Equation~\eqref{eq:wtc:csi5} as an implication of~\eqref{eq:wtc:csi1} and~\eqref{eq:wtc:csi2}.

Observe that $((M,\emptyset),\mbf{X},\mbf{Y}_{\mcf{P}(\mcf{Y}|\mcf{X})})$ is a regular collection, and therefore there exists 
\begin{itemize}[leftmargin=*]
\item a DRV  $U: \left\{ \begin{array}{c}  (U,M) \markov \mbf{X} \markov \mbf{Y}_{\mcf{P}(\mcf{Y}|\mcf{X})}\\ |\mcf{U}| < O(-n \varepsilon_n \logt\varepsilon_n) \end{array} \right\}$,
\item a positive real number $\nu_n = O(-\sqrt{\varepsilon_n} \logt \varepsilon_n)$, and
\item  a set $\mcf{\tilde U} \subseteq \mcf{U}$ such that 
\begin{align}
p_{U}(\mcf{\tilde U}) \geq 1 - O(2^{-n\varepsilon_n})
\end{align}
and 
\begin{align}
\inf_{w \in \mcf{P}(\mcf{Y}|\mcf{X})} \Pr \left( (\mbf{Y}_w,M) \notin \mcf{D}_+(U,w;\nu_n)\middle|U=u \right)& \notag \\
&\hspace{-70pt} \leq 16 \cdot 2^{-n \varepsilon_n},
\end{align}
for all $u \in \mcf{\tilde U}$,
\end{itemize} 
where $\varepsilon_n = n^{\frac{-1}{|\mcf{X}||\mcf{Y}|+1}}$ and $$\mcf{D}_+(u,w;\nu_n) = \mcf{D}_{\stab,(M)}(u,w;\nu_n) \cap \mcf{D}_{\stab,(\emptyset)}(u,w;\nu_n),$$ by Corollary~\ref{cor:mt}. 
Using this new methodology we will establish necessary conditions under both secrecy metrics. Proof these conditions are sufficient can be found in our earlier work~\cite{graves2017wiretap}. 

First, let us consider the weak information leakage. 
\begin{theorem}
\begin{align}
r \leq c\left( \frac{\ell}{1-\delta} \right) + O(-\sqrt{\varepsilon_n} \logt \varepsilon_n)
\end{align}
for any $(\delta,\ell)$ code subject to weak information leakage. 
\end{theorem}
\begin{IEEEproof}
First, repeating the derivation of~\eqref{eq:app:1:rmvU1} from~\eqref{eq:app:1:nec'}, we obtain
\begin{align}
\Pr \left( nr \leq \mathbb{I}_{U}(\mbf{Y};M) + 5 n \nu_n + n \varepsilon_n \right) \geq 1 - \delta - O(2^{-n \varepsilon_n}) \label{eq:wtc:pe_nec}
\end{align}
from Equation~\eqref{eq:wtc:pe}. The set of $u \in \mcf{U}^+ \defn  \set{ u : nr \leq \mathbb{I}_{U}(\mbf{Y};M) + 5 n \nu_n + n \varepsilon}$ will play a critical role in how error alters the leakage condition. In fact, starting from Equation~\eqref{eq:wtc:ws} and using basic information inequalities we have
\begin{align}
n\ell&> \mathbb{I}(\mbf{Z};M) \geq - \logt|\mcf{U}| + \sum_{u} \mathbb{I}(\mbf{Z};M|u)p_{U}(u)  \\
&\geq - \logt|\mcf{U}|  + \sum_{u \in \mcf{U}^{+} } \mathbb{I}(\mbf{Z};M|u)p_{U}(u)\\
&\geq - 2\logt|\mcf{U}| +  \mathbb{I}(\mbf{Z};M|U \in \mcf{U}^+)\Pr \left( U \in \mcf{U}^+ \right) \\
&\geq - 2\logt|\mcf{U}| - O(2^{-n\varepsilon_n}n\logt|\mcf{Y}|)   \notag \\
&\hspace{50pt} + \mathbb{I}(\mbf{Z};M|U \in \mcf{U}^+)(1-\delta) \label{eq:wtc:2:1}.
\end{align}
Hence, from Equations~\eqref{eq:wtc:pe_nec} and~\eqref{eq:wtc:2:1} we obtain 
\begin{align}
nr &\leq \mathbb{I}(\mbf{Y};M| U \in \mcf{U}^+) + O(-n\sqrt{\varepsilon_n} \logt \varepsilon_n) \\
n \frac{\ell}{1-\delta} &\geq \mathbb{I}(\mbf{Z};M|U \in \mcf{U}^+) \notag \\
&\hspace{20pt} - \frac{1}{1-\delta} O(-n \varepsilon_n \logt \varepsilon_n + n 2^{-n\varepsilon_n}).
\end{align}
Furthermore the preceding equations take on the form of Equations~\eqref{eq:wtc:csi1} and~\eqref{eq:wtc:csi2} since $(U,M) \markov \mbf{X} \markov \mbf{Y}_{\mcf{P}(\mcf{Y}|\mcf{X})}$, and therefore 
\begin{align}
r &\leq c\left( \frac{\ell}{1-\delta} \right) + O(-\sqrt{\varepsilon_n} \logt \varepsilon_n).
\end{align}
\end{IEEEproof}

Thus the capacity increases if the tolerable leakage increases or tolerable error increases. Only in the special case where the leakage is restricted to be $0$ is there a strong converse. The necessary proof also hints at a simple direct proof where one code is constructed which has a probability of error near $1$, but also does not have any information leakage, while a second code is constructed with a low error rate and $\frac{\ell}{1-\delta}$ bits per symbol of information leakage. Which of these two codes will be used to transmit the information will then be selected at random prior to transmission.

 This result stands in contrast to the more modern metric which exhibits an ``all or nothing'' dichotomy of the region.
\begin{theorem}
\begin{align}
r - O(-\sqrt{\varepsilon_n} \logt \varepsilon_n) &< \begin{cases} 
c(0)  & \text{ if } \delta + \ell < 1\\
\max_{p_{X}} \mathbb{I}(Y;X)   &\text{ o.w.}
\end{cases},
\end{align}
for any $(\delta,\ell)$ code subject to bounded variational distance. 
\end{theorem}
\begin{IEEEproof}
As in the proof for weak information leakage, meeting the reliability criterion requires
\begin{align}
\Pr \left( nr \leq \mathbb{I}_{U}(\mbf{Y};M) + 5 n \nu_n + n \varepsilon_n \right) \geq 1 - \delta - O( 2^{-n \varepsilon_n}). \label{eq:wtc:vl:pe}
\end{align}
Now, starting from Equation~\eqref{eq:wtc:vl} we can derive new bounds as follows
\begin{align}
\ell&> \hspace{-10pt}  \sum_{\substack{\mbf{z},m :\\ p_{\mbf{Z}|M}(\mbf{z}|m) \geq p_{\mbf{Z}}(\mbf{z})}}   p_{\mbf{Z},M}(\mbf{z},m) - p_{\mbf{Z}}(\mbf{z})p_{M}(m) \\
&\geq  \hspace{-10pt}   \sum_{\substack{ \mbf{z},m : \\ p_{\mbf{Z}|M}(\mbf{z}|m) \geq 2^{n\nu_n} p_{\mbf{Z}}(\mbf{z})}}  \hspace{-30pt}  p_{\mbf{Z},M}(\mbf{z},m) - p_{\mbf{Z},M}(\mbf{z},m) 2^{-n\nu_n} \\
&\geq   \hspace{-10pt} \sum_{\substack{ (\mbf{z},m,u) : \\ (\mbf{z},m) \in \mcf{D}_+(u,p_{Z|X};\nu_n)   \\ 2^{-\mathbb{H}(\mbf{Z}|M,u) - n\nu_n} \geq 2^{-\mathbb{H}(\mbf{Z}|u) + 2n\nu_n}}} \hspace{-30pt} (1 - 2^{-n\nu_{n}}) p_{\mbf{Z},M,U}(\mbf{z},m,u) \label{eq:wtc:vl:nm}\\
&\geq (1 - 2^{-n\nu_{n}})  \Pr (  \mathbb{I}_U(\mbf{Z};M) \geq 3n \nu_n ) - O(2^{-n\varepsilon_n}), \label{eq:wtc:vl:huh}
\end{align}
where~\eqref{eq:wtc:vl:nm} is by our new methodology, while~\eqref{eq:wtc:vl:huh} is because
\begin{align}
&\Pr \left( \begin{array}{ c}  (\mbf{Z}, M) \in \mcf{D}_{+}(U,p_{Z|X};\nu_n) \\   \mathbb{I}_U(\mbf{Z};M) \geq 3n \nu_n \end{array} \right)\notag \\
&\geq \Pr \left(\mathbb{I}_U(\mbf{Z};M) \geq 3n \nu_n  \right) - \Pr \left( (\mbf{Z}, M) \notin \mcf{D}_{+}(U,p_{Z|X};\nu_n) \right)  \notag \\
&\geq \Pr \left(\mathbb{I}_U(\mbf{Z};M) \geq 3n \nu_n  \right) - \Pr \left( U \notin \mcf{\tilde U} \right) \notag \\
&\hspace{10pt} - \Pr \left( (\mbf{Z}, M) \notin \mcf{D}_{+}(U,p_{Z|X};\nu_n) \middle| U \in \mcf{\tilde U}  \right)  \notag \\
&= \Pr \left(\mathbb{I}_U(\mbf{Z};M) \geq 3n \nu_n  \right)  - O(2^{-n\varepsilon_n}) . \notag
\end{align}
Combining Equation~\eqref{eq:wtc:vl:pe} and~\eqref{eq:wtc:vl:huh} yields
\begin{align}
\Pr \left( \begin{array}{r l} nr &\leq  \mathbb{I}_{U}(\mbf{Y};M) + 5 n \nu_n + n \varepsilon_n  \\  0 & >  \mathbb{I}_U(\mbf{Z};M) - 3n \nu_n    \end{array} \right) &\notag\\
&\hspace{-160pt} \geq 1 - \delta - \ell - \frac{\ell2^{-n\nu_n} }{1-2^{-n\nu_n}} - O\left( \frac{2^{-n\varepsilon_n}}{1-2^{-n\nu_n}}\right) \\
&\hspace{-160pt} = 1 - \delta - \ell  - O(2^{-n\varepsilon_n}).
\end{align}
Therefore if $1- \delta - \ell > 0$ then for large enough $n$ there exists a $u \in \mcf{U}$ such that
\begin{align}
nr &\leq  \mathbb{I}(\mbf{Y};M|U=u) + 5 n \nu_n + n \varepsilon_n  \\ 
0 & >  \mathbb{I}(\mbf{Z};M|U=u) - 3n \nu_n.
\end{align}
These equations take the form of Equations~\eqref{eq:wtc:csi1} and~\eqref{eq:wtc:csi2} and therefore
\begin{align}
r &\leq c(0) + 8 \nu_n +  \varepsilon_n = c(0) + O(-\sqrt{\varepsilon_n} \logt \varepsilon_n).
\end{align}

On the other hand, if $1- \delta - \ell > 0$ and $\delta < 1$ then Equation~\eqref{eq:wtc:vl:pe} still directly implies 
\begin{align}
r &\leq \max_{p(x)} \mathbb{I}(Y;X) + 5n \nu_n + n \varepsilon_n
\end{align}
as discussed in the first example. 
\end{IEEEproof}

\subsection{Converse for error exponents: keyed authentication} 
\label{sec:app:conee}

For this next example we consider a communication model recently employed by Lai et al.~\cite{lai2009authentication}, and\footnote{To be more precise, this model is a special case of the model found in~\cite{gungor2016basic}.} and Gungor and Koksal~\cite{gungor2016basic}. Here the source and destination must now maintain reliable communications in the presence of an interloper who has the ability to modify any transmitted information. In order for communications to be considered reliable, the destination must be able to detect when the interloper has modified the transmission.

More formally, this model has two different modes of operation depending on if the interloper intercedes or not. If the interloper does not intercede then the source is connected to the destination through a DMC $w_s \in \mcf{P}(\mcf{Y}|\mcf{X})$. However, if the interloper does intercede, then the source is connected first to the interloper through a DMC $w_i \in \mcf{P}(\mcf{Y}|\mcf{X})$, and the interloper is then connected to the destination through a noiseless channel. In this case, the interloper may observe the entire $n$-length transmission sequence before arbitrarily choosing the value of $\mbf{Y}$ that the destination will observe. In order to aid the source and destination in the detection of this interloper, they pre-share a secret key, $K$, which is chosen uniformly from the set of all possible keys\footnote{\cite{lai2009authentication} is only concerned with keys for which $n^{-1}\mathbb{H}(K)$ vanished asymptotically as $n\rightarrow \infty$. We, nor~\cite{gungor2016basic}, will make such a restrictions in our derivations.}. 

Our goal will be to establish information theoretic necessary conditions on the existence of a reliable encoder $F:\mcf{M} \times \mcf{K} \rightarrow \mcf{X}^n$, and decoder $\Phi: \mcf{Y}^n \times \mcf{K} \rightarrow \mcf{M} \cup \mbf{!}$, where $\mbf{!}$ is a declaration of intrusion attempt. The probability of error of the message $\mcf{M}$ will be completely ignored, and instead we focus solely on the necessary conditions which arise due to the need to detect intrusion. To that end let $\mcf{S}(k) \defn \set{\mbf{y} : \Phi(\mbf{y},k) \neq \mbf{!}}$ denote the set of sequences of $\mbf{y}$ which will be deemed authentic for given key $k \in \mcf{K}$. With this the probability of intrusion given intercession (that is, the false authentication probability) can be written as
\begin{align}
2^{- \beta}   &\defn \hspace{-8pt}  \sup_{\psi \in \mcf{P}(\mcf{Y}^n|\mcf{Y}^n)} \sum_{\substack{\mbf{y},\mbf{y}_{w_i}, k: \\ \mbf{y} \in \mcf{S}(k)} } \hspace{-3pt} \psi(\mbf{y}|\mbf{y}_{w_i}) p_{\mbf{Y}_{w_i},K}(\mbf{y}_{w_i},k). \label{eq:conee:exp:op}
\end{align} 
Indeed, in order for the decoder to falsely authenticate a transmission from the interloper, the interloper must choose a value $\mbf{y}$ for which the decoder does not declare intrusion, in other words the interloper must choose a $y \in \mcf{S}(k)$. Thus, $\psi$ represents the attacking strategy of the interloper, in which the interloper receives the value $\mbf{y}_{w_i}$ and then decides to transmit $\mbf{y}$ with probability $\psi(\mbf{y}|\mbf{y}_{w_i})$. 

The best previously established results were by Simmons~\cite{Auth} and Maurer~\cite{Maurer00a} who demonstrated that
\begin{align}
\beta &< \mathbb{I}(K;\mbf{Y}_{w_s}) \\
\beta &< \mathbb{H}(K|\mbf{Y}_{w_i}).
\end{align} 
We will improve on these bounds by applying our new methodology to Equation~\eqref{eq:conee:exp:op}. In particular our new bounds will show how the variation in the channel limits the value of $\beta$.

To do this we assume some code exists with a $2^{-\beta}$ probability of intrusion. Next we set $\mbf{X} = F(M,K)$ and observe that $((K,\mbf{X},\emptyset), \mbf{X}, \mbf{Y}_{\mcf{P}(\mcf{Y}|\mcf{X})})$ is a regular collection, and DRV $$T\defn \sum_{ t \in \mcf{P}_n(\mcf{X})} t \cdot \idc{ \mbf{X} \in \set{ \mbf{x} : p_{\mbf{x}} = t}},$$ has the properties $\logt|\mcf{T}| = O(\logt n)$ and $T \ll  \mbf{X}$. Therefore there exists:
\begin{itemize}[leftmargin=*] 
\item a DRV $U: \left\{ \begin{array}{c} U \gg T \\  \logt|\mcf{U}| =  O(\logt n - n \varepsilon_n \logt \varepsilon_n) \\   (U,M_{[1:l]}) \markov \mbf{X} \markov \mbf{Y}_{\mcf{P}(\mcf{Y}|\mcf{X})} \end{array} \right\}$, 
\item a real number $\nu_n = O(n^{-1}\logt n -\sqrt{\varepsilon_n} \logt \varepsilon_n )$, and  
\item a set $\mcf{\tilde U}$ such that 
\begin{align}
p_{U}(\mcf{\tilde U}) \geq 1 - O( 2^{-n\varepsilon_n})
\end{align}
and 
\begin{align}
\Pr \left( ( \mbf{Y}_{w},K,\mbf{X}) \notin \mcf{D}_+(U,w;\nu_n) \middle| U = u \right) <  16 \cdot 2^{-n \varepsilon_n} \label{eq:conee:exp:wat1}
\end{align}
for all $u \in \mcf{\tilde U}$ and $ w\in \mcf{P}(\mcf{Y}|\mcf{X})$,
\end{itemize}
where $$\mcf{D}_{+}(u,w;\nu_n) \defn \mcf{D}_{\stab,(\emptyset)}(u,w;\nu_n) \cap \mcf{D}_{\stab,(K)}(u,w;\nu_n)$$ and $\varepsilon_n \defn n^{-\frac{1}{|\mcf{X}||\mcf{Y}|+1}},$ by Corollary~\ref{cor:mt}.

Because of this we will obtain the following theorem. 
\begin{theorem}
Let $n$ be large enough so that $(n+1)^{-|\mcf{X}||\mcf{Y}} \geq 17 \cdot 2^{-n\varepsilon_n}$. Then
\label{thm:conee:1}
\begin{align}
\beta &\leq \hspace{-25pt}  \inf_{\substack{ (u,w) \in \mcf{\tilde U} \times \mcf{P}(\mcf{Y}|\mcf{X}) : \\  \Pr \left( \mbf{Y}_w \in \mcf{S}(K) |u \right) > 17 \cdot 2^{-n\varepsilon_n} } }  \hspace{-28pt} \mathbb{I}(K;\mbf{Y}_w|u)-h(u) + O(-n\sqrt{\varepsilon_n} \logt \varepsilon_n)  \label{eq:conee:thm:1}\\
\beta &\leq   \hspace{-5pt} \min_{(u,w) \in \mcf{\tilde U} \times \mcf{P}_n(\mcf{Y}|t_u) }  \hspace{-6pt} \mathbb{H}(K|\mbf{Y}_{w},u) + n\mathbb{D}(w||w_i|t_u ) - h(u)  \notag \\
&\hspace{120pt} + O(-n\sqrt{\varepsilon_n} \logt \varepsilon_n) \label{eq:conee:thm:2} ,
\end{align}
where $t_{u}$ denotes the $t$ such that $p_{T|U}(t|u) =1$ (that is the type of the distribution of $\mbf{X}$).
\end{theorem}
Before proof, we take a moment to discuss some of the implications of Theorem~\ref{thm:conee:1}. Note that restricting the code to a $u \in \mcf{U}$ is akin to restricting all of the codewords to be a particular type. So much like constant composition codes allow us to consider this type fixed, we will for the sake of discussion assume that the code is restricted to a single value of $u \in \mcf{\tilde U}$. In this case Theorem~\ref{thm:conee:1} simplifies to 
\begin{align}
\beta &\leq \hspace{-20pt} \inf_{\substack{w \in \mcf{P}(\mcf{Y}|\mcf{X}) :\\ \Pr \left( \mbf{Y}_w \in \mcf{S}(K) \right) > 17\cdot 2^{-n\varepsilon_n}} }   \hspace{-20pt} \mathbb{I}(K;\mbf{Y}_w) + O(-n\sqrt{\varepsilon_n} \logt \varepsilon_n)    \label{eq:conee:thm:1b}\\
\beta &\leq \min_{ w \in \mcf{P}_n(\mcf{Y}|t_u )  }   \mathbb{H}(K|\mbf{Y}_{w}) + n\mathbb{D}(w||w_i|t_u) \notag \\
&\hspace{100pt} + O(-n\sqrt{\varepsilon_n} \logt \varepsilon_n)  \label{eq:conee:thm:2b}.
\end{align}

From ~\eqref{eq:conee:thm:1b} we see that the mutual information between the key and the observation for any empirical channels (that is $p_{\mbf{y}|\mbf{x})}$) over which the correct key could be identified with non-zero probability will upper bound the probability of false authentication. 
%upper bounds the  by $\mathbb{I}(K;\mbf{Y}_w)$ for the worst DMC $w$ such that $\Pr \left( \mbf{Y}_w \in \mcf{S}(K) \right) > 16\cdot 2^{-n\varepsilon_n}$. 
This is compounded by the fact that $\Pr \left( \mbf{Y}_w \in \mcf{S}(K) \right) > 17\cdot 2^{-n\varepsilon_n}$ does not imply $\mathbb{I}(K;\mbf{Y}_w) \approx \mathbb{H}(K)$ since the sets $\mcf{S}(k)$ are not necessarily disjoint for different values of $k$. This is unfortunate since choosing a code more robust to channel deviations may disproportionately increase the probability of false authentication. 

Next, from \eqref{eq:conee:thm:2b} we see that the probability of false authentication will be constrained by the entropy of the key given the adversaries observation over a number of different empirical channels to the adversary. Moreover, the KL divergence term can be thought of as relating to a correction term to account for the probability of the empirical channel being $w$ while the actual channel is $w_i$. Intuitively, one may think of the probability the empirical channel $w_i$ occurs as $2^{-n\mathbb{D}(w_i||w|t_u)}$, while the probability of false authentication given empirical channel $w_i$ occurs as $2^{-n\mathbb{H}(K|\mbf{Y}_{w_i})}$. The stated bound then clearly follows.

Both of these equations are clearly less than the existing equations, which follows simply because the infimum of~\eqref{eq:conee:thm:1} generally includes $w_s$, while the infimum of the set for~\eqref{eq:conee:thm:2} contains $w_i$. We now prove Theorem~\ref{thm:conee:1}.

\begin{IEEEproof}
To prove both Equation~\eqref{eq:conee:thm:1} and~\eqref{eq:conee:thm:2} it will be important to first note that Equation~\eqref{eq:conee:exp:op} directly implies
\begin{align}
2^{-\beta} \geq \sum_{\substack{\mbf{y},\mbf{y}_{w_i}, k: \\ \mbf{y} \in \mcf{S}(k)}} \psi^*(\mbf{y}|\mbf{y}_{w_i}) p(\mbf{y}_{w_i},k|u) p(u), \label{eq:conee:p:st}
\end{align}
for all $\psi^* \in \mcf{P}(\mcf{Y}^n|\mcf{Y}^n)$ and $u \in \mcf{\tilde U}$. To prove both Equation~\eqref{eq:conee:thm:1} and~\eqref{eq:conee:thm:2}, specific values of $\psi^*$ and $u$ will be chosen\footnote{Readers familiar with authentication problems should recognize these chosen values as relating to ``substitution'' and ``impostor'' attacks. Both of these attacks are encompassed by the general framework presented here.}.

To derive Equation~\eqref{eq:conee:thm:1}, first fix a $w \in \mcf{P}(\mcf{Y}|\mcf{X})$, and a $u \in \mcf{\tilde U}$ such that $\Pr\left( \mbf{Y}_w \in \mcf{S}(K) |U =u \right) > 17 \cdot 2^{-n\varepsilon_n}$  and set $\psi^*(\mbf{y} |\mbf{y}_{w_i})$ equal to $p_{\mbf{Y}_w|U}(\mbf{y}|u)$ in Equation~\eqref{eq:conee:p:st}. Doing so, we may derive Equation~\eqref{eq:conee:thm:1} as follows:
\begin{align}
&2^{-\beta} \notag\\
&\geq \sum_{\substack{\mbf{y},\mbf{y}_{w_i},k: \\ \mbf{y} \in \mcf{S}(k)}} p_{\mbf{Y}_w|u}(\mbf{y}|u) p(\mbf{y}_{w_i},k,u) \\
&= \sum_{\substack{\mbf{y},\mbf{x},k :\\ \mbf{y} \in \mcf{S}(k)}} p_{\mbf{Y}_w,K,\mbf{X},U}(\mbf{y},k,\mbf{x},u) \frac{ p_{\mbf{Y}_w|U}(\mbf{y}|u)}{p_{\mbf{Y}_w|K,U}(\mbf{y}|k,u)} \\
&\geq \hspace{-10pt} \sum_{\substack{\mbf{y},\mbf{x},k:\\ \mbf{y} \in \mcf{S}(k) \\ (\mbf{y},\mbf{x},k) \in \mcf{D}_{+}(u,w;\delta)}}   \hspace{-25pt} p_{\mbf{Y}_w,\mbf{X},K|U}(\mbf{y},\mbf{x},k|u) p_{U}(u) 2^{- \mathbb{I}(\mbf{Y}_w;K|u) - 2n \nu_n  }  \label{eq:conee:1:d3} \\
&\geq p_{U}(u) 2^{-\mathbb{I}(\mbf{Y}_w;K|u) - 2n \nu_n -n\varepsilon_n } \label{eq:conee:1:d4} 
\end{align}
where~\eqref{eq:conee:1:d3} is by our new methodology since $u \in \mcf{\tilde U}$, and~\eqref{eq:conee:1:d4} is because 
\begin{align}
&\Pr \left( \begin{array}{c}  \mbf{Y}_w \in \mcf{S}(K) \\  (\mbf{Y}_w,\mbf{X},K) \in \mcf{D}_+(U,w;\nu_n) \end{array} \middle|U = u \right) \notag \\
&\geq \Pr\left( \mbf{Y}_w \in \mcf{S}(K) |U =u \right)\notag \\ &\hspace{40pt} - \Pr\left( (\mbf{Y}_w,\mbf{X},K) \notin \mcf{D}_+(U,w;\nu_n) |U =u \right) \notag \\
&\geq \Pr\left( \mbf{Y}_w \in \mcf{S}(K) |U =u \right) - 16 \cdot 2^{-n\varepsilon_n} \geq 2^{-n\varepsilon_n} . \notag
\end{align}

Moving on to proving Equation~\eqref{eq:conee:thm:2}. First understand that if $\mcf{S}(k)$ are not pairwise disjoint (i.e., $\mcf{S}(k) \cap \mcf{S}(k') \neq \emptyset$ for some $k \in \mcf{K}$ and $k' \in \mcf{K}\setminus \set{k}$), then the RHS of Equation~\eqref{eq:conee:p:st} is 
$$\geq \sum_{\substack{\mbf{y} ,\mbf{y}_{w_i},\mbf{x},k:\\ \mbf{y} \in \mcf{\tilde S}(k)}}   \psi^*(\mbf{y}|\mbf{y}_{w_i}) p(\mbf{y}_{w_i},\mbf{x},k,u),$$
for any collection of sets $\{ \mcf{\tilde S}(k) \}_{k \in \mcf{K}}$ that are pairwise disjoint and for which  $\mcf{\tilde S}(k)\subseteq \mcf{S}(k)$ for all $k$. As a point to note later, changing the decoder in such a way would not change the values of $\mathbb{H}(K|\mbf{Y}_w,u)$ or $\mathbb{D}(w||w_i|t_u)$. As such we will proceed with the assumption that $\{ \mcf{S}(k)\}_{k \in \mcf{K}}$ are pairwise disjoint from the outset, which is valid since we are looking for a lower bound on the probability of false authentication, 

Now once again fix a $u \in \mcf{\tilde U}$ and $w \in \mcf{P}_n(\mcf{Y}|t_u)$, but this time set $\psi^*$ to be any distribution such that
\begin{align}
\psi^*(\mcf{S}(k)|\mbf{y}_{w_i}) %&= p_{K|\mbf{Y}_w,U}(k|\mbf{y}_{w_i},u) \\
&=  \frac{p_{\mbf{Y}_w|K,U}(\mbf{y}_{w_i}|k,u) p_{K|U}(k|u)}{p_{\mbf{Y}_w|U}(\mbf{y}_{w_i}|u)}   ,
\end{align}
for all $(\mbf{y}_{w_i},k)  : \{ p_{\mbf{Y}_w|U}(\mbf{y}_{w_i}|u) \neq 0\}$. Since we assume the sets $\mcf{S}(k)$ are disjoint we may perform the summation over $\mbf{y}$ to obtain that
\begin{align}
2^{-\beta} &\geq \sum_{\mbf{y}_{w_i},\mbf{x},k}  p_{K|\mbf{Y}_w,U}(k|\mbf{y}_{w_i},u)  w_{i}^n(\mbf{y}_{w_i}|\mbf{x}) p(\mbf{x},k,u)
\end{align}
is always necessary. Furthermore since all summands are positive, we may restrict the summation to only consider $\mbf{y}_{w_i}$ such that $p_{\mbf{y}_{w_i}|\mbf{x}} = w$, hence giving 
\begin{align}
2^{-\beta} 
%&\geq \hspace{-10pt} \sum_{\substack{\mbf{y}_{w_i},\mbf{x},k :  \\ p_{\mbf{y}_{w_i}|\mbf{x}} = w   }}  \hspace{-10pt} \frac{p_{\mbf{Y}_w|K,U}(\mbf{y}_{w_i}|k,u) p_{K|U}(k|u)}{p_{\mbf{Y}_w|U}(\mbf{y}_{w_i}|u)}  w_i^n(\mbf{y}_{w_i}|\mbf{x}) p(\mbf{x},k,u) \label{eq:app:2:hencegiving}.
&\geq \hspace{-10pt} \sum_{\substack{\mbf{y}_{w_i},\mbf{x},k :  \\ p_{\mbf{y}_{w_i}|\mbf{x}} = w   }}  \hspace{-10pt} p_{K|\mbf{Y}_w,U}(k|\mbf{y}_{w_i},u)  w_i^n(\mbf{y}_{w_i}|\mbf{x}) p(\mbf{x},k,u) \label{eq:app:2:hencegiving}.
\end{align}
Now applying our new methodology yields that the RHS of Equation~\eqref{eq:app:2:hencegiving} is 
\begin{align}
&\geq p(u) 2^{-\mathbb{H}(K|\mbf{Y}_w,u) -3n \nu_n }   \hspace{-20pt} \sum_{\substack{\mbf{y}_{w_i},\mbf{x},k : \\ (\mbf{y}_{w_i},\mbf{x},k) \in \mcf{D}_+(u,w;\nu_n)  \\ p_{\mbf{y}_{w_i}|\mbf{x}} = w }} \hspace{-20pt}  w_i^n(\mbf{y}_{w_i}|\mbf{x}) p(\mbf{x},k|u) \label{eq:conee:2:d2} 
\end{align}
which may be continued to derive Equation~\eqref{eq:conee:thm:2} as
\begin{align}
&= p(u) 2^{-\mathbb{H}(K|\mbf{Y}_w,u) - \mathbb{D}(w||w_i|t_u) -3n\nu_n}   \notag \\
&\hspace{50pt} \cdot \sum_{\substack{\mbf{y}_{w_i},\mbf{x},k : \\ (\mbf{y}_{w_i},\mbf{x},k) \in \mcf{D}_+(u,w;\nu_n)  \\ p_{\mbf{y}_{w_i}|\mbf{x}} = w }} \hspace{-20pt}  w^n(\mbf{y}_{w_i}|\mbf{x}) p(\mbf{x},k|u)  \label{eq:conee:2:d3}\\
%&\geq  p(u) 2^{-\mathbb{H}(K|\mbf{Y}_w,u)  - \mathbb{D}(w||w_i|u_v) -3n\nu_n}   \left( n^{-|\mcf{X}||\mcf{Y}|}  - 14 \cdot 2^{-n\varepsilon_n} \right) \label{eq:conee:2:d4}.
&\geq  p(u) 2^{-\mathbb{H}(K|\mbf{Y}_w,u)  - \mathbb{D}(w||w_i|t_u) -3n\nu_n - n\varepsilon_n} \label{eq:conee:2:d4}.
\end{align}
where~\eqref{eq:conee:2:d3} is because $w_i^n(\mbf{y} |\mbf{x}) = 2^{-n \mathbb{D}(w||w_i|t_u)} w^n(\mbf{y}|\mbf{x})$ for all $(\mbf{y},\mbf{x}) : p_{\mbf{y}|\mbf{x}} = w$ (see~\cite[Lemma~2.3]{csiszar2004information}), and~\eqref{eq:conee:2:d4} is because $\Pr( \mbf{Y}_w \in \set{ \mbf{y} : p_{\mbf{y}|\mbf{x}} = w } |\mbf{X} = \mbf{x} ) \geq n^{-|\mcf{X}||\mcf{Y}|} \geq 17 \cdot 2^{-n\varepsilon_n} $ for all $w \in \mcf{P}_n(\mcf{Y}|t_u)$ (see~\cite[Lemma~2.6]{CK}) and therefore
\begin{align}
&\Pr \left( \begin{array}{c} (\mbf{Y}_w,\mbf{X}) \in \set{ (\mbf{y},\mbf{x}) : p_{\mbf{y}|\mbf{x}} = w }  \\  (\mbf{Y}_w,\mbf{X},K) \in \mcf{D}_+(U,w;\nu_n) \end{array} \middle|U = u \right) \notag \\
&\geq \Pr\left( (\mbf{Y}_w,\mbf{X}) \in \set{ (\mbf{y},\mbf{x}) : p_{\mbf{y}|\mbf{x}} = w }  |U =u \right)\notag \\ 
&\hspace{10pt} - \Pr\left( (\mbf{Y}_w,\mbf{X},K) \notin \mcf{D}_+(U,w;\nu_n) |U =u \right)  \\
&\geq  (n+1)^{-|\mcf{X}||\mcf{Y}|} - 16 \cdot 2^{-n\varepsilon_n}   \geq 2^{-n\varepsilon_n},
\end{align} 
for all $u \in \mcf{\tilde U}$. 
\end{IEEEproof}

\section{Proof of Theorem~\ref{lem:daco}}\label{sec:daco}

To prove Theorem~\ref{lem:daco}, we construct here the subset
$\mcf{A}^\dagger \subseteq \mcf{X}^n$ with non-negligible probability
for which the quasi-image of
$\mbf{X}\,|\set{\mbf{X} \in \mcf{A}^\dagger}$ by
$p_{Y|X} \in \mcf{P}(\mcf{Y}|\mcf{X})$ is stable. Our construction is
based on information spectrum slicing~\cite{han2003}. In particular,
the set $\mcf{A}^\dagger$ will be the pre-image of the union of a few
entropy spectrum slices of $Y$. Therefore, before proving
Theorem~\ref{lem:daco} we will first need to build up a few
definitions related to the entropy spectrum. From there we will derive
a few lemmas which will help to streamline the proof of
Theorem~\ref{lem:daco}.

\subsection{Information Spectrum Slicing} \label{sec:partition} 

Here we build a few results that relate to information (or entropy)
spectrum introduced by Han~\cite{han2003}. Keep in mind that the
overarching goal of this work is to create quasi-images with nearly
uniform distribution. The entropy spectrum provides a language with
which we can succinctly discuss these variations.

\begin{define}
\label{def:infspec}
For DRV $Y$ arbitrarily distributed over $\mcf{Y}$, the set
\begin{equation*}
\mcf{S}_{Y}(s; \lambda, t) 
\defn 
\set{ y \in \mcf{Y}:  \left\lfloor \min\left( \frac{h_{Y}(y) }{\lambda} , t\right) \right \rfloor = s},
\end{equation*}
for a given $s \in \mathbb{N}$, $\lambda \in (0,\infty)$, and
$t \in \mathbb{N}_+$, is the $(s;\lambda,t)$-slice of $Y$.
\end{define}
\begin{remark}
  $\mcf{S}_Y(s; \lambda,t) = \emptyset$ for all $s \notin
  [0:t]$. Moreover as we care only about the support set
  $\set{y \in \mcf{Y}: p_Y(y) > 0}$ of $Y$, we will interpret
  $\bigcup_{i=0}^{t}\mcf{S}_Y(i; \lambda,t)$ as the support set.
\end{remark}
The terminology of $\mcf{S}_{Y}(s;\lambda,t)$ being a
$(s;\lambda,t)$-slice is supported by the following lemma.
\begin{lemma}
For every $s \in [0:t-1]$ and $y\in \mcf{S}_{Y}(s;\lambda,t)$,
\label{lem:bk_prob}
\begin{align}
 s \lambda \leq h_{Y}(y)  < (s+1)\lambda  \label{eq:bk_prob:1}.
\end{align}
%and
%\begin{align}
%& s \lambda - \rho(s) \leq  h_{Y|Y \in \mcf{S}_{Y}(s;\lambda,t)}(y)  <  (s+1) \lambda - \rho(s) ,\label{eq:bk_prob:2}
%\end{align}
For $s = t$, the lower bound in~\eqref{eq:bk_prob:1} still holds.
\end{lemma}
\begin{IEEEproof}
  The %first
  double inequality is simply a restatement of
  Definition~\ref{def:infspec} for $s \in [0:t-1]$, and as such the
  lower bound still clearly holds for $s =t$.
%The second double inequality follows directly from the first since the $s$ for which $y \in \mcf{S}_{Y}(s;\lambda,t)$ is unique, hence $P_{Y}(y) = P_{Y|Y \in \mcf{S}_{Y}(s;\lambda,t)}(y) P_{Y}(\mcf{S}_{Y}(s;\lambda,t))$ for all $y  \in \mcf{S}_{Y}(s;\lambda,t)$. 
\end{IEEEproof}
\begin{cor}
\label{lem:bk_size}
For every $s \in [0: t-1]$,
\begin{equation} \label{eq:cardbound1}
s \lambda  - \varrho(s)  \leq  \logt |\mcf{S}_{Y}(s;\lambda,t) | < (s+1)\lambda -\varrho(s),
\end{equation}
$\varrho(s) \defn - \logt p_{Y}(\mcf{S}_{Y}(s;\lambda,t))$.
In addition, 
\begin{equation}\label{eq:cardbound2}
t \lambda  - \varrho(t)   \leq  \logt |\mcf{S}_{Y}(t;\lambda,t) | < (t+1)\lambda
\end{equation}
if $t> \lambda^{-1} \logt |\mcf{Y}|$.
\end{cor}
\begin{IEEEproof}
  For $s \in [0:t-1]$, \eqref{eq:bk_prob:1} implies
\begin{equation*}%\label{eq:bk_size:1}
  2^{-(s+1)\lambda} |\mcf{S}_{Y}(s;\lambda,t)|  < \sum_{y \in
    \mcf{S}_{Y}(s;\lambda,t)} p_{Y}(y) \leq 2^{-s\lambda}
  |\mcf{S}_{Y}(s;\lambda,t)|,
\end{equation*}
which in turn yields~\eqref{eq:cardbound1} since
$\sum_{y \in \mcf{S}_{Y}(s;\lambda,t)} p_{Y}(y) =
p_{Y}(\mcf{S}_{Y}(s;\lambda,t))$.

For the case of $s=t$, the lower bound in~\eqref{eq:cardbound2} is
still valid by the same argument above. The upper bound, on the other
hand, is because
$\logt |\mcf{S}_{Y}(t;\lambda,t)| \leq \logt |\mcf{Y}| < t \lambda$.
\end{IEEEproof}

\subsection{Supporting Lemmas} \label{lem:daco support} 

Next, we begin to employ information spectrum slicing to derive a
number of lemmas which help to simplify and streamline the proof of
Theorem~\ref{lem:daco}. All lemmas here are under the assumption that
$(\emptyset,\mbf{X},\mbf{Y})$ form a regular collection of
DRVs. Furthermore, throughout the proofs we let $\mcf{A}$ denote the
support set of $\mbf{X}$ (i.e.,
$\mcf{A} \defn \set{ \mbf{x} \in \mcf{X}^n: p_{\mbf{X}}(\mbf{x}) > 0
}$). This is important since an image is independent of $\mbf{X}$'s
distribution, and instead is solely a function of the support set. On
the other hand, the quasi-image is clearly a function of the
distribution.

To reduce notational clutter, positive real numbers $\lambda$ and $t$
are to be chosen such that $\lambda t > n \logt |\mcf{Y}|$, and we
simply write $\mcf{S}_{\mbf{Y}}(i) =
\mcf{S}_{\mbf{Y}}(i;\lambda,t)$. One important aspect to observe
through the proofs will be the interplay between quasi-images of
distributions and the images of their support sets. To this end, as we
will see, it will be helpful to let
\begin{align*}
\eta_{s} \defn p_{\mbf{Y}}\left(\bigcup_{i=0}^s \mcf{S}_{\mbf{Y}}(i)\right)
\end{align*}
for $s \in [0:t]$, and $\eta_{s'} = 0$ for all integers $s' < 0$ and
$\eta_{s'} = 1$ for all integers $s' \geq t$. Note then that
$0=\eta_{-1} \leq \eta_0 \leq \eta_1 \leq \cdots \leq \eta_{t} =
1$. In addition $\eta_{s-1} = \eta_s$ implies
$\mcf{S}_{\mbf{Y}}(s) = \emptyset$. Of direct importance to the
following lemmas is that $\bigcup_{i=0}^s \mcf{S}_{\mbf{Y}}(i)$ is an
$\eta_s$-quasi image of $\mbf{X}$ by $p_{Y|X}$. As we will show
later, it is in fact the unique minimum such $\eta_s$-quasi image.

First, we determine bounds on the image size, and the probability of,
the set $\mcf{A}'\subseteq \mcf{A}$ for which a minimum $\alpha$-quasi
image of $\mbf{X}$ by $p_{Y|X}$ is also an $\epsilon$-image of
$\mcf{A}'$ by $p_{Y|X}$.
\begin{lemma} 
\label{lem:ApsubA} 
Given $\alpha$, $\epsilon \in (0,1)$ and
$p_{Y|X}\in \mcf{P}(\mcf{Y}|\mcf{X})$, let $\mcf{B}\subset\mcf{Y}^n$
be a minimum $\alpha$-quasi image of $\mbf{X}$ by
$p_{Y|X}$. If
\begin{equation*}
\mcf{A}' \defn \set{ \mbf{x} \in \mcf{A} : p_{Y|X}^n (\mcf{B}|\mbf{x}) \geq \epsilon},
\end{equation*}
then 
\begin{equation*}
p_{\mbf{X}}(\mcf{A}' ) \geq  \frac{\alpha - \epsilon}{1 - \epsilon} ,
\end{equation*}
and
\begin{equation*}
\logt \g{Y|X}{\mcf{A}'}{1-\beta} \leq \logt \bg{Y|X}{\mbf{X}}{\alpha} + n\tau_n(\epsilon , \beta),
\end{equation*}
for each $\beta \in (0,1)$.
\end{lemma}
\begin{IEEEproof}
  Not only is $\mcf{B}$ a minimum $\alpha$-quasi image of $\mbf{X}$
  by $p_{Y|X}$, but clearly $\mcf{B}$ is an $\epsilon$-image of
  $\mcf{A}'$ by $p_{Y|X}$. Hence
\begin{align*}
\logt \bg{Y|X}{\mbf{X}}{\alpha} 
&= \logt|\mcf{B}| 
\geq 
\logt \g{Y|X}{\mcf{A}'}{\epsilon} \\
&\geq 
\logt \g{Y|X}{\mcf{A}'}{1-\beta} - n\tau_n(\epsilon , \beta)
\end{align*}
by Lemma~\ref{lem:6.6}. Furthermore as $\mcf{B}$ is a minimum
$\alpha$-quasi image of $\mbf{X}$ by $p_{Y|X}$, we have
\begin{align}
\alpha &\leq p_{\mbf{Y}}(\mcf{B}) 
\leq \sum_{\mbf{x} \in \mcf{A}} p_{Y|X}^n (\mcf{B}|\mbf{x}) p_{\mbf{X}}(\mbf{x}) 
\notag \\
&= \sum_{\mbf{x} \in \mcf{A}' } p_{Y|X}^n (\mcf{B}| \mbf{x})p_{\mbf{X}}(\mbf{x})  
%\notag \\ &\hspace{20pt} 
+ \sum_{\mbf{x} \in \mcf{A} \setminus \mcf{A}' } p_{Y|X}^n(\mcf{B}| \mbf{x}) p_{\mbf{X}}(\mbf{x}) 
\notag \\
&\leq p_{\mbf{X}}(\mcf{A}') + (1 - p_{\mbf{X}}(\mcf{A}' )) \epsilon \label{eq:ApsubA:end}
\end{align}
where~\eqref{eq:ApsubA:end} is because $\mcf{A}'$ contains all
$\mbf{x} \in \mcf{A}$ (recall, this is the support set of $\mbf{X}$)
such that $p^n_{Y|X}(\mcf{B}|\mbf{x}) \geq \epsilon$. In turn,
\eqref{eq:ApsubA:end} then implies
$p_{\mbf{X}}(\mcf{A}') \geq (\alpha - \epsilon)/(1- \epsilon)$.
\end{IEEEproof}

\begin{lemma}
\label{lem:barg_bound}
For each $s \in [0:t]$, $\bigcup_{i=0}^{s} \mcf{S}_{\mbf{Y}}(i)$ is
the unique minimum $\eta_{s}$-quasi image of $\mbf{X}$ by $p_{Y|X}$.
Furthermore
\begin{equation}\label{eq:barg_ub}
\logt \bar g^n_{Y|X}(\mbf{X},\eta_s)  < s\lambda+ \lambda + \logt(t + 1),
\end{equation}
and
\begin{equation}\label{eq:barg_lb}
\logt \bar g^n_{Y|X}(\mbf{X},\eta_s)  \geq  s\lambda + \logt  p_{\mbf{Y}}(\mcf{S}_{\mbf{Y}}(s)) .
\end{equation}
\end{lemma}
\begin{IEEEproof}
  Assume first that $\bigcup_{i=0}^{s} \mcf{S}_{\mbf{Y}}(i)$ is the
  unique minimum $\eta_{s}$-quasi image of $\mbf{X}$ by
  $p_{Y|X}$. Under this assumption, it follows that
\begin{align*}
|\mcf{S}_{\mbf{Y}}(s)| \leq \bg{Y|X}{\mbf{X}}{\eta_s} = \sum_{i=0}^s | \mcf{S}_{\mbf{Y}}(i)|,
\end{align*}
and thus~\eqref{eq:barg_ub} and~\eqref{eq:barg_lb} follow by
Corollary~\ref{lem:bk_size}.

Therefore what remains to be proven is that
$\bigcup_{i=0}^{s} \mcf{S}_{\mbf{Y}}(i)$ is the unique minimum
$\eta_{s}$-quasi image of $\mbf{X}$ by $p_{Y|X}$. Clearly
$\bigcup_{i=0}^{s} \mcf{S}_{\mbf{Y}}(i)$ is an $\eta_{s}$-quasi image
by definition. Hence assume, in hopes of a contradiction, that there
exists a
\[
  \mcf{B}\subseteq \mcf{Y}^n : \left\{ \begin{array}{c} \mcf{B} \neq
    \bigcup_{i=0}^{s} \mcf{S}_{\mbf{Y}}(i) \\ p_{\mbf{Y}}(\mcf{B})
    \geq \eta_s\\ |\mcf{B}| \leq | \bigcup_{i=0}^{s}
    \mcf{S}_{\mbf{Y}}(i) | \end{array} \right\}.
\]
For $s=t$, $\mcf{B}$ cannot be a minimum $\eta_{t}$-quasi image of
$\mbf{X}$ by $p_{Y|X}$, since $\bigcup_{i=0}^{t} \mcf{S}_{\mbf{Y}}(i)$
contains the entire support set of $\mbf{Y}$, which is also the
minimum $1$-quasi image by definition. On the other hand, for
$s \in [0:t-1]$,
\begin{align}
& \hspace{-10pt} p_{\mbf{Y}}(\mcf{ B}) \notag \\
 &= p_{\mbf{Y}} \left(\bigcup_{i=0}^{s} \mcf{S}_{\mbf{Y}}(i) \right) - 
   p_{\mbf{Y}} \left(\bigcup_{i=0}^{s} \mcf{S}_{\mbf{Y}}(i)\setminus
   \mcf{ B} \right) 
\notag \\ &\hspace{20pt}
+ p_{\mbf{Y}} \left(\mcf{ B} \setminus \cup_{i=0}^{s}
             \mcf{S}_{\mbf{Y}}(i) \right)
\notag \\
& =  \eta_{s} - \sum_{i=0}^{s} \sum_{\mbf{y} \in\mcf{S}_{\mbf{Y}}(i) \setminus \mcf{B}}
\hspace{-10pt}p_{\mbf{Y}}(\mbf{y})
%\notag \\ &\hspace{20pt}
+ \sum_{i = s+ 1}^{t} \sum_{\mbf{y} \in  \mcf{S}_{\mbf{Y}}(i) \cap \mcf{B}} \hspace{-10pt}p_{\mbf{Y}}(\mbf{y}) 
             \label{eq:barg_bound:1}\\
&\leq  \eta_{s} 
- \left( \left|\bigcup_{i=0}^{s} \mcf{S}_{\mbf{Y}}(i) \right| - 
  \left|\bigcup_{i=0}^{s} \mcf{S}_{\mbf{Y}}(i) \cap \mcf{B}\right| \right) 2^{-s\lambda} %\\
\notag \\ &\hspace{20pt} 
+ \left( \abs{\mcf{B}} - |\bigcup_{i=0}^{s} \mcf{S}_{\mbf{Y}}(i)  \cap \mcf{B}| \right) 2^{-(s+1) \lambda}  
  \notag\\
&< \eta_{s} \label{eq:barg_bound:2}
\end{align}
where~\eqref{eq:barg_bound:1} is because the support set of $Y$ is
$\bigcup_{i=0}^{t} \mcf{S}_{\mbf{Y}}(i)$, and~\eqref{eq:barg_bound:2}
is because $|\mcf{B}| \leq |\bigcup_{i=0}^{s} \mcf{S}_{\mbf{Y}}(i)|$
by assumption. But this is a contradiction since $\mcf{B}$ is an
$\eta_s$-quasi image of $\mbf{X}$ by $p_{Y|X}$.
\end{IEEEproof}

Next we need to introduce an entropy spectrum equivalent to the
inequality 
$|\mathbb{H}(\mbf{Y}) - \mathbb{H}(\mbf{Y}|U)| \leq \logt
|\mcf{U}|$. This is done in two parts.
\begin{lemma}
\label{lem:dacosupport}
Fix any $U: \{U \markov \mbf{X} \markov \mbf{Y}\}$ and  any $u \in
\mcf{U}$.  Let $\mcf{A}_u$ be the support set of $\mbf{X}|\{ U = u\}$.
If $\logt \g{Y|X}{\mcf{A}_u}{1-\beta} \leq c$ for some
$c\in \mathbb{R}_+$ and $\beta \in (0,1)$, then
\begin{equation}\label{eq:daco:1}
\Pr \left( h_{\mbf{Y}|U}(\mbf{Y}|U) \leq \mu  +  c \middle| U=u \right) \geq 1 - 2^{-\mu } - \beta
\end{equation}
for each $\mu \in \mathbb{R}_+$.
\end{lemma}
\begin{IEEEproof}
  First observe that an upper bound on the minimum $(1-\beta)$-image
  of $\mcf{A}_u$ by $p_{Y|X}$ also yields an upper bound on the minimum
  $(1-\beta)$-quasi image of $\mbf{X}_{u} \defn \mbf{X}|\{ U = u\}$ by
  $p_{Y|X}$. In other
  words,
\begin{equation} \label{eq:bgXu}
\bg{Y|X}{\mbf{X}_{u}}{1-\beta}\leq  \g{Y|X}{\mcf{A}_u}{1-\beta} \leq
2^c,
\end{equation}
which follows from the assumption that
$U \markov \mbf{X} \markov \mbf{Y}$ and from the definitions of images
and quasi-images.
%since the support set of $\mbf{X}_u$ is a subset of $\mcf{A}$.

Now letting $\mcf{B}$ denote a minimum $(1-\beta)$-quasi image of
$\mbf{X}_u$ by $p_{Y|X}$, the result follows since
\begin{align*}
&\hspace*{-10pt} \Pr \left( h_{\mbf{Y}|U}(\mbf{Y}|U) > \mu + c  \middle| U = u \right) \notag \\
&\leq 1 - \Pr \left( \mbf{Y} \in \mcf{B}\middle|U=u \right) 
\notag \\ &\hspace{20pt}
+ \Pr \left( h_{\mbf{Y}|U}(\mbf{Y}|U) > \mu + c, \mbf{Y} \in \mcf{B}  \middle| U=u \right) 
  \notag \\
&\leq \beta + \sum_{\mbf{y} \in \mcf{B}: p_{\mbf{Y}|U}(\mbf{y}|u) < 2^{-\mu - c} } p_{\mbf{Y}|U}(\mbf{y}|u)  
  \notag \notag \\
&< \beta + 2^{-\mu}, 
\end{align*}
where the last inequality results from applying~\eqref{eq:bgXu}. 
%and the assumption that $\abs{\mcf{B}} \leq 2^{c}$.
\end{IEEEproof}

\begin{lemma}
\label{lem:daco=taco}
\begin{equation*}
h_{\mbf{Y}|U}(\mbf{y}|u)\geq h_{\mbf{Y}}(\mbf{y}) + \logt p_{U}(u)
\end{equation*}
for any $u \in \mcf{U}$ and $U : \{ U \markov \mbf{X} \markov \mbf{Y}\}$.
\end{lemma}
\begin{IEEEproof}
  We can upper bound $p_{\mbf{Y}|U}(\mbf{y}|u)$ in terms of
  $p_{\mbf{Y}}(\mbf{y})$, for any $u \in \mcf{U}$ and
  $U : \{ U \markov \mbf{X} \markov \mbf{Y}\}$, as follows
\begin{align}
%& \hspace*{-5pt} 
p_{\mbf{Y}|U}(\mbf{y}|u) 
& = \sum_{\mbf{x} \in \mcf{A} } p_{Y|X}^n(\mbf{y}|\mbf{x}) p_{\mbf{X} | U}(\mbf{x}|u) 
\notag \\
&\leq \frac{1}{p_{U}(u)} \sum_{\mbf{x} \in \mcf{A}} p_{Y|X}^n(\mbf{y}|\mbf{x}) p_{\mbf{X}}(\mbf{x})  
  \notag \\
&= \frac{ p_{\mbf{Y}}(\mbf{y})}{p_{U}(u)}. \label{eq:daco=taco:1}
\end{align}
%Therefore
% \begin{align}
% p_{\mbf{Y}|U}(\mbf{y}|u) &\leq p_{\mbf{Y}}(\mbf{y}) 2^{- \logt p_{U}(u)},
% \end{align}
Taking $-\logt$ on both sides of~\eqref{eq:daco=taco:1} proves the
lemma.
\end{IEEEproof}

\subsection{Proof of Theorem~\ref{lem:daco}}\label{sec:daco proof}

Recall from the statement of Theorem~\ref{lem:daco} that $n \geq 27$,
$|\mcf{Y}|\geq 2$, and
$\alpha \in \left(\frac{\logt n}{n}, \frac{1}{8 \ln 2} \right)$.  For
the proof, we set
\begin{align*}
\lambda =2 (\logt n)( 1 - n^{-1} \logt n)^{-1} (\logt \abs{\mcf{Y}}) \leq 4 |\mcf{Y}| \logt n 
\end{align*}
and 
\begin{align*}
t = 2\lambda ^{-1} n \logt |\mcf{Y}| = \frac{n}{\logt n}-1,
\end{align*}
for which $t \lambda = 2n \logt \abs{\mcf{Y}} >
\logt(|\mcf{Y}|^n)$. Once again, we write
$\mcf{S}_{\mbf{Y}}(i) =
\mcf{S}_{\mbf{Y}}(i;\lambda,t)$ in order to simply notation.

\begin{IEEEproof}
  First we will identify the set $\mcf{A}^\dagger$, and then
  prove~\eqref{eq:daco:property2} before
  proving~\eqref{eq:daco:property1}. In the act of
  proving~\eqref{eq:daco:property2}, we will by necessity prove the
  existence of the positive real numbers
  $\delta = O(-\sqrt{\alpha} \logt \alpha)$ and $r$ described in the
  statement of the theorem.

  To establish $\mcf{A}^\dagger$, observe there exists at least one
  $s^* \in \left[0:t\right]$ such that
  $p_{\mbf{Y}}(\mcf{S}_{\mbf{Y}}(s^*)) \geq (t+1)^{-1} = n^{-1}\logt
  n$ since $\sum_{i=0}^t p_{\mbf{Y}}(\mcf{S}_{\mbf{Y}}(i)) =
  1$. Furthermore $s^* \neq t$, since
  $p_{\mbf{Y}}(\mbf{y}) \leq \abs{\mcf{Y}}^{-2n}$ for each
  $\mbf{y} \in \mcf{S}_{\mbf{Y}}(t)$, and thus
  $p_{\mbf{Y}}(\mcf{S}_{\mbf{Y}}(t)) \leq \abs{\mcf{Y}}^{-n} < n^{-1}
  \logt n$. The theorem follows by setting
\begin{equation} \label{eq:daco:adag}
\mcf{A}^\dagger = \mcf{A}^+ \setminus \mcf{A}^-
\end{equation}
where
\begin{align}
\mcf{A}^+ 
&\defn \set{ \mbf{x} \in \mcf{A} : p_{Y|X}^n\left(
    \bigcup_{i=0}^{s^*} \mcf{S}_{\mbf{Y}}(i) \middle| \mbf{x} \right) \geq 2^{-n\alpha}  } 
\label{eq:daco:a+},\\
\mcf{A}^- 
&\defn \set{ \mbf{x} \in \mcf{A} : p_{Y|X}^n \left(
  \bigcup_{i=0}^{s^-} \mcf{S}_{\mbf{Y}}(i) \middle| \mbf{x} \right) \geq  2^{-n\alpha}   } 
\label{eq:daco:a-}, \\
s^- 
&\defn \lfloor s^* - n \lambda^{-1}  \tilde \delta \rfloor ,
\notag \\
\tilde \delta 
&\defn  \tau_n(2^{-n\alpha}, 2^{-n\alpha}) + \alpha + 7.19|\mcf{Y}| \frac{\logt n}{n}  .
\notag 
\end{align} 
Note that if $s^- < 0$ in the above definition, we have
$\bigcup_{i=0}^{s^-} \mcf{S}_{\mbf{Y}}(i) = \emptyset$ and hence
$\mcf{A}^- = \emptyset$.

%\begin{align} 
%&\Pr \left( \left| h(\mbf{Y}|U) - s^* \lambda_n \right| > O(n\sqrt{\alpha}\ln\alpha) + h(U)  \middle|U=u \right)  
%\notag \\
%&\hspace{150pt} < 3 \cdot 2^{-n\alpha} \label{eq:daco:prop1}.
%\end{align}
%Equation~\eqref{eq:daco:prop1} will follow from

Having identified the set $\mcf{A}^\dagger$, we move on to
proving~\eqref{eq:daco:property2}, and to that end consider any DRV
$U : \{ U \markov \mbf{X} \markov \mbf{Y}\}$ and
$u \in \mcf{U} : \{ \Pr ( \mbf{X} \in \mcf{A}^\dagger | U=u) = 1
\}$. Notice that such a $U$ always exists, e.g., consider $U$ as the
indicator of $\mcf{A}^\dagger$.  Assume for the moment that
\begin{align}
\Pr \left( h(\mbf{Y}|U) > s^* \lambda + n \tilde \delta \middle| U=u \right) 
&\leq 2\cdot 2^{-n\alpha } 
\label{eq:daco:prove1}
\\
\Pr \left( h(\mbf{Y}|U) \leq  s^-\lambda -h(U) \middle|U=u \right) 
&< 2^{-n\alpha }.
\label{eq:daco:prove2}
\end{align}
Clearly, applying the union bound
with~\eqref{eq:daco:prove1},~\eqref{eq:daco:prove2} gives
\begin{align}
\Pr \left( |h(\mbf{Y}|U) -  s^* \lambda |  < n \delta  + h(U) \middle| U= u \right) < 3\cdot 2^{-n\alpha } 
  \label{eq:daco:prop1}.
\end{align}
for a positive real number $\delta = O(-\sqrt{\alpha} \logt \alpha)$.
This therefore validates~\eqref{eq:daco:property2}, since $s^*\lambda$ is a constant less than $\lambda t = 2n\logt|\mcf{Y}|$. A detailed verification of the
order term of $\delta$ is provided in Appendix~\ref{app:ot:daco:prop1p}. 

Let us now turn to confirming~\eqref{eq:daco:prove1}
and~\eqref{eq:daco:prove2}. To verify~\eqref{eq:daco:prove1},
first observe
\begin{align}
& \hspace*{-5pt}
\logt \g{Y|X}{\mcf{A}^+}{1-2^{-n\alpha}} 
\notag  \\ &
\leq \logt \bg{Y|X}{\mcf{\mbf{X}}}{\eta_{s^*}}  + n\tau_{n}(2^{-n\alpha},2^{-n\alpha}) \label{eq:daco:bb:-1}  
 \\ &
< s^* \lambda + \lambda + \logt(t+1) + n\tau_n(2^{-n\alpha},2^{-n\alpha}) \label{eq:daco:bb} 
\\
&\leq  s^* \lambda +  4 |\mcf{Y}|\logt n + \logt \left( \frac{n}{\logt n} \right) +
  n\tau_n(2^{-n\alpha},2^{-n\alpha})   \notag \\
&\leq  s^* \lambda +  7.19 |\mcf{Y}|\logt n + n\tau_n(2^{-n\alpha},2^{-n\alpha})
  \notag \\
&=  s^* \lambda +  n\tilde \delta - n\alpha . \label{eq:daco:rotd}  
\end{align}
where~\eqref{eq:daco:bb:-1} results by applying Lemma~\ref{lem:ApsubA}
because of the definition of $\mcf{A}^+$ and the fact that
$\bigcup_{i=0}^{s^*}\mcf{S}_{\mbf{Y}}(i)$ is a minimum
$\eta_{s^*}$-quasi image of $\mbf{X}$ by $p_{Y|X}$,
while~\eqref{eq:daco:bb} is by
Lemma~\ref{lem:barg_bound}. Equation~\eqref{eq:daco:prove1} now
directly follows from~\eqref{eq:daco:rotd} and
Lemma~\ref{lem:dacosupport} since the support set of
$\mbf{X}|\set{U=u}$ is a subset of $\mcf{A}^+$.

On the other hand~\eqref{eq:daco:prove2} can be derived as follows
\begin{align}
&\hspace*{-10pt} \Pr\left( h(\mbf{Y}|U) \leq s^- \lambda - h(U) \middle|U=u \right)\notag \\
&\leq \Pr\left( h(\mbf{Y})\leq s^- \lambda \middle|U=u \right) \label{eq:daco:prove2:-3}\\
&\leq p_{\mbf{Y}|U} \left(\bigcup_{i=0}^{s^-} \mcf{S}_{\mbf{Y}}(i)
  \middle |u \right) \notag \\
&= \sum_{\mbf{x} \in \mcf{A}^\dagger} p^n_{Y|X} \left(
  \bigcup_{i=0}^{s^-} \mcf{S}_{\mbf{Y}}(i) \middle|
  \mbf{x} \right)p_{\mbf{X}|U}(\mbf{x}|u) \notag \\
& <  2^{-n\alpha} \label{eq:daco:prove2:0}
\end{align}
where~\eqref{eq:daco:prove2:-3} is by Lemma~\ref{lem:daco=taco},
and~\eqref{eq:daco:prove2:0} is because $\mcf{A}^-$ contains all
$\mbf{x}\in \mcf{A}$ such that
$p^n_{Y|X}(\cup_{i=0}^{s^-} \mcf{S}_{\mbf{Y}}(i)|\mbf{x}) \geq
2^{-n\alpha}$ yet $\mcf{A}^\dagger \cap \mcf{A}^- = \emptyset$.

Having identified the set $\mcf{A}^\dagger$, the positive real numbers
$\delta$ and $r$, and proven~\eqref{eq:daco:property2}, we now move on
to prove~\eqref{eq:daco:property1}. 
To do so, 
we start by noting that if $\mcf{A}^- \neq \emptyset$, 
\begin{align}
&\logt g^n_{Y|X}(\mcf{A}^-,1-2^{-n\alpha})
\notag \\ &\hspace{10pt}
\leq \logt \bar g^n_{Y|X}(\mbf{X},\eta_{s^-} ) + n \tau_{n}(2^{-n\alpha},2^{-n\alpha}) \label{eq:bkpmg:-2}\\
&\hspace{10pt}
< s^- \lambda  + \lambda + \logt(t+1) + n \tau_{n}(2^{-n\alpha},2^{-n\alpha})  \label{eq:bkpmg:-1} \\
&\hspace{10pt}
\leq s^* \lambda  - n\alpha \label{eq:bkpmg} 
\end{align}
where~\eqref{eq:bkpmg:-2} is due to Lemma~\ref{lem:ApsubA} and the
definition of $\mcf{A}^-$ and
$\bigcup_{i=0}^{s^-}\mcf{S}_{\mbf{Y}}(i)$ being a minimum
$\eta_{s^-}$-quasi image of $\mbf{X}$ by $p_{Y|X}$;
while~\eqref{eq:bkpmg:-1} is due to Lemma~\ref{lem:barg_bound};
finally~\eqref{eq:bkpmg} follows by applying the inequality that
$s^*\lambda \geq s^-\lambda + n\tilde \delta$ and then the same chain
of inequalities from~\eqref{eq:daco:bb}
through~\eqref{eq:daco:rotd}. Also, we have
$g^n_{Y|X}(\mcf{A}^-,1-2^{-n\alpha}) = 0$ trivially if
$\mcf{A}^- = \emptyset$,

Now, let $\mcf{B}^-$ be a minimum $(1 - 2^{-n\alpha})$-image of
$\mcf{A}^-$ by $p_{Y|X}$.
%We will show that the probability of the intersection of $\mcf{A}^-$ and $\mcf{A}^+$ is necessarily small because $\mcf{B}^-$ is much smaller than $\mcf{S}_{\mbf{Y}}(s^*)$. Because the intersection is small, this in turn implies that the probability of $\mcf{A}^\dagger$ is bounded away from zero. 
A lower bound on the probability of
$\mcf{S}_{\mbf{Y}}(s^*)\setminus \mcf{B}^-$ can be constructed as
follows:
\begin{align}
&\hspace{-20pt}p_{\mbf{Y}}(\mcf{S}_{\mbf{Y}}(s^*) \setminus \mcf{B}^-) \notag \\
&= p_{\mbf{Y}}(\mcf{S}_{\mbf{Y}}(s^*)  )  -
  p_{\mbf{Y}}(\mcf{S}_{\mbf{Y}}(s^*)  \cap \mcf{B}^-)   \notag \\
&\geq \frac{\logt n}{n} -   2^{-s^* \lambda}\left| \mcf{S}_{\mbf{Y}}(s^*) \cap \mcf{B}^- \right| \label{eq:finaldacodown:-2} \\
& \geq \frac{\logt n}{n} -  2^{-s^* \lambda}
  g^n_{Y|X}(\mcf{A}^-,1-2^{-n\alpha})  \notag \\%\label{eq:finaldacodown:-1} \\
&\geq  \frac{\logt n}{n} -  2^{-n\alpha} \label{eq:finaldacodown}
\end{align}
where~\eqref{eq:finaldacodown:-2} is because
$p_{\mbf{Y}}(\mbf{y}) \leq 2^{-s^* \lambda}$ for all
$\mbf{y} \in \mcf{S}_{\mbf{Y}}(s^*)$, and~\eqref{eq:finaldacodown} is
due to~\eqref{eq:bkpmg} when $\mcf{A}^- \neq \emptyset$ and due to
$g^n_{Y|X}(\mcf{A}^-,1-2^{-n\alpha}) = 0$ when
$\mcf{A}^- = \emptyset$. But this implies a lower bound on the
probability of $\mcf{A}^\dagger$ since now
\begin{align}
&\hspace*{-10pt} \frac{\logt n}{n} - 2^{-n\alpha} \notag \\
&\leq p_{\mbf{Y}} (\mcf{S}_{\mbf{Y}}(s^*) \setminus \mcf{B}^-)  \notag
\\
&= \sum_{\mbf{x} \in \mcf{A}^- } p^n_{Y|X}(\mcf{S}_{\mbf{Y}}(s^*) \setminus \mcf{B}^- | \mbf{x} ) p_{\mbf{X}}(\mbf{x}) \notag \\
&\hspace{5pt} +\sum_{\mbf{x} \in \mcf{A}^\dagger } p^n_{Y|X}( \mcf{S}_{\mbf{Y}}(s^*)\setminus \mcf{B}^- | \mbf{x} ) p_{\mbf{X}}(\mbf{x}) \notag \\
&\hspace{5pt} + \sum_{\mbf{x} \in \mcf{A} \setminus \set{\mcf{A}^\dagger \cup \mcf{A}^-}  }  p^n_{Y|X}(\mcf{S}_{\mbf{Y}}(s^*) \setminus \mcf{B}^- | \mbf{x}) p_{\mbf{X}}(\mbf{x})  \label{eq:daco:fin:0}\\
&\leq 2^{-n\alpha} + p_{\mbf{X}}(\mcf{A}^\dagger) + 2^{-n\alpha} \label{eq:daco:fin:1}
\end{align}
where each term in~\eqref{eq:daco:fin:1} bounds the corresponding term
in~\eqref{eq:daco:fin:0}. In particular, the bound on the first term
in~\eqref{eq:daco:fin:0} is due to the fact that each
$\mbf{x} \in \mcf{A}^-$ satisfies
$1 - p^n_{Y|X}(\mcf{B}^- | \mbf{x})\leq 2^{-n\alpha}$. On the other
hand, the bound on the third term in~\eqref{eq:daco:fin:0} is because
$\mcf{A}^\dagger \cup \mcf{A}^- = \mcf{A}^+ \cup \mcf{A}^- $ contains
all $\mbf{x} \in \mcf{A}$ such that
$p^n_{Y|X} \left( \mcf{S}_{\mbf{Y}}(s^*) \middle| \mbf{x} \right) \geq
2^{-n\alpha}$. Solving~\eqref{eq:daco:fin:1} for
$p_{\mbf{X}}(\mcf{A}^\dagger)$ and simplifying, we have
\begin{align*}
p_{\mbf{X}}(\mcf{A}^\dagger)
& \geq \frac{\logt n}{n} - 3 \cdot 2^{-n\alpha}   \\
&\geq \frac{\logt n}{n} - \frac{3}{n} = \frac{1}{n} \logt \frac{n}{8}
%\label{eq:Aminus},
\end{align*}
since $\alpha \geq n^{-1} \logt n$.
\end{IEEEproof}

\section{Proof of Theorem~\ref{thm:mt}; Information stable partitioning}\label{sec:mt}
Throughout this section we will once again assume
$(\emptyset,\mbf{X},\mbf{Y})$ are a regular collection of DRVs, and we
will let $\mcf{A}$ denote the support set of $\mbf{X}$. Also, as
before, we first present a few key lemmas.

\subsection{Supporting lemmas}

The first lemma repeatedly applies Theorem~\ref{lem:daco} in order to
construct a DRV $V$ which provides stability when conditioned
upon. That is, we apply Theorem~\ref{lem:daco} to $\mcf{A}$ obtaining
a subset which gives stability. This subset is then removed from
$\mcf{A}$, and Theorem~\ref{lem:daco} is applied to the remaining set
again to obtain a new subset. This process is then repeated a number
of times. The random variable $V$ is then an index to the stable
subset that $\mbf{X}$ belongs.

\begin{lemma}
\label{lem:part1}
Given any regular collection $(\emptyset, \mbf{X},\mbf{Y})$, positive
real number
$\alpha \in \left( \frac{\logt n}{n}, \frac{1}{8 \ln 2} \right)$, and
$\zeta \in \mathbb{N}_+$, there exists:
\begin{itemize}[leftmargin=*]
\item a DRV $V : \left\{ \begin{array}{c} \mcf{V} = [0:\zeta-1] \\ 
                           V \ll \mbf{X} \\
                           h_{V}(0) > \frac{\zeta-1}{n} \logt
                           \frac{n}{8} \end{array}   \right\}$, \vspace{3pt}
\item positive real number $\delta = O(-\sqrt{\alpha} \logt \alpha)$, and 
\item function $r: \mcf{V} \rightarrow   \mathbb{R}_+$ such that 
\begin{align*}
\max_{v} r(v) < 2n \logt |\mcf{Y}|
\end{align*}
and
\begin{align}
\Pr \left( \abs{h(\mbf{Y}|U) - r(V)} > n \delta + h(U) \middle| U=u \right) 
  <  3 \cdot 2^{-n\alpha}, \label{eq:lem:part1:2}
\end{align}
for all DRV
$U: \left\{ \begin{array}{c} U \gg V \\ U \markov \mbf{X} \markov
    \mbf{Y} \end{array} \right\}$ and
$u \in \mcf{U}:\{p_{V|U}(0|u) = 0\}$.
\end{itemize}
\end{lemma}
\begin{IEEEproof}
%  For a given $\mbf{X}$ with support set $\mcf{A}$
By Theorem~\ref{lem:daco} there
  exists a set $\mcf{A}_1^\dagger \subseteq \mcf{A}$ where
\begin{align}
p_{\mbf{X}}(\mcf{A}_1^\dagger) \geq \frac{1}{n} \logt \frac{n}{8},\label{eq:part1:daco:idkwhattocallthis1}
\end{align}
and positive real numbers $\delta_1 = O(-\sqrt{\alpha} \logt \alpha)$ and $r_1< 2n \logt |\mcf{Y}|$ such that
\begin{align*}
\Pr \left( \abs{h(\mbf{Y}|U) - r_1} > n \delta_1 + h(U) \middle| U=u \right) <  3 \cdot 2^{-n\alpha}
\end{align*}
for all DRV $U : \{ U \markov \mbf{X} \markov \mbf{Y} \}$ and
$u \in \mcf{U}: \{ \Pr \left( \mbf{X} \in \mcf{A}_1^\dagger \middle| U
  =u \right) = 1 \}$. Now for each $i \in [2:\zeta-1]$, given the
recursively defined
$\mbf{X}_i \defn \mbf{X} | \{ \mbf{X} \notin \bigcup_{j=1}^{i-1}
\mcf{A}_j^\dagger \}$ there exists a
$\mcf{A}_i^\dagger\subseteq \mcf{A}\setminus \bigcup_{j=1}^{i-1}
\mcf{A}_j^\dagger$ such that
\begin{align}
p_{\mbf{X}_i}(\mcf{A}_i^\dagger) \geq \frac{1}{n} \logt \frac{n}{8}, \label{eq:part1:daco:idkwhattocallthis2}
\end{align}
and positive real numbers $\delta_i = O(-\sqrt{\alpha} \logt \alpha)$
and $r_i < 2n \logt|\mcf{Y}|$ where
\begin{align*}
\Pr \left( \abs{h(\mbf{Y}|U) - r_i} > n \delta_i + h(U) \middle| U=u \right) <  3 \cdot 2^{-n\alpha}
\end{align*}
for all DRV $U : \{ U \markov \mbf{X} \markov \mbf{Y} \}$ and
$u \in \mcf{U}: \{ \Pr \left( \mbf{X} \in \mcf{A}_i^\dagger \middle| U
  =u \right) = 1 \}$.

Now, define the following:
\begin{itemize}[leftmargin=*]
\item DRV
\begin{align*}
V = \sum_{i=1}^{\zeta-1} i \cdot \idc{ \mbf{X} \in \mcf{A}^\dagger_i}
\end{align*}
\item positive real number
  $\delta = \max_{i \in [1:\zeta-1]} \delta_i$ (for which clearly
  $\delta = O(-\sqrt{\alpha} \logt \alpha)$), and
\item function $r(i) = r_i$ for $i \in [1:\zeta-1]$ and $r(0) = 3$
  (for which clearly $\max_{v} r(v) < 2n \logt |\mcf{Y}|$).
\end{itemize}
Thus $\mcf{V} = [0:\zeta-1]$ and $\mbf{X} \gg V$ 
% since $V$ is a deterministic function of $\mbf{X}$ and
% $\mcf{A}^\dagger_i$ are mutually exclusive
by construction. Before proving the upper bound on $h_{V}(0)$, note
$V=i$ implies $\mbf{X} \in \mcf{A}_i^\dagger$ for $i \in [1:\zeta-1]$,
and if $V=0$ then
$\mbf{X} \in \mcf{A} \setminus \bigcup_{i=1}^{\zeta-1}
\mcf{A}_i^\dagger$. Hence,
\begin{align}
p_{V}(0) 
&= 
\Pr \left( \mbf{X} \in \mcf{A} \setminus \bigcup_{i=1}^{\zeta-1}
  \mcf{A}^\dagger_i \right) \notag \\
% &=
% \Pr \letf( \mbf{X} \in \mcf{A} \setminus \bigcup_{i=1}^{\zeta-1}
%   \mcf{A}^\dagger_i \middle|  \mbf{X} \in \mcf{A} \setminus \cup_{i=1}^{\zeta-2} \mcf{A}^\dagger_i ) \notag\\
% &\hspace{25pt} \cdot \Pr( \mbf{X} \in \mcf{A}\setminus \cup_{i=1}^{\zeta-2} \mcf{A}^\dagger_i) \\
&= 
\prod_{j=1}^{\zeta-1} \Pr \left(\mbf{X}\in  \mcf{A} \setminus
  \bigcup_{i=1}^{j} \mcf{A}^\dagger_i \middle| \mbf{X} \in \mcf{A}
  \setminus \bigcup_{i=1}^{j-1} \mcf{A}^\dagger_i \right)  \notag \\
&= \prod_{j=1}^{\zeta-1} \left(1-p_{\mbf{X}_{j}}(\mcf{A}_j^\dagger )
  \right) 
\notag \\
&\leq  \prod_{j=1}^{\zeta-1} \left(1 - \frac{1}{n} \logt \frac{n}{8} \right)
  \label{eq:part1:daco}\\
&= 2^{(\zeta-1) \logt\left(1 - n^{-1} \logt (n/8) \right)} 
  <
  2^{-\frac{\zeta-1}{n}  \logt \frac{n}{8}} \notag 
\end{align}
where~\eqref{eq:part1:daco} is due
to~\eqref{eq:part1:daco:idkwhattocallthis1}
and~\eqref{eq:part1:daco:idkwhattocallthis2}.

Finally we prove~\eqref{eq:lem:part1:2}. If given a DRV
$U: \left\{ \begin{array}{c} U \gg V\\ U \markov \mbf{X} \markov
    \mbf{Y} \end{array} \right\}$, then for all
$u\in \mcf{U} : \{ p_{V|U}(0|u) = 0\}$ there must exist an
$i \in [1:\zeta-1]$ for which $p_{V|U}(i|u) = 1$ since $V$ is a
deterministic function of $U$. In turn then
$\Pr \left( \mbf{X} \in \mcf{A}^\dagger_i\middle|U =u \right) = 1$
since $V=i$ implies $\mbf{X} \in \mcf{A}_i^\dagger$. Hence it also
follows that
\begin{align*}
3 \cdot 2^{-n\alpha}& > \Pr \left( \abs{h(\mbf{Y}|U) - r_i} > n \delta_i + h(U) \middle| U=u \right) \\
&\geq\Pr \left( \abs{h(\mbf{Y}|U) - r(V) } > n \delta + h(U) \middle| U=u \right),
\end{align*}
establishing~\eqref{eq:lem:part1:2}.
\end{IEEEproof}

Notice that Lemma~\ref{lem:part1} only applies to regular collections
of the form $(\emptyset, \mbf{X},\mbf{Y})$. The next corollary is the
first step in generalizing the result to regular collections of the
form $(M,\mbf{X},\mbf{Y})$. This generalization is achieved by
leveraging Lemma~\ref{lem:part1} with the fact that
$(M,\mbf{X},\mbf{Y})|\{ M = m\}$ is a regular collection for all
$m \in \mcf{M}$.
\begin{cor}
\label{cor:part1}
Given any regular collection $(M, \mbf{X},\mbf{Y})$, positive real
number
$\alpha \in \left( \frac{\logt n}{n}, \frac{1}{8 \ln 2} \right)$, and
$\zeta \in \mathbb{N}_+$, there exists:
\begin{itemize}[leftmargin=*]
\item a DRV $V : \left\{ \begin{array}{c} \mcf{V} = [0:\zeta-1] \\ V \ll (\mbf{X},M) \\ h_{V}(0) > \frac{\zeta-1}{n} \logt \frac{n}{8} \end{array}   \right\}$,\vspace{2pt}
\item positive real number $\delta = O(-\sqrt{\alpha} \logt \alpha)$, and 
\item function $r: \mcf{V} \times \mcf{M} \rightarrow   \mathbb{R}_+$ such that 
\begin{align*}
\sup_{(v,m) \in \mcf{V} \times \mcf{M}} r(v,m) < 2n \logt |\mcf{Y}|   
\end{align*}
and
\begin{align}
\hspace{-8pt} \Pr \left( \abs{h(\mbf{Y}|U,M) \! - \! r(V,M)} > n \delta + h(U|M) \middle| (U,M)=u,m \right) &\notag \\
&\hspace{-90pt} <  3 \cdot 2^{-n\alpha}, \label{eq:s:cor:part1:2}
\end{align}
for all DRV
$U: \left\{ \begin{array}{c} U \gg V \\ (U,M) \markov \mbf{X} \markov
    \mbf{Y} \end{array} \right\}$ and
$u \in \mcf{U} : \{p_{V|U}(0|u) = 0 \} $.
\end{itemize}
\end{cor}
\begin{IEEEproof}
  Let us be given a regular collection $(M ,\mbf{X},\mbf{Y})$. For
  each $m \in \mcf{M}$, let $(\mbf{X}_{(m)},\mbf{Y}_{(m)})$ be DRVs
  defined by setting their distributions according to
\begin{align*}
p_{\mbf{X}_{(m)},\mbf{Y}_{(m)}}(\mbf{x},\mbf{y}) = p_{\mbf{X},\mbf{Y}|M}(\mbf{x},\mbf{y}|m).
\end{align*} 
It is clear then that the set
$(\emptyset, \mbf{X}_{(m)},\mbf{Y}_{(m)})$ is also a regular
collection. Thus for each $m \in \mcf{M}$, there exists:
\begin{itemize}[leftmargin=*]
\item a DRV
  $V_{(m)} : \left\{ \begin{array}{c} \mcf{V}_{(m)} = [0:\zeta-1] \\
      V_{(m)} \ll \mbf{X}_{(m)} \\ h_{V_{(m)}}(0) > \frac{\zeta-1}{n}
      \logt \frac{n}{8} \end{array} \right\} $, \vspace{3pt}
\item a positive real number $ \delta_{(m)} = O(-\sqrt{\alpha} \logt \alpha)$, and 
\item function $r_{(m)} : [0:\zeta -1] \rightarrow \mathbb{R}_+$ such that 
\begin{align*}
\max_{v \in [0:\zeta-1]} r_{(m)}(v) < 2n\logt |\mcf{Y}|
\end{align*}
and
\begin{align}
&\hspace{-7pt}\Pr \left(\hspace{-4pt}  \begin{array}{l} \abs{h(\mbf{Y}_{(m)}|U) - r_{(m)}(V_{(m)})}  >n  \delta_{(m)} + h(U) \end{array} \hspace{-4pt} \middle|U = u\right) \notag \\
&\hspace{110pt} <  3 \cdot 2^{-n\alpha}, \label{eq:p:cor:part1:-10b}
\end{align}
for all
$U: \left\{ \begin{array}{c} U \gg V_{(m)} \\ U \markov \mbf{X}_{(m)}
    \markov \mbf{Y}_{(m)} \end{array} \right\} $ and
$u \in \mcf{U} : \set{ p_{V_{(m)}|U}(0|u) = 0 }$
\end{itemize}
by Lemma~\ref{lem:part1}.  

The DRV $V$ in the corollary statement is
constructed from the set of $V_{(m)}$, $m \in \mcf{M}$ by setting
\begin{align}
p_{\mbf{Y},\mbf{X},V|M}(\mbf{y},\mbf{x},v|m) = p_{\mbf{Y}_{(m)},\mbf{X}_{(m)},V_{(m)} }(\mbf{y},\mbf{x},v) \label{eq:p:cor:part1:1}
\end{align}
for each $m \in \mcf{M}$.  Let us now verify the properties of $V$ so
constructed. Clearly
$\mcf{V} = \bigcup_{m \in \mcf{M}} \mcf{V}_{(m)} = [0:\zeta-1]$. Next,
that $V$ is a deterministic function of $(\mbf{X},M)$ follows
from~\eqref{eq:p:cor:part1:1} and $V_{(m)}$ being a deterministic
function of $\mbf{X}_{(m)}$ for each $m \in \mcf{M}$. Finally,
\begin{align*}
p_{V}(0) &= \sum_{m \in \mcf{M}} p_{M}(m) p_{V|M}(0|m) =  \sum_{m \in \mcf{M}} p_{M}(m) p_{V_{(m)}}(0)
  \notag \\
&< \sum_{m \in \mcf{M}} p_{M}(m) 2^{-\frac{\zeta-1}{n} \logt \frac{n}{8}} = 2^{-\frac{\zeta-1}{n} \logt \frac{n}{8}}.
\end{align*}

Further, let $\delta = \sup_{m \in \mcf{M}} \delta_{(m)}$, for which
clearly $\delta = O(-\sqrt{\alpha} \logt \alpha)$ since
$\delta_{(m)}= O(-\sqrt{\alpha} \logt \alpha)$ for all
$m \in \mcf{M}$. Also let function $r(v,m) = r_{(m)}(v)$ for each
$v\in \mcf{V}$ and $m\in \mcf{M}$. Clearly
$\sup_{v,m} r(v,m) < 2n\logt |\mcf{Y}|$, since
$r_{(m)}(v) < 2n\logt|\mcf{Y}|$ for all $m\in \mcf{M}$ and
$v\in \mcf{V}$.

To confirm~\eqref{eq:s:cor:part1:2}, fix any DRV
$U: \left\{ \begin{array}{c} U \gg V \\ (U,M) \markov \mbf{X} \markov
    \mbf{Y} \end{array} \right\}$ and any
$u \in \mcf{U} :\{p_{V|U}(0|u) = 0\}$. Define DRV $U_{(m)}$ by setting
\begin{align}
 p_{\mbf{Y}_{(m)},\mbf{X}_{(m)},V_{(m)} ,U_{(m)} }(\mbf{y},\mbf{x},v,u) = p_{\mbf{Y},\mbf{X},V,U|M}(\mbf{y},\mbf{x},v,u|m)  \label{eq:p:cor:part1:2}
\end{align}
for each $m \in \mcf{M}$. Notice that $U_{(m)} \gg V_{(m)}$ and
$U_{(m)} \markov \mbf{X}_{(m)} \markov \mbf{Y}_{(m)}$ and
$p_{V_{(m)}|U_{(m)}}(0|u) = 0$ clearly follow
from~\eqref{eq:p:cor:part1:2} since $U \gg V$ and
$(U,M) \markov \mbf{X} \markov \mbf{Y}$ and $p_{V|U}(0|u) =
0$. Furthermore \eqref{eq:p:cor:part1:-10b} must hold because
$U_{(m)}$ satisfies these requirements. That is,
\begin{align}
  &\Pr \left(\hspace{-4pt}  \begin{array}{l} \abs{h(\mbf{Y}_{(m)}|U_{(m)} ) \!\! -\!\! r_{(m)}(V_{(m)})} \! > \! n  \delta_{(m)} + h(U_{(m)}) \end{array} \hspace{-4pt} \middle|U = u\right) \notag \\
  &\hspace{150pt} <  3 \cdot 2^{-n\alpha}. \label{eq:p:cor:part1:-10b2}
\end{align}
But, at the same time
\begin{align*}
h_{\mbf{Y}_{(m)}|U_{(m)}}(\mbf{y}|u) &= h_{\mbf{Y}|U,M}(\mbf{y}|u,m)
  \\
h_{U_{(m)}}(u) &= h_{U|M}(u|m)
\end{align*}
for each $\mbf{y}, u,m \in \mcf{Y}^n \times \mcf{U} \times \mcf{M}$,
by~\eqref{eq:p:cor:part1:2}. Hence it follows that
\begin{align}
&\Pr \left( |h(\mbf{Y}|U,M) \!\! -\!\!  r(V,M)| \!\! >\!\! n \delta + h(U|M) \middle| (U,M) = (u,m) \right) \notag \\
&=\sum_{\substack{\mbf{y},v: \\ | h_{\mbf{Y}|U,M} (\mbf{y}|u,m) - r(v,m)|  \notag\\
~~~\geq n \delta + h_{U|M}(u|m)}} p_{\mbf{Y},V|U,M}(\mbf{y},v|u,m) \notag\\
&=\sum_{\substack{\mbf{y},v: \\ | h_{\mbf{Y}_{(m)}|U_{(m)}} (\mbf{y}|u) - r_{(m)}(v)|  \notag\\
~~~\geq n \delta + h_{U_{(m)}}(u)}} p_{\mbf{Y}_{(m)},V_{(m)}|U_{(m)}}(\mbf{y},v|u) \notag\\
&= \Pr \left( |h(\mbf{Y}_{(m)}|U_{(m)}) \!\!-\!\! r_{(m)}(V)| \!>\! n \delta \!+\! h(U_{(m)}) \middle| U_{(m)} = u \right) \label{eq:p:cor:part1:huh}.
\end{align}
Equation~\eqref{eq:s:cor:part1:2} is therefore confirmed from
combining~\eqref{eq:p:cor:part1:huh} and~\eqref{eq:p:cor:part1:-10b2}
since $\delta \geq \delta_{(m)}$ for all $m \in \mcf{M}$.

\end{IEEEproof}

Next, we remove the dependence of the function $r$ upon $M$ since this
will allow for $h(\mbf{Y}|M)$ to stabilize to $\mathbb{H}(\mbf{Y}|M)$.
\begin{lemma}
\label{lem:part2}
Given any regular collection $(M, \mbf{X},\mbf{Y})$, positive real
number
$\alpha \in \left( \frac{\logt n}{n}, \frac{1}{8 \ln 2} \right)$, and
$\zeta \in \mathbb{N}_+$, there exists:
\begin{itemize}[leftmargin=*]
\item a DRV $V: \left\{ \begin{array}{c} |\mcf{V}| \leq 2\zeta n\logt
                          |\mcf{Y}| \\ V \ll (\mbf{X},M) \\ h_{V}(v_0)
                          > \frac{\zeta-1}{n} \logt \frac{n}{8}~
                          \text{for a } v_0 \in \mcf{V}\end{array}   \right\}$,\vspace{3pt}
\item positive real number $\delta = O(-\sqrt{\alpha} \logt \alpha)$, and 
\item function $r: \mcf{V} \rightarrow   \mathbb{R}_+$ such that 
\begin{align}
\hspace{-10pt} \Pr \left( \abs{h(\mbf{Y}|U,M)  -  r(V)} > n \delta + h(U) \middle| U=u\right)  <  4 \cdot 2^{-n\alpha}, \label{eq:s:lem:part2:1}
\end{align}
for all DRV
$U: \left\{ \begin{array}{c} U \gg V\\ (U,M) \markov \mbf{X} \markov
    \mbf{Y} \end{array} \right\}$ and
$u \in \mcf{U} : \{p_{V|U}(v_0|u) = 0 \} $.
\end{itemize}
\end{lemma}
\begin{IEEEproof}
  For any regular collection $(M,\mbf{X},\mbf{Y})$ let $\tilde V$,
  $\tilde \delta$, and $\tilde r$ be the DRV, constant, and function
  respectively guaranteed by Corollary~\ref{cor:part1}. Now define the
  following:
\begin{itemize} 
\item DRV 
%\begin{align}
$\hat V= \left \lfloor  \tilde r(\tilde V,M) \right \rfloor$, %\label{eq:p:lem:part2:vdef}
%\end{align}
\item DRV 
%\begin{align} 
$V = \begin{cases}(\hat V,\tilde V) &\text{ if } \tilde V \neq 0 \\ v_0 &\text{ o.w.} \end{cases}$,
%\end{align}  
\item constant 
\begin{align}
\delta = \tilde \delta + \alpha + \frac{1}{n} = O(-\sqrt{\alpha}
\logt \alpha) \label{eq:lem:part2:p:ot} 
\end{align}
(see Appendix~\ref{app:ot:lem:part2:p:ot} for
  verification of the order term),  and 
\item function $r(V) = \hat V$.
\end{itemize}

First, let us verify the properties of $V$. Clearly
$|\mcf{V}| \leq |\mcf{\hat V}||\mcf{\tilde V}| + 1$, where the
additional term is to account for $v_0$. The upper bound on
$|\mcf{V}|$ now follows since $|\mcf{\hat V}| < 2n\logt|\mcf{Y}|$ and
$|\mcf{\tilde V}|= \zeta$. Next, $(\mbf{X},M) \gg V$ clearly follows
from $(\tilde V,M) \gg V$ and $(\mbf{X},M) \gg \tilde V$. Finally,
$p_{V}(v_0) = p_{\tilde V}(0) < 2^{-\frac{\zeta-1}{n} \logt
  \frac{n}{8}}$.

Now, to prove~\eqref{eq:s:lem:part2:1} fix any DRV
$U : \set{ \begin{array}{c} U \gg V \\ (U,M) \markov \mbf{X} \markov
    \mbf{Y} \end{array} }$, and
$u\in \mcf{U}: \{ p_{V|U}(v_0|u) = 0 \}$, and observe that
\begin{align}
& \hspace*{-5pt}\Pr \left(   |r(V) -  h(\mbf{Y}|M,U)|  >n\tilde \delta + n \alpha + 1 + h(U)   \middle| U=  u \right) \notag\\
&\leq \Pr \left(  |h(\mbf{Y}|M,U) - \tilde r(\tilde V,M)| >n \tilde \delta  + h(U|M)  \middle|U = u \right) \notag \\
&\hspace{10pt} + \Pr \left(  |r(V) -  \tilde r(\tilde V,M) | > 1  \middle| U = u \right) \notag \\
&\hspace{10pt}  + \Pr \left( h(U|M) > h(U) + n\alpha \middle| U= u  \right) \label{eq:p:lem:part2:pred0}.
\end{align}
Indeed, assume the predicates of all three probability terms on the
right hand side (RHS) of~\eqref{eq:p:lem:part2:pred0} fail, then it would also follow that
\begin{align*}
&\hspace{-10pt} | r(V) -  h(\mbf{Y}|M,U)| \notag\\
&\leq |r(V) -  \tilde r(\tilde V,M)| + |h(\mbf{Y}|M,U) - \tilde
  r(\tilde V,M)| \notag \\%\label{eq:p:lem:part2:pred1} \\
&\leq n\tilde \delta   +h(U|M)  + 1 \notag \\ %\label{eq:p:lem:part2:pred2} \\
&\leq n\tilde \delta + n \alpha  + h(U) + 1  = n \delta + h(U). %\label{eq:p:lem:part2:pred3}
\end{align*}
% where~\eqref{eq:p:lem:part2:pred1} is the trianlge inequality, and~\eqref{eq:p:lem:part2:pred2} and~\eqref{eq:p:lem:part2:pred3} are consequences of the predicates of the probability terms on the RHS of~\eqref{eq:p:lem:part2:pred0} being false. 
% If (not A) and (not B) and (not C) imply (not D) then 1- p(D) = p(not D) \geq p(not A and not B and not C) = 1 - p(A or B or C) \geq 1 - p(A) - p(B) - p(C) 
% hence p(A) + p(B) + p(C) \geq p(D)
Equation~\eqref{eq:s:lem:part2:1} therefore follows from
combining~\eqref{eq:p:lem:part2:pred0} with the following three
to-be-proven inequalities:
\begin{align}
\Pr \left(  \!|h(\mbf{Y}|M,U) \!\!-\!\! \tilde r(\tilde V,M)|  \!\!  > \!\!n \tilde \delta \!\! + \!\! h(U|M)  \middle|U \!\!= \!\!u \! \right)  &\!\! < \!\! 3\! \cdot \!2^{-n\alpha }  \label{eq:p:lem:part2:prove1},\\
\Pr \left( h(U|M) > h(U) + n\alpha \middle| U = u  \right) &\leq 2^{-n\alpha} \label{eq:p:lem:part2:prove2}, \\
\Pr \left(  |r(V) -  \tilde r(\tilde V,M)| > 1 \middle| U = u \right) &=0  \label{eq:p:lem:part2:prove3}  .
\end{align}

We finish the proof by
confirming~\eqref{eq:p:lem:part2:prove1}--\eqref{eq:p:lem:part2:prove3}. First,
\eqref{eq:p:lem:part2:prove1} is a property directly obtained
from~\eqref{eq:s:cor:part1:2} of Corollary~\ref{cor:part1}.  Next,
\eqref{eq:p:lem:part2:prove2} can be derived as follows:
\begin{align*}
&\hspace{-15pt}\Pr \left( h(U|M) > h(U) + n\alpha  \middle| U =u  \right) \notag \\
&= \sum_{m: p(u|m) < 2^{-n\alpha} p(u) } p(m|u)   \\
&= \sum_{m: p(u|m) < 2^{-n\alpha} p(u) } \frac{p(u|m) p(m)}{p(u)}  \\
&< \sum_{m} 2^{-n\alpha} p(m) = 2^{-n\alpha}.  
\end{align*}
Finally, for proving~\eqref{eq:p:lem:part2:prove3},
\begin{equation*}%\label{eq:p:lem:part2:prove2:-1}
| r(V)  - \tilde r(\tilde V,M) | \leq 1,
\end{equation*}
follows by the definitions of $r$ and $\tilde r$ for all $v \neq v_0$. 
\end{IEEEproof}

\subsection{Proof of Theorem~\ref{thm:mt}}
\begin{IEEEproof}
  Let us be given a regular collection
  $(M_{[1:l]} ,\mbf{X},\mbf{Y}_{[1:k]})$. For each $i \in [1:k]$ and
  $j \in [1:l]$, there exists
\begin{itemize}[leftmargin=*]
\item a DRV
  $V_{i,j} : \left\{ \begin{array}{c} |\mcf{V}_{i,j}| \leq 2n^3\logt
      |\mcf{Y}| \\ V_{i,j} \ll (\mbf{X},M_j) \\
      h_{V_{i,j}}(v_{0,i,j}) > \frac{n^2-1}{n} \logt
      \frac{n}{8} \end{array} \right\} $, \vspace{3pt}
\item positive real number
  $\tilde \delta_{i,j} = O(-\sqrt{\alpha} \logt \alpha)$, and
\item function $r_{i,j} : \mcf{V}_{i,j} \rightarrow \mathbb{R}_+$ such
  that
\begin{align}
\hspace{-7pt}\Pr \left(\abs{h(\mbf{Y}_{i}|U,M_j) - r_{i,j}(V_{i,j})} >n \tilde \delta_{i,j} + h(U)  \middle|U = u\right) & \notag \\
&\hspace{-70pt} <  4 \cdot 2^{-n\alpha}, \label{eq:mt:-10}
\end{align}
for all
$U: \left\{ \begin{array}{c} U \gg V_{i,j} \\ (U,M_j) \markov \mbf{X}
    \markov \mbf{Y}_{[1:k]} \end{array} \right\} $, and
$u \in \mcf{U}_{i|j} \defn \set{ u \in \mcf{U} :
  p_{V_{i,j}|U}(v_{0,i,j}|u) = 0 }$
\end{itemize}
by Lemma~\ref{lem:part2}\footnote{Set $\zeta = n^2$ in the lemma.}.

The DRV $V$ in the theorem statement can now be defined as
\begin{align*} 
V = \bigotimes_{i \in [1:k],j \in [1:l]} V_{i,j}.
\end{align*} 
Indeed, let us quickly verify the properties of $V$. First,
$(\mbf{X},M_{[1:l]}) \gg V$ since $(\mbf{X},M_{[1:l]}) \gg V_{i,j}$
for each $i,j$. Second,
\begin{equation*}%\label{eq:mt:qsize}
  |\mcf{V}| \leq ( 2n^3\logt|\mcf{Y}| )^{lk}
\end{equation*}
since $|\mcf{V}_{i,j}| \leq 2n^3\logt|\mcf{Y}|$, while $|\mcf{V}|$ is
at most $\prod_{i \in [1:k], j \in [1:l]} |\mcf{V}_{i,j} |$.

Going forth, it will be important to note that $V \gg V_{i,j}$, thus
to every $v \in \mcf{V}$, $i \in [1:k]$ and $j \in [1:l]$ there exists
a $v_{i,j}$ such that $p_{V_{i,j}|V}(v_{i,j}|v) =1$. Furthermore if
$U\gg V$, then similarly for each $i \in [1:k]$ and $j \in [1:l]$
there exists a $v_{i,j}$ such that $p_{V_{i,j}|U}(v_{i,j}|u) =1$.

To prove the remaining properties, fix a DRV
$U : \left\{ \begin{array}{c} U \gg V\\ (U,M_{[1:l]}) \markov
    \mbf{X}\markov \mbf{Y}_{[1:k]}\end{array} \right\}$, and let
$u \in \mcf{U}_{i|j}$. The probability of the event
$\{U \in \mcf{U}_{i|j}\}$ is determined by the probability that
$V_{i,j} = v_{0,i,j}$; in specific
\begin{align}
p_{U}\left(\mcf{U} \setminus \mcf{ U}_{i|j}\right)  
&= \sum_{u : p_{V_{i,j}|U}(v_{0,i,j}|u) = 1} p_{U}(u) \label{eq:thm:mt:p:pruij1} \\
&= \sum_{u : p_{V_{i,j}|U}(v_{0,i,j}|u) = 1}
  p_{V_{i,j},U}(v_{0,i,j},u) \notag \\%\label{eq:thm:mt:p:pruij2} \\
&\leq p_{V_{i,j}}(v_{0,i,j}) \notag \\ %\label{eq:thm:mt:p:pruij3}\\
&< 2^{-\frac{(n^2-1)}{n} \logt \frac{n}{8} }  \notag \\ %\label{eq:thm:mt:p:pruij4}\\
&\leq 2^{-\frac{n}{2} \logt \frac{n}{8} } \label{eq:thm:mt:p:pruij5}
\end{align}
where~\eqref{eq:thm:mt:p:pruij1} is because $U \gg V \gg V_{i,j}$,
while~\eqref{eq:thm:mt:p:pruij5} is because $n \geq 27$.

Finally, note that $r_{i,j}(V_{i,j})$ is a constant given $U = u$
since $U \gg V \gg V_{i,j}$. Hence there exists a
$\delta = O(-\sqrt{\alpha} \logt \alpha)$ such that
\begin{align}
\hspace{-7pt}\Pr \left(\abs{h(\mbf{Y}_{i}|U,M_j) -\mathbb{H}(\mbf{Y}_i|U,M_j)} >n  \delta + 3 h(U)  \middle|U = u\right) & \notag \\
&\hspace{-70pt} <  4 \cdot 2^{-n\alpha}, \label{eq:thm:mt:p:stab}
\end{align}
for all $i \in [1:k]$, $j \in [1:l]$ and $u \in \mcf{U}_{i|j}$ due to
the combination of~\eqref{eq:mt:-10} and Corollary~\ref{cor:daco
  entropy}. A more detailed verification of which can be found in
Appendix~\ref{app:ot:thm:mt:p:stab}.
\end{IEEEproof}

\section{Proof of Theorem~\ref{lem:mt_aug:1}} \label{sec:mt:aug:1} 

The proof works by identifying two distinct subsets of elements in
$\mcf{M}_j$ for each $j \in [1:l]$. The first subset contains those
$m_j \in \mcf{M}_j$ such that $h(m_j)$ is small and the subset
contains those $m_j \in \mcf{M}_j$ such that $h(m_j)$ is large. $M_j$
can be stabilized conditioned on the first subset, but not conditioned
on the second subset. Luckily we may ignore the second subset since
its probability of occurrence is small.

\subsection{Supporting Lemma}
We streamline the arguments of the proof by first considering the
following lemma:
\begin{lemma} 
\label{lem:only_lemma} 
For any DRVs $U$, $V$, and $\alpha \in \mathbb{R}_+$,
\begin{align*}
\Pr \left( \abs{h(V|U) - h(V)} > \alpha \right) <
  (|\mcf{U}|+1)2^{-\alpha} .
\end{align*}
\end{lemma}
\begin{IEEEproof}
By the union bound, it is clear that
\begin{align}
\Pr \left(\abs{ h_{V|U}(V|U) - h(V)} >  \alpha \right)  &\leq
                                                          p_{V,U}(\mcf{A}^+)
                                                          +
                                                          p_{V,U}(\mcf{A}^-)
\label{eq:apx1}
\end{align}
where 
\begin{align*}
\mcf{A}^- &\defn \set{(v,u) \in  \mcf{V}\times \mcf{U}:
  h(v|u) <  h(v) -  \alpha  }\\
\mcf{A}^+ &\defn \set{(v,u) \in  \mcf{V}\times \mcf{U}:
  h(v|u) > h(v) + \alpha }. 
\end{align*} 
Thus the lemma is verified by~\eqref{eq:apx1} if we can show that
$p_{V,U}(\mcf{A}^+) < |\mcf{U}|2^{-\alpha}$ and
$p_{V,U}(\mcf{A}^-) \leq 2^{-\alpha}$.

First, observe that if there exists $v \in \mcf{V}$ such that
$(v,u) \in \mcf{A}^-$, then
\begin{equation}
p_{U}(u) < 2^{-\alpha} .
\label{eq:ql1}
\end{equation}
Indeed, for all $(v,u) \in
\mcf{A}^-$ it must hold that
\begin{align*}
p_V(v) \geq p_{V|U}(v|u) p_{U}(u) > 2^{\alpha}p_V(v)
p_{U}(u). 
\end{align*}
The upper bound on $p_{V,U} (\mcf{A}^-)$ now follows
from~\eqref{eq:ql1} as below:
\begin{align*}
p_{V,U} (\mcf{A}^-) 
&= \sum_{(v,u) \in \mcf{A}^- } p_{V|U}(v|u) p_{U}(u)
\\
&< \sum_{(v,u) \in \mcf{A}^- } p_{V|U}(v|u) 2^{-\alpha} 
\leq |\mcf{U}|2^{-\alpha}.
\end{align*}
The upper bound on $p_{V,U} (\mcf{A}^+)$ follows similarly in that
\begin{align*}
p_{V,U} (\mcf{A}^+)
&= \sum_{(v,u) \in \mcf{A}^+ } p_{V|U} (v|u) p_{U}(u)
\\
&< \sum_{(v,u) \in \mcf{A}^+ } p_{V} (v) p_{U}(u) 2^{-\alpha} 
\leq 2^{-\alpha}.
\end{align*}
\end{IEEEproof}

\subsection{Proof of Theorem~\ref{lem:mt_aug:1}}
\begin{IEEEproof}
  Let
  $Q_{j} = \left \lfloor \min \left( h(M_j) , \psi \right) \right
  \rfloor$ for each $j \in [1:l]$. We first identify the DRV $Q$ in
  the theorem statement as
\begin{align*}
Q = \bigotimes_{j=1}^l Q_j.
%\label{eq:mt_aug:1:1}
\end{align*}
We start by verifying the properties of $Q$. First,
\begin{align*}
|\mcf{Q}|\leq \prod_{j = 1}^l |\mcf{Q}_j| \leq (\psi+1)^l.
\end{align*}
Second, $Q \ll (M_{[1:l]})$ since $Q_j \ll M_j$ for each $j \in [1:l]$.

Next, choose any DRV $U: \{ U \gg Q\}$, and let
$\tilde \rho = 2\rho + \logt(|\mcf{U}|+1)$.  Fix $j \in [1:l]$, let
$q_j(u)$ be the unique element in $\mcf{Q}_j$ that
$p_{Q_j|U}(q_j |u) = 1$ for each $u \in \mcf{U}$. Then identify the
following sets:
\begin{align}
\mcf{U}_{j,\stab} &= \mcf{U}_{j,*} \cap \set{ u \in \mcf{U} : q_j(u) <
                    \psi } \notag \\
\mcf{U}_{j,\sat} &= \mcf{U}_{j,*} \cap \set{ u \in \mcf{U} : q_j(u) =
                   \psi } 
\label{eq:mt_aug:1:4b}
\end{align}
where
\begin{equation*}
\mcf{U}_{j,*} = \set{ u \in \mcf{U} : \Pr  \left( | h(M_j|U) -  h(M_j) | >
                \tilde \rho  | U=u \right) < 2^{-\rho} }.
\end{equation*}
We proceed to verify~\eqref{eq:thm:mt_aug:1},
\eqref{eq:thm:mt_aug:2stab}, and~\eqref{eq:thm:mt_aug:2sat} stated in
the theorem.

First, observe that $\mcf{U}_{j,\stab} \cup \mcf{U}_{j,\sat} = \mcf{U}_{j,^*}$.
By Lemma~\ref{lem:only_lemma}, we have 
\begin{align}
2^{-2\rho}  
&> \Pr \left( | h(M_j) - h(M_j|U) | > 2\rho + \logt(|\mcf{U}|+1)
  \right)  \notag \\ %\label{eq:mt_aug:1:5}\\
&=\sum_{u} \Pr \left( | h(M_j|U) - h(M_j) | >  \tilde \rho
  \middle|U=u \right) p_{U}(u)  \notag \\
&\geq \sum_{u \notin \mcf{U}_{j,*}} 2^{-\rho} p_U(u) \label{eq:mt_aug:1:p:3}\\
&=   2^{-\rho}  ( 1 -  p_U(\mcf{ U}_{j,*})) \notag 
\end{align}
where~\eqref{eq:mt_aug:1:p:3} is because 
$$\Pr \left( |h(M_j|U) - h(M_j) | > \tilde \rho \middle|U=u \right)
\geq 2^{-\rho}$$ for all $u \notin \mcf{U}_{j,*}$ by definition.
Thus~\eqref{eq:thm:mt_aug:1} is established, i.e.,
\begin{align*}
p_U(\mcf{U}_{j,*}) > 1-2^{-\rho}.
\end{align*}

Next, to prove~\eqref{eq:thm:mt_aug:2stab}, we must show that there
exists a $\beta = O(\rho + 2^{-\rho} \psi)$ such that
\begin{align}
\Pr  \left( | h(M_j|U) - \mathbb{H}_{U}(M_j) | >   \beta + 3 \logt|\mcf{U}|  | U = u \right) < 2^{-\rho}\label{eq:mt_aug:1:newstab:fin}
\end{align}
for all $u \in \mcf{U}_{j,\stab}$. To that end, observe first that
% \begin{align}
% \Pr  \left( |Q_j -  h(M_j|U) | > \tilde \rho + 1   | U = u \right) < 2^{-\rho}\label{eq:mt_aug:1:newstab2}.
% \end{align}
% Indeed
\begin{align}
&\hspace*{-10pt}
  \Pr  \left( | h(M_j|U) - Q_j| >  \tilde \rho  + 1  | U = u \right) \notag \\
&\leq \Pr  \left( | h(M_j|U) - h(M_j) | >  \tilde \rho  | U = u \right) \notag \\
& \hspace{10pt} + \Pr  \left( | h(M_j) - Q_j | >  1  | U = u \right) \label{eq:mt_aug:1:post_fix1} \\
&<  2^{-\rho} \label{eq:mt_aug:1:post_fix2} 
\end{align}
where the first probability in~\eqref{eq:mt_aug:1:post_fix1} is upper
bounded by $2^{-\rho}$ if $u \in \mcf{U}_{j,\stab}$ and the second
probability in~\eqref{eq:mt_aug:1:post_fix1} is exactly $0$, and
thus~\eqref{eq:mt_aug:1:post_fix2} results.
% where~\eqref{eq:mt_aug:1:post_fix1} is because if the predicates of the probabilities fail then
% \begin{align}
% | Q_j -  h(M_j|U)  | &\leq |Q_j - h(M_j)| + |h(M_j) - h(M_j|U)| \notag \\
% &\leq \tilde \rho + 1 \notag,
% \end{align}
% while Equation~\eqref{eq:mt_aug:1:post_fix2} is because
% $u \in \mcf{U}_{j,\stab} \subseteq
% \mcf{U}_{j,*}$.
Equation~\eqref{eq:mt_aug:1:newstab:fin} follows directly
from~\eqref{eq:mt_aug:1:post_fix2} and Corollary~\ref{cor:daco
  entropy} by the fact that the support set, $\mcf{M}_{j,u}$, of
$M_j | \{U=u\}$ satisfies $\logt|\mcf{M}_{j,u}| < \psi$ for each 
$u \in \mcf{U}_{j,\stab}$. A detailed calculation of the order term of
$\beta$ can be found in Appendix~\ref{app:mt_aug:1:newstab:fin}.

On the other hand, for each $u \in \mcf{U}_{j,\sat}$, note that
$h(m_j) \geq \psi$ for all $m_j \in \mcf{M}_{j,u}$. Then since
$u \in \mcf{U}_{j,\sat} \subseteq \mcf{U}_{j,*}$, we must have
\begin{align}
\Pr  \left( h(M_j|U) <  \psi - \tilde \rho  | U = u \right) < 2^{-\rho}\label{eq:mt_aug:1:newsat:fin}
\end{align}
for all $u \in \mcf{U}_{j,\sat}$ by the definition of $\mcf{U}_{j,*}$.
Replacing $\tilde \rho$ in~\eqref{eq:mt_aug:1:newsat:fin} with
$\beta + 3 \logt|\mcf{U}|$ gives~\eqref{eq:thm:mt_aug:2sat} because
$\beta + 3 \logt|\mcf{U}| > \tilde \rho$ (see
Appendix~\ref{app:mt_aug:1:newstab:fin}).

Finally, if $M_j$ is uniform over $\mcf{M}_j$ then
$h(M_j) = \logt \mcf{M}_j$. Thus, by re-defining $\mcf{U}_{j,\stab} =
\mcf{U}_{j,*}$ and  $\mcf{U}_{j,*} = \emptyset$,
$p_U\left(\mcf{U}_{j,\stab} \right) \geq 1- 2^{-\rho}$
and 
\begin{align}
\Pr \left( |h(M_j|U) - \logt|\mcf{M}_j| |   >  \tilde \rho | U=u \right) < 2^{-\rho}
 \label{eq:needtoaddthis}
\end{align}
for all $u \in \mcf{U}_{j,*}$ by
definition. Equation~\eqref{eq:thm:mt_aug:2stabprime} can then be
obtained from~\eqref{eq:needtoaddthis} by replacing $\tilde \rho$ with
$\beta + 3 \logt|\mcf{U}|$ as above.
\end{IEEEproof}

\section{Proof of Theorem~\ref{lem:mt_aug:2}}\label{sec:thm:mt_aug:2}

To prove Theorem~\ref{lem:mt_aug:2}, we construct a finite but
``dense'' subset of conditional distributions in
$\mcf{P}(\mcf{Y}|\mcf{X})$ that stability for all conditional
distributions in the subset implies stability for the whole
$\mcf{P}(\mcf{Y}|\mcf{X})$. It is clear that we will need to consider
all possible product distributions from $\mcf{P}_{Y|X}$. To simplify
the notation required, given any two conditional distributions
$w, \tilde w \in \mcf{P}_{Y|X}$, we will write
\begin{align*}
w_n(\mbf{y}|u) 
&\defn \sum_{\mbf{x}} w^n(\mbf{y}|\mbf{x}) p_{\mbf{X}|U}(\mbf{x}|u) 
\notag \\
\tilde w_n(\mbf{y}|u) 
&\defn \sum_{\mbf{x}} \tilde w^n(\mbf{y}|\mbf{x}) p_{\mbf{X}|U}(\mbf{x}|u)
  %\label{eq:mt_aug:2:-1:wdef}
\end{align*}
throughout this section.

\subsection{Supporting lemmas}

\begin{lemma}\label{lem:olalt}
Let $(\emptyset,\mbf{X},\mbf{Y}_{\mcf{P}(\mcf{Y}|\mcf{X})})$ be a
regular collection of DRVs, then
\begin{equation*}
\logt \frac{w_n(\mbf{y}|u)}{\tilde w_n(\mbf{y}|u)} 
\leq \hspace{8pt}
n \hspace{-8pt}\sup_{\substack{\hat w  \in \mcf{P}(\mcf{Y}|\mcf{X}) \\ \hat p \in
      \mcf{P}(\mcf{X}) }}\mathbb{D}_{\hat w} ( w ||\tilde w|\hat p ) +
  2 |\mcf{X}||\mcf{Y}| \logt n, 
%n\gamma(w||\tilde w) + 2 |\mcf{X}||\mcf{Y}| \logt n,
\end{equation*}
% where
% \begin{equation*}
% \gamma ( w||\tilde w) \defn  
%   \sup_{\substack{\hat w  \in \mcf{P}(\mcf{Y}|\mcf{X}) \\ \hat p \in
%       \mcf{P}(\mcf{X}) }}\mathbb{D}_{\hat w} ( w ||\tilde w|\hat p ) ,
% \end{equation*}
for all $U: \{ U\markov \mbf{X} \markov
\mbf{Y}_{\mcf{P}(\mcf{Y}|\mcf{X})}\}$, $u \in \mcf{U}$, and $\mbf{y}
\in \mcf{Y}^n : \{ \tilde w_n(\mbf{y}|u) > 0 \}$.
\end{lemma}
\begin{IEEEproof}
Let 
\begin{align*}
\zeta(\hat w,\hat p,\mbf{y}) 
&\defn 
\sum_{\mbf{x} : \substack{ p_{\mbf{y}|\mbf{x} } = \hat w \\
  p_{\mbf{x}} = \hat p } } \hat w^n(\mbf{y}|\mbf{x})
  p_{\mbf{X}|U}(\mbf{x}|u) \\
(\hat w^{(\mbf{y})}\!,\! \hat p^{(\mbf{y})}) 
&\defn \!\! 
\argmax_{\substack{ (\hat w,\hat p): \\ \hat p \in \mcf{P}_n(\mcf{X})
  \\ \hat w \in \mcf{P}_n(\mcf{Y}|\hat p) }}   \zeta(\hat w,\hat
  p,\mbf{y}) 2^{-n \mathbb{D}(\hat w||w|\hat p)}.
\end{align*}
We will prove
\begin{align}
\logt \frac{w_n(\mbf{y}|u)}{\tilde w_n(\mbf{y}|u)} \!  
&\leq \! 
n \mathbb{D}_{\hat w^{(\mbf{y})}} ( w ||\tilde w|\hat p^{(\mbf{y})} )
  \! + \! 2|\mcf{X}||\mcf{Y}| \logt n \label{eq:olalt:divdiff},
\end{align}
of which the lemma is a clear consequence.
% (note the maximum is justified since there are a finite number of
% empirical distributions).

Towards proving~\eqref{eq:olalt:divdiff}, recognize 
\begin{align}
w_n(\mbf{y}|u) &= \sum_{\substack{ (\hat w,\hat p): \\ \hat p \in
  \mcf{P}_n(\mcf{X}) \\ \hat w \in \mcf{P}_n(\mcf{Y}|\hat p) }}
  \zeta(\hat w,\hat p,\mbf{y})  \, 2^{-n \mathbb{D}(\hat w||w|\hat p)}  \label{eq:olalt:wn}  \\
\tilde w_n(\mbf{y}|u) &= \sum_{\substack{ (\hat w,\hat p): \\ \hat p
  \in \mcf{P}_n(\mcf{X}) \\ \hat w \in \mcf{P}_n(\mcf{Y}|\hat p) }}
  \zeta(\hat w,\hat p,\mbf{y}) \,  2^{-n \mathbb{D}(\hat w||\tilde w|\hat p)},\label{eq:olalt:hatwn}
\end{align}
since
\begin{equation*}
w^n(\mbf{y}|\mbf{x}) = \hat w^n(\mbf{y}|\mbf{x}) \, 2^{-n \mathbb{D}(\hat w || w | \hat p)}
\end{equation*}
for any
$(\mbf{y},\mbf{x}) : \{ p_{\mbf{y},\mbf{x}}(a,b) = \hat w(a|b) \hat
p(b) ~~ \forall (a,b) \in \mcf{Y} \times \mcf{X}\}$
by~\cite[Lemma~2.3]{csiszar2004information}.  Further notice
that~\eqref{eq:olalt:wn} and~\eqref{eq:olalt:hatwn} imply
\begin{equation}
w_n(\mbf{y}|u)
\leq |\mcf{P}_n(\mcf{Y},\mcf{X})| \zeta(\hat w^{(\mbf{y})},\hat
p^{(\mbf{y})},\mbf{y})  2^{-n \mathbb{D}(\hat w^{(\mbf{y})}||w|\hat
  p^{(\mbf{y})})} \label{eq:olalt:divdiff:10}
\end{equation} 
and
\begin{equation}
  \tilde w_n(\mbf{y}|u) 
  \geq \zeta(\hat w^{(\mbf{y})},\hat p^{(\mbf{y})},\mbf{y})  2^{-n \mathbb{D}(\hat w^{(\mbf{y})}||\tilde w|\hat p^{(\mbf{y})})}, 
  \label{eq:olalt:divdiff:11}
\end{equation}
respectively.  Using the fact~\cite[Lemma~2.1]{CK} that
$|\mcf{P}_n(\mcf{Y},\mcf{X})| \leq (n+1)^{|\mcf{X}||\mcf{Y}|} \leq
n^{2|\mcf{X}||\mcf{Y}|}$, we can arrive at~\eqref{eq:olalt:divdiff}
from~\eqref{eq:olalt:divdiff:10} and~\eqref{eq:olalt:divdiff:11}.
\end{IEEEproof}

\begin{lemma}
\label{lem:olalt2}
Let $(M,\mbf{X},\mbf{Y}_{\mcf{P}(\mcf{Y}|\mcf{X}) })$ be a regular
collection of DRVs, and $\delta,\alpha \in \mathbb{R}_+$, and real
number $\epsilon> \frac{2 |\mcf{X}||\mcf{Y}|}{n} \logt n$. If
\begin{align} 
 \Pr\left( | h(\mbf{Y}_{\tilde w}|M) - \mathbb{H}(\mbf{Y}_{\tilde w}|M)  | >  n\delta   \right)  
<   2^{-n \alpha} ,\label{eq:olalt2:thm:1}
\end{align}
for $\tilde w \in \mcf{P}(\mcf{Y}|\mcf{X})$, then  
\begin{align}
  \Pr\left( | h(\mbf{Y}_{w}|M) - \mathbb{H}(\mbf{Y}_w|M) |  
  >  n \tilde \delta   \right)   <   2^{-n\epsilon} + 2^{-n(\alpha - \epsilon)},  
  \label{eq:olalt:thm:2} 
\end{align}
where 
\begin{align*}
\tilde \delta &= (2+ 2^{-n\epsilon} + 2^{-n(\alpha-\epsilon)}) ( \delta + \epsilon) \notag \\
&+ (2^{-n\epsilon} \!+\! 2^{-n(\alpha-\epsilon)}) \left[
  \logt|\mcf{Y}| \!- \!\frac{2}{n}\logt( 2^{-n\epsilon}
  \!+\!2^{-n(\alpha - \epsilon)}) \right], \notag
\end{align*} 
for all $w \in \mcf{P}(\mcf{Y}|\mcf{X})$ such that 
\begin{align}
\sup_{\substack{\hat p \in \mcf{P}(\mcf{X}) \\ \hat w \in \mcf{P}(\mcf{Y}|\mcf{X})}} \mathbb{D}_{\hat w}(w||\tilde w|\hat p) 
\leq \epsilon - 
  \frac{2 |\mcf{X}||\mcf{Y}|}{n} \logt n . \label{eq:olalt2:thm:cond}
\end{align}
\end{lemma}
\begin{IEEEproof}
  Given~\eqref{eq:olalt2:thm:1}, consider any
  $w \in \mcf{P}(\mcf{Y}|\mcf{X})$ that
  satisfies~\eqref{eq:olalt2:thm:cond}. In this case,
\begin{align}
h_{\mbf{Y}_{\tilde w}|M}(\mbf{y}|m) - h_{\mbf{Y}_{w}|M}(\mbf{y}|m) 
= \logt \frac{w_n(\mbf{y}|m)}{\tilde w_n(\mbf{y}|m)} \leq n \epsilon 
\label{eq:olalt2:new}
\end{align}
for all $m \in \mcf{M}$ and
$\mbf{y}\in \mcf{Y}^n : \{ w^n(\mbf{y}|m) > 0\}$ by
Lemma~\ref{lem:olalt}. Because of~\eqref{eq:olalt2:new}, it must also
follow that
\begin{align}
|h_{\mbf{Y}_{w}|M}(\mbf{y}|m) - \mathbb{H}(\mbf{Y}_{\tilde w}|M) |  \leq n\delta + n \epsilon,\label{eq:olalt2:1}
\end{align}
for all $\mbf{y} \notin \mcf{B}^*(m) \cup \mcf{B}^-(m)$, where
\begin{align*}
\mcf{B}^*(m) &\defn \set{ \mbf{y} : |h_{\mbf{Y}_{\tilde w}|M}(\mbf{y}|m) -  \mathbb{H}(\mbf{Y}_{\tilde w}|M)  | > n\delta } \\
\mcf{B}^-(m) &\defn \set{ \mbf{y} : \logt \frac{w_n(\mbf{y}|m)}{\tilde w_n(\mbf{y}|m)} < - n\epsilon }.
\end{align*}
Thus, we will have 
\begin{align}
\Pr \left( |h_{\mbf{Y}_{w}|M}(\mbf{y}|m) - \mathbb{H}(\mbf{Y}_{\tilde w}|M) |  > n(\delta + \epsilon) \right) & \notag \\
&\hspace{-60pt} <2^{-n( \alpha - \epsilon) } + 2^{-n \epsilon}\label{eq:olalt2:probbound-10}
\end{align}
if we can show that  
\begin{align}\label{eq:olalt2:probbound1}
\Pr \left( \mbf{Y}_{w} \in  \mcf{B}^*(M) \right) < 2^{-n(\alpha - \epsilon) }
\end{align}
and
\begin{align}\label{eq:olalt2:probbound2}
\Pr \left( \mbf{Y}_{w} \in \mcf{ B}^-(M) \right) < 2^{-n \epsilon}.
\end{align}
Equation~\eqref{eq:olalt:thm:2} is then a direct result of
combining~\eqref{eq:olalt2:probbound-10} and Corollary~\ref{cor:daco
  entropy}, since $\mathbb{H}(\mbf{Y}_{\tilde w}|M)$ is a constant
positive real number.

The derivations of~\eqref{eq:olalt2:probbound1}
and~\eqref{eq:olalt2:probbound2} though are straightforward. First,
for~\eqref{eq:olalt2:probbound1},
\begin{align}
& \hspace{-10pt} \Pr\left(\mbf{Y}_{w} \in \mcf{B}^*(M) \right) \notag \\
&= \sum_{m} p_{M}(m) \sum_{\mbf{y}\in\mcf{B}^*(m)  } w_n(\mbf{y}|m)
  \notag \\
&\leq \sum_{m} p_{M}(m) \sum_{\mbf{y}\in\mcf{B}^*(m)  } \tilde w_n(\mbf{y}|m) 2^{n\epsilon} 
  \label{eq:olalt2:probbound:1} \\
&\leq 2^{n \epsilon} \Pr\left(\mbf{Y}_{\tilde w}  \in \mcf{B}^*(M) \right)   <  2^{-n ( \alpha - \epsilon ) } 
  \label{eq:olalt2:probbound:2}
\end{align}
where~\eqref{eq:olalt2:probbound:1} is because
of~\eqref{eq:olalt2:new}, and~\eqref{eq:olalt2:probbound:2} is because
$\Pr\left(\mbf{Y}_{\tilde w} \in \mcf{B}^*(M) \right) <
2^{-n\alpha}$i, which in turn is consequence of the hypothesis
in~\eqref{eq:olalt2:thm:1}. Second for~\eqref{eq:olalt2:probbound2},
similarly,
\begin{align}
\Pr\left(\mbf{Y}_{w} \in \mcf{B}^-(M) \right) &= \sum_{m} P_{M}(m)
                                                \sum_{\mbf{y}\in\mcf{B}^-(m)
                                                } w_n(\mbf{y}|m)
                                                \notag \\
&< \sum_{m} P_{M}(m) \sum_{\mbf{y}\in\mcf{B}^-(m)  } \tilde w_n(\mbf{y}|m) 2^{-n\epsilon} \label{eq:mt_aug:2:probbound:3} \\
&\leq 2^{-n \epsilon} \notag %\label{eq:mt_aug:2:probbound:4}
\end{align}
where~\eqref{eq:mt_aug:2:probbound:3} is because $\mbf{y} \in \mcf{B}^-(m)$. 
\end{IEEEproof}

\subsection{Proof of Theorem~\ref{lem:mt_aug:2}}
\begin{IEEEproof}
  Given an
  $\epsilon \in \left( \frac{4|\mcf{X}||\mcf{Y}|}{n} \logt n, 1 \right)$,
  we will create a finite set of distributions
  $\mcf{ P}^{(\epsilon)}(\mcf{Y}|\mcf{X})$ with cardinality 
\begin{equation}\label{eq:mt_aug:2:prapp:size}
|\mcf{P}^{(\epsilon)}(\mcf{Y}|\mcf{X})| \leq \left(|\mcf{Y}| \left(1 +\left\lfloor \frac{4|\mcf{Y}|^2}{\epsilon} \right\rfloor \right) \right) ^{|\mcf{X}||\mcf{Y}|}
\end{equation} 
that can approximates $\mcf{P}(\mcf{Y}|\mcf{X})$ in the following
sense. For every $w \in \mcf{P}(\mcf{Y}|\mcf{X})$, there exists a
$\tilde w \in \mcf{P}^{(\epsilon)}(\mcf{Y}|\mcf{X})$ such that
\begin{equation}\label{eq:mt_aug:2:1}
\sup_{\substack{\hat p \in \mcf{P}(\mcf{X})\\ \hat w \in \mcf{P}(\mcf{Y}|\mcf{X})} } \mathbb{D}_{\hat w}(w||\tilde w|\hat p) 
\leq  \epsilon - \frac{2|\mcf{X}||\mcf{Y}|}{n} \logt n .
\end{equation}
Achieving this approximation, the theorem then follows directly from
Lemma~\ref{lem:olalt2} with
$\mcf{\tilde P} = \mcf{ P}^{(\epsilon)}(\mcf{Y}|\mcf{X})$.

To form $\mcf{P}^{(\epsilon)}(\mcf{Y}|\mcf{X})$, we will first create
a parameterized set of distributions over $\mcf{Y}$, and then consider
the union of these distributions over all possible parameters. We will
use the union to approximate $\mcf{P}(\mcf{Y})$. Finally
this is extended to a set conditional distributions by observing that
any $w \in \mcf{P}(\mcf{Y}|\mcf{X})$ is simply an order collection of
$w_x \in \mcf{P}(\mcf{Y})$. To begin, let
$\tilde \epsilon = \frac{\epsilon}{4|\mcf{Y}|^2}$, and for each $y \in
\mcf{Y}$ construct the
set of distributions $\mcf{P}^{(\epsilon)}(\mcf{Y};y)$ 
such that 
contains all distributions in $\mcf{P}(\mcf{Y})$ in the form:
\begin{align*}
p(\tilde y) = 
\begin{cases}
j_{\tilde y}\tilde \epsilon \text{ for some }
  j_{\tilde y}\in \left[0:\lfloor 1/\tilde \epsilon \rfloor  \right] &
  \text{ if } \tilde y \in \mcf{Y}\setminus \set{y}\\ 
1 -\sum_{\hat y \in \mcf{Y}\setminus \set{y}} p(\hat y) & \text{ if } \tilde y = y.
\end{cases}
\end{align*}
% This set contains distributions such that $w(\tilde y) = j_{\tilde y} \tilde \epsilon$, for some $j_{\tilde y} \in [0:\lfloor \frac{1}{\tilde \epsilon} \rfloor ] $ for all $\tilde y \in \mcf{Y}\setminus \set{y}$, while $w(y) = 1-\sum_{\tilde y \in \mcf{Y}\setminus\set{y}} w(\tilde y)$. 
Next, % considering all possible parameters $y \in \mcf{Y}$, let
form
\begin{align*}
\mcf{P}^{(\epsilon)}(\mcf{Y}) &\defn \bigcup_{y \in \mcf{Y}} \mcf{P}^{(\epsilon)}(\mcf{Y};y).
\end{align*}
Finally, we can define the approximating set as
\begin{align*}
& \hspace{-10pt} \mcf{P}^{(\epsilon)}(\mcf{Y}|\mcf{X}) 
\\
&\defn 
  \set{ w \in \mcf{P}(\mcf{Y} | \mcf{X}) : w(\cdot|x) \in
                                        \mcf{P}^{\epsilon}(\mcf{Y})
                                        \text{ for each } x \in \mcf{X} }.
\end{align*}
Note that these definitions yield~\eqref{eq:mt_aug:2:prapp:size} since
we have in progression
\begin{align*}%\label{eq:prapp:b1}
|\mcf{P}^{(\epsilon)}(\mcf{Y};y)| 
&\leq (1+\lfloor 1/\tilde \epsilon \rfloor)^{\abs{\mcf{Y}}} \\
|\mcf{P}^{(\epsilon)}(\mcf{Y})| 
&\leq \sum_{y \in \mcf{Y}} |\mcf{P}^{(\epsilon)}(\mcf{Y};y)| \\
|\mcf{P}^{(\epsilon)}(\mcf{Y}|\mcf{X})| 
&\leq |\mcf{P}^{(\epsilon)}(\mcf{Y})|^{|\mcf{X}|}. 
\end{align*}

What remains is to establish~\eqref{eq:mt_aug:2:1}, which can
be done by providing upper bounds on the maximum of
$\logt \frac{w(y|x)}{\tilde w(y|x)}$, since clearly
\begin{equation} \label{eq:mt_aug:2:wind}
\sup_{\hat w,\hat p} \mathbb{D}_{\hat w}(w||\tilde w|\hat p) \leq
\max_{(y,x) : w(y|x) >0} \logt \frac{w(y|x)}{\tilde w(y|x)}.
\end{equation}
For every $x \in \mcf{X}$, let $y_{x}$ denote a maximal element in
$\mcf{Y}$ that $w(y_x|x) = \max_{\tilde y \in \mcf{Y}}w (\tilde
y|x)$. Observe that for every $w \in \mcf{P}(\mcf{Y}|\mcf{X})$ there
exists a $\tilde w \in \mcf{P}^{(\epsilon)}(\mcf{Y}|\mcf{X})$ that
has the following properties:
\begin{itemize}
\item $\tilde w(y|x) \geq w(y|x)$ for all $(y,x) \neq (y_x,x)$, and
\item $\tilde w(y_x|x) \geq w(y_x|x) - |\mcf{Y}|\tilde \epsilon$ for
  all $x \in \mcf{X}$.
\end{itemize}
Hence
\begin{align}
\sup_{\hat w,\hat p} \mathbb{D}_{\hat w}(w||\tilde w|\hat p) 
&\leq \max_{(y_x,x)} \logt \frac{w(y_x|x)}{w(y_x|x) - |\mcf{Y}|\tilde
  \epsilon} \label{eq:thm13_1} \\
&\leq \logt \frac{|\mcf{Y}|^{-1}}{|\mcf{Y}|^{-1} - (4|\mcf{Y}|)^{-1}\epsilon} \leq  \frac{\epsilon}{2} 
\label{eq:thm13_2}\\
&\leq \epsilon - \frac{2|\mcf{X}||\mcf{Y}|}{n} \logt n \label{eq:thm13_3}
\end{align}
where~\eqref{eq:thm13_1} is obtained by combining the two
aforementioned properties of $\tilde w$ and~\eqref{eq:mt_aug:2:wind},
while~\eqref{eq:thm13_2} is because
$w(y_x|x) \geq \abs{\mcf{Y}}^{-1}$, and~\eqref{eq:thm13_3} is because
$-\logt(1-x) > 2x$ for $x\in \left[0:1/2\right)$.
\end{IEEEproof}

\section{Concluding remarks} \label{sec:conclusion}

Our contribution is simply the construction of a DRV that provides information stability, and the examples demonstrating how such a result can be applied. This work is a self contained collection and refinement of our previous works~\cite{graves2014equating,graves2015equal,graves2016information,graves2017wiretap}. Perhaps more accurately, our current work is to our past work as Theseus' final ship is to his starting ship. Every theorem and proof has been drastically changed as to provide results that could be more immediately applicable. In this regards we must thank all reviewers of our previous work, without whose adroit criticisms this work would not be possible. 

There are multiple possible future directions this work may proceed in. As mentioned previous the error bounds can most likely be improved by simply finding an alternative to the blowing up lemma. They would be drastically improved if a replacement for Theorem~\ref{lem:mt_aug:2} were to be found. Next, the work needs to be extended to consider continuous distributions. This seems as if it would be a natural extension since our methods are built similarly to those of the information spectrum. There are also some concerns regarding independent sources that must be addressed. In particular the provided DRV introduces correlation between independent $M_{[1:l]}$, and thus care must be taken when applying these methods to models where achieving such correlation is impossible.

\appendix 

\subsection{Proof of Lemma~\ref{lem:daco entropy}}\label{app:daco entropy}
\begin{IEEEproof}
Let 
%\begin{equation*}
$\mcf{B} 
= \set{ (y,u) \in \mcf{Y} \times \mcf{U} : |h_{Y|U }(y|u) - c| >
  \epsilon }$
%\end{equation*}
and $Q = \idc{(Y,U) \in \mcf{B}}$. From the hypothesis of the lemma,
we have $p_Q(1) < \mu < \frac{1}{2}$.  The conditional entropy
$\mathbb{H} (Y|U)$ can always be expanded in the following manner:
\begin{equation}
\mathbb{H}(Y|U) = \zeta_{Q=0} + \zeta_{Q=1} \label{eq:daco entropy:1}
\end{equation}
where
\begin{align}
\zeta_{Q=0} 
&\defn  
 \sum_{y,u} p_{Y,U, Q}(y,u, 0) h_{Y, Q|U}(y, 0|u)\label{eq:daco entropy:out} \\
\zeta_{Q=1} 
& \defn  
  \sum_{y,u} p_{Y,U,Q}(y,u,1) h_{Y,Q|U}(y,1|u) \label{eq:daco entropy:in}.
\end{align} 
Note that~\eqref{eq:daco entropy:out}  and~\eqref{eq:daco entropy:in}
result from that $p_{Y, Q|U}(y, 0|u) = \begin{cases} p_{Y|U}(y|u) &
  \text{if } (y,u) \notin \mcf{B} \\ 0 & \text{if } (y,u) \in
  \mcf{B} \end{cases}$ and $p_{Y, Q|U}(y, 1|u) = \begin{cases} p_{Y|U}(y|u) &
  \text{if } (y,u) \in \mcf{B} \\ 0 & \text{if } (y,u) \notin
  \mcf{B} \end{cases}$, respectively. 
Clearly 
\begin{equation}\label{eq:daco entropy:out:fin}
(1-\mu)(c-\epsilon) < (1-p_{Q}(1)) \left( c - \epsilon  \right) \leq \zeta_{Q=0} \leq c +  \epsilon
\end{equation}
follows by directly inserting the condition for $(y,u)$'s inclusion in
$\mcf{B}$ into~\eqref{eq:daco entropy:out} and performing the
summation. On the other hand, 
\begin{align}%\label{eq:daco entropy:in:fin}
0 \leq \zeta_{Q=1} 
%&= p_{Y,U}(\mcf{B}) \mathbb{H}(Y |U,Q=1 ) 
%  \notag \\
%& \hspace*{10pt} - 
%   \sum_{(y,u) \in \mcf{B}} p_{Y,U}(y,u) 
%   \logt p_{Q|U}(1|u) 
%\\
&= p_{Q}(1) \mathbb{H}(Y |U,Q=1 ) 
 \notag \\
& \hspace*{10pt} - 
  \sum_{u} p_{Q|U}(1|u)p_{U}(u) 
   \logt p_{Q|U}(1|u)  \notag \\
&\leq p_{Q}(1) \mathbb{H}(Y |U,Q=1 ) + \mathbb{H}(Q|U) \notag \\
&\leq \mu \logt \frac{|\mcf{Y}|}{\mu^2}  \label{eq:daco entropy:p:sub1}
\end{align}
where~\eqref{eq:daco entropy:p:sub1} results from the bound
$\mathbb{H}(Y|U,Q=1 ) \leq \logt |\mcf{Y}|$ and the bound
\begin{align*}
\mathbb{H}(Q|U) &\leq H(Q)  %= - p_{Q}(1) \logt p_{Q}(1)  - p_{Q}(0) \logt p_{Q}(0) \\
\leq -2 \mu \logt \mu 
\end{align*}
due to $\mu < \frac{1}{2}$.

Substituting~\eqref{eq:daco entropy:out:fin},~\eqref{eq:daco entropy:p:sub1}, 
and $p_{Q}(1) < \mu$ into~\eqref{eq:daco entropy:1} yields
\begin{equation} \label{eq:H(Y|U)bound}
(1-\mu)(c  -  \epsilon) < \mathbb{H}(Y|U)  
< c+ \epsilon +  \mu \logt \frac{|\mcf{Y}|}{\mu^2}   
\end{equation} 
The proof can therefore be concluded from~\eqref{eq:H(Y|U)bound} by
demonstrating that
\begin{equation}\label{eq:daco entropy:final}
c \leq   \epsilon   - \logt(1-\mu)  + \logt |\mcf{Y}|  < \epsilon +  \logt\frac{|\mcf{Y}|}{\mu}.
\end{equation}
Equation~\eqref{eq:daco entropy:final} can be verified by observing
\begin{align}
1-\mu < p_{Q}(0) 
= \sum_{(y,u) \notin \mcf{B}} p_{U}(u) 2^{-h(y|u)} %\notag \\
&\leq 2^{-c + \epsilon +  \logt |\mcf{Y}|}
\label{eq:daco entropy:final:2}
\end{align}
where the last inequality is made by substituting in
$2^{-h(y|u)}\leq 2^{-c + \epsilon}$, since $(y,u) \notin \mcf{B}$, and
then summing over all possible $(y,u)$. Solving for $c$
in~\eqref{eq:daco entropy:final:2} gives the first inequality
in~\eqref{eq:daco entropy:final}, while the second inequality is
because $\mu < \frac{1}{2}$ which gives rise to
$-\logt (1-\mu) < -\logt \mu$.
\end{IEEEproof}

\subsection{Proof of Corollary~\ref{cor:mt}}
\label{app:p:cor:mt}
\begin{IEEEproof}
Let us be given a regular collection $(M_{[1:l]},\mbf{X},\mbf{Y}_{\mcf{P}(\mcf{Y}|\mcf{X})})$ and DRV $T: \left\{  (T,M_{[1:l]}) \markov \mbf{X} \markov \mbf{Y}_{\mcf{P}(\mcf{Y}|\mcf{X})} \right\}$. First we will use Theorem~\ref{lem:mt_aug:2} to obtain a set $\mcf{\tilde P} \subseteq \mcf{P}(\mcf{Y}|\mcf{X})$ which extends stability. Next we apply Theorems~\ref{thm:mt} and~\ref{lem:mt_aug:1} to $(M_{[1:l]},\mbf{X},\mbf{Y}_{\mcf{\tilde P}})$ obtaining DRVs $V$ and $Q$ respectively. The DRV $U$ described in the corollary is then constructed by setting $U = (V,Q,T)$. From there we shall construct the set $\mcf{\tilde U}$ and derive the properties of both $U$ and $\mcf{\tilde U}$.

To begin, let $\varepsilon_n \defn n^{-\frac{1}{|\mcf{X}||\mcf{Y}|+1}}$. By Theorem~\ref{lem:mt_aug:2} there exists a set $\tilde P\subseteq \mcf{P}(\mcf{Y}|\mcf{X})$, where
\begin{align}
|\mcf{\tilde P}| \leq  \tilde k  \defn \left( |\mcf{Y}|\left( 1 + \left \lfloor \frac{4|\mcf{Y}|^2}{\varepsilon_n}    \right \rfloor \right) \right)^{|\mcf{X}||\mcf{Y}|} %=  O(n \varepsilon_n),
\end{align}
which has the following property; for all $w \in \mcf{P}(\mcf{Y}|\mcf{X})$ there exists a corresponding $\tilde w_w \in \mcf{\tilde P}$ such that if 
\begin{align}
\Pr \left( | h(\mbf{Y}_{\tilde w_w} |M) - \mathbb{H}(\mbf{Y}_{\tilde w_w}|M)| > n \hat \delta \right) < 2^{-n\hat \alpha} \label{eq:cor:from2}
\end{align}
for some DRV $M: \{M \markov \mbf{X} \markov \mbf{Y}_{\mcf{P}(\mcf{Y}|\mcf{X})}\}$ and positive real numbers $\hat \delta$ and $\hat \alpha$, then 
\begin{align}
\Pr \left( | h(\mbf{Y}_{w} |M) - \mathbb{H}(\mbf{Y}_{w}|M)| > n \check \delta \right) < 2^{-n\varepsilon_n} + 2^{-n(\hat \alpha - \varepsilon_n)}, \label{eq:cor:from2out}
\end{align}
where 
\begin{align} 
\check \delta &= (2+2^{-n\varepsilon_n} + 2^{-n(\hat \alpha - \varepsilon_n)})(\hat \delta + \varepsilon_n) \notag\\
&+ \!\! (2^{-n\varepsilon_n}\! \!\!+\!\! 2^{-n(\hat \alpha - \varepsilon_n)})( \logt|\mcf{Y}| \!\!-\!\! \frac{2}{n} \logt (2^{-n\varepsilon_n} \!\!\!+\!\! 2^{-n(\hat \alpha - \varepsilon_n)}) ). \notag
\end{align}

%= 2^{O( l n \varepsilon_n \logt n) }
Now for $(M_{[1:l]},\mbf{X},\mbf{Y}_{\tilde P})$, there exists:
\begin{itemize}[leftmargin=*] 
\item a DRV $  V : \left\{ \begin{array}{c} |\mcf{V}| \leq (2n^3\logt |\mcf{Y}| )^{l \tilde k}  \\  V \ll (\mbf{X},M_{[1:l]}) \end{array}  \right\}$,\vspace{3pt}
\item a real number $\tilde \delta = O(-\sqrt{2\varepsilon_n} \logt (2\varepsilon_n) ) $, and \vspace{3pt} 
\item for each DRV $U : \left\{ \begin{array}{c} U\gg  V\\ (U,M_{[1:l]}) \markov \mbf{X} \markov \mbf{Y}_{\mcf{\tilde P}} \end{array}  \right\}$, $\tilde w \in \mcf{\tilde P}$, and $j \in [1:l]$ there exists a set $\mcf{U}_{\tilde w|j} \subseteq \mcf{U}$ such that 
\begin{align}
p_{U}(\mcf{ U}_{\tilde w|j}) \geq 1 - 2^{-\frac{n}{2} \logt \frac{n}{8} } \label{eq:cor:mt:mt+2:s},
\end{align}  
and
\begin{align}
\Pr \left( |h(\mbf{Y}_{\tilde w}|M_j,U) - \mathbb{H}_{U}(\mbf{Y}_{\tilde w}|M_j)| > n \tilde \delta + 3 h(U)    \middle| U=u\right) &\notag \\
&\hspace{-100pt}<  4 \cdot 2^{-2n\varepsilon_n} \label{eq:cor:mt:mt+2:p-1}
\end{align}  
for all $u \in \mcf{U}_{\tilde w|j}$,
\end{itemize} 
by Theorem\footnote{Setting $\alpha = 2 \varepsilon_n$}~\ref{thm:mt}.
The existence of a $\delta = O(-\sqrt{\varepsilon_n} \logt \varepsilon_n)$ so that 
\begin{align}
\Pr \left( |h(\mbf{Y}_w|M_j,U) - \mathbb{H}_{U}(\mbf{Y}_w|M_j)| > n\delta + 7 h(U)   \middle| U=u\right) \notag &\\
&\hspace{-100pt}< 5\cdot 2^{-n\varepsilon_n} \label{eq:cor:mt:mt+2:p}
\end{align} 
for each $ w \in \mcf{P}(\mcf{Y}|\mcf{X})$ and $u \in \mcf{U}_{\tilde w_{w}|j}$ follows since Equation~\eqref{eq:cor:mt:mt+2:p-1} takes the form of Equation~\eqref{eq:cor:from2}, and Equation~\eqref{eq:cor:from2} implies Equation~\eqref{eq:cor:from2out}. A verification of Equation~\eqref{eq:cor:mt:mt+2:p} can be found in Appendix~\ref{app:cor:mt:mt+2:p}.

Similarly for any DRVs $M_{[1:l]}$ there exists:
\begin{itemize}[leftmargin=*]
\item DRV $ Q : \left\{ \begin{array}{c} |\mcf{Q}| \leq (n^{2}+1)^l  \\  Q \ll M_{[1:l]} \end{array} \right\}$, \vspace{3pt} 
\item positive real number  $\beta = O( n \varepsilon_n + n^{2} 2^{-n\varepsilon_n} )$, and 
\item for each DRV $U: \{U \gg  Q\}$ and $j \in [1:l]$, there exists sets $\mcf{U}_{j,\stab} \subseteq \mcf{U}$ and $\mcf{U}_{j,\sat} \subseteq \mcf{U}$, such that 
\begin{align}
p_{U}(\mcf{U}_{j,\stab} \cup \mcf{U}_{j,\sat}) \geq 1 - 2^{-n\varepsilon_n}   \label{eq:cor:mt:1:s},
\end{align}  
and 
\begin{align}
\Pr \left( |h(M_j|U) \!-\! \mathbb{H}_{U}(M_j)| \! >  \! \beta \!+ \!3\logt|\mcf{U}| \middle| U=u\right) &\!< \! 2^{-n\varepsilon_n}  \label{eq:cor:mt:1:p1}
\end{align}  
for all $u \in \mcf{U}_{j,\stab}$, and
\begin{align}
\Pr \left(  h(M_j|U) - n^2 < - \beta - 3\logt|\mcf{U}| \middle| U=u\right) < 2^{-n\varepsilon_n }   \label{eq:cor:mt:1:p2}
\end{align}  
for all $u \in \mcf{U}_{j,\sat}$,
\end{itemize}
by Theorem\footnote{Setting $\psi = n^{2}$ and $\rho = n\varepsilon_n$}~\ref{lem:mt_aug:1}.

Now set $U = (V,Q,T)$, and let us confirm the properties of $U$. First, $U \gg T$ is a direct consequence of the definition of $U$. Second,
\begin{align}
\logt |\mcf{U}| &= \logt |\mcf{V}||\mcf{Q}||\mcf{T}| \\%= n^{2l} (4 |\mcf{Y}|n)^{l\tilde k} |\mcf{T}| \\
&\leq \logt |\mcf{T}| + 3l\tilde k \logt n \notag\\
&\hspace{10pt} + l \logt(n^2+1)  + (l\tilde k) \logt (2 \logt |\mcf{Y}|) \notag \\
&\leq \logt |\mcf{T}| + l(3\tilde k + 4) \logt n +  (l\tilde k) \logt (2 \logt |\mcf{Y}|) \notag \\
&= O( \logt|\mcf{T}| - l n \varepsilon_n \logt \varepsilon_n ), \label{eq:cor:mt:p:uot}
\end{align}
with a detailed analysis of the order term appearing in Appendix~\ref{app:ot:cor:mt:p:uot}. 
Finally $(U,M_{[1:l]}) \markov \mbf{X} \markov \mbf{Y}_{\mcf{P}(\mcf{Y}|\mcf{X})}$ since  $Q \ll (\mbf{X},M_{[1:l]})$, $V \ll (\mbf{X},M_{[1:l]})$ and $(T,M_{[1:l]}) \markov \mbf{X} \markov \mbf{Y}_{\mcf{P}(\mcf{Y}|\mcf{X})})$.

It is important to observe that Equation~\eqref{eq:cor:mt:mt+2:p} applies since $U$ satisfies the properties just listed. Still Equation~\eqref{eq:cor:mt:mt+2:p} depends on $h(U)$, and neither Equation~\eqref{eq:cor:mt:mt+2:p} nor Equation~\eqref{eq:cor:mt:1:p1} and~\eqref{eq:cor:mt:1:p2} can bound the distance between $h(M_j)$ and $h(M_j|U)$. For this reason let $\mcf{\check U}$ be the set of all $u \in \mcf{U}$ such that
\begin{align}
h(u)< n\varepsilon_n +  \logt |\mcf{U}| \label{eq:cor:mt:hatu:p}
\end{align}  
and for each $j \in [1:l]$ let $\mcf{\hat U}_j$ be the set of all $u \in \mcf{U}$ such that 
\begin{align}
\Pr \left( |h(M_j|U) - h(M_j)|  > 2 n\varepsilon_n + \logt(|\mcf{U}|+1)   \middle| U = u \right) &\notag \\
&\hspace{-70pt} < 2^{-n \varepsilon_n }   \label{eq:cor:mt:check:p}.
\end{align}  
For these sets we have
\begin{align}
p_{U}(\mcf{\check U}) \geq 1 -2^{-n\varepsilon_n }   \label{eq:cor:mt:hatu:s}
\end{align}  
since
\begin{align}
1 &= \sum_{u} p_{U}(u) \leq p_{U}(\mcf{\check U}) + \sum_{u \notin \mcf{\check U}} 2^{-n\varepsilon_n }|\mcf{U}|^{-1} \\
&\leq p_{U}(\mcf{\check U}) + 2^{-n\varepsilon_n  }.
\end{align} 
Likewise
\begin{align}
p_{U}( \mcf{\hat U}_{j} ) \geq 1 - 2^{-n \varepsilon_n }  \label{eq:cor:mt:check:s}
\end{align}  
which follows by Lemma~\ref{lem:only_lemma} and Markov's inequality.

Now we are in a position to construct the set $\mcf{\tilde U}$ from the corollary statement and prove the associated properties. In particular 
\begin{align}
\mcf{\tilde U} &\defn  \cap_{\tilde w \in \mcf{\tilde P}, j \in [1:l]} \left( \mcf{U}_{\tilde w|j}  \cap \left( \mcf{U}_{j,\stab} \cup \mcf{U}_{j,\sat} \right) \cap  \mcf{\hat U}_j \cap \mcf{\check U}\right),
\end{align}
and note that for each $j \in [1:l]$ and each $w \in \mcf{P}(\mcf{Y}|\mcf{X})$ one of the following two cases must occur; either $u \in  \mcf{U}_{\tilde w_w|j} \cap \mcf{U}_{j,\stab} \cap \mcf{\hat U}_j \cap \mcf{\check U}$ or $u \in  \mcf{U}_{\tilde w_w|j} \cap \mcf{U}_{j,\sat} \cap \mcf{\hat U}_j \cap \mcf{\check U}$. In the first case
\begin{align}
&\Pr \left(\!\!\! \begin{array}{l r l} 
&|h(\mbf{Y}_w|M_j,U) - \mathbb{H}_{U}(\mbf{Y}_w|M_j)| &> n  \nu_n \\
\text{or } &|h(M_j|U) - \mathbb{H}_{U}(M_j)| &> n  \nu_n \\
\text{or } & |h(M_j) - h(M_j|U)| &> n  \nu_n \\
\end{array}   \middle| U = u \right) \notag \\
&\hspace{150pt} < 8\cdot 2^{-n\varepsilon_n} ,
\end{align}
where 
\begin{align}
 \nu_n &= \max \left( \delta  + \frac{7 h(U)}{n},  \frac{\beta+3\logt|\mcf{U}|}{n} ,    2 \varepsilon_n + \frac{\logt(|\mcf{U}|+1)}{n} \right) \\
&\leq \max \left( \delta  + 7 \varepsilon_n + 7 \frac{\logt|\mcf{U}|}{n},  \frac{\beta+3\logt|\mcf{U}|}{n}, \dots \right. \notag \\
&\hspace{80pt}   \left.  2 \varepsilon_n + \frac{\logt(|\mcf{U}|+1)}{n}  \right) \label{eq:cor:mt:p:hatnuout}
\end{align} 
by Equations~\eqref{eq:cor:mt:mt+2:p},~\eqref{eq:cor:mt:1:p1},~\eqref{eq:cor:mt:hatu:p} and~\eqref{eq:cor:mt:check:p}.  Similarly in the second case  
\begin{align}
&\Pr \left( \!\!\! \begin{array}{l r l} 
&\hspace{-15pt}|h(\mbf{Y}_w|M_j,U) - \mathbb{H}_{U}(\mbf{Y}_w|M_j)| &> n  \nu_n \\
\text{or } &h(M_j|U)  - n^2  &< -n  \nu_n  \\
\text{or } & |h(M_j) - h(M_j|U)| &> n  \nu_n \\
\end{array}   \middle| U = u \right) \notag \\
&\hspace{170pt} <  8 \cdot 2^{-n\varepsilon_n}  
\end{align}
by Equations~\eqref{eq:cor:mt:mt+2:p},~\eqref{eq:cor:mt:1:p2},~\eqref{eq:cor:mt:hatu:p} and~\eqref{eq:cor:mt:check:p}. Furthermore 
\begin{align}
\nu_n = O(n^{-1} \logt|\mcf{T}| -l \sqrt{\varepsilon_n} \logt \varepsilon_n)  \label{eq:cor:mt:p:nuot}
\end{align}
as detailed in Appendix~\ref{app:ot:cor:mt:p:nuot}, and
\begin{align}
\Pr \left( (\mbf{Y}_w,M_{[1:l]}) \notin \mcf{D}_{\stab,(M_j)}(U,w; \nu_n) \middle| U = u \right) &< 8 \cdot 2^{-n\varepsilon_n}
\end{align}  
in the first case, while
\begin{align}
\Pr \left( (\mbf{Y}_w,M_{[1:l]}) \notin \mcf{D}_{\sat,(M_j)}(U,w; \nu_n) \middle| U = u \right) &\notag\\
&\hspace{-70pt} < 8 \cdot 2^{-n\varepsilon_n}.
\end{align} 
in the second case.

Finally
\begin{align}
p_{U}(\mcf{\tilde U}) &\geq 1 - l \tilde k  2^{-\frac{n}{2} \logt \frac{n}{8}} - (2l+1)2^{-n\varepsilon_n}  \\
&\geq 1 - O(l2^{-n \varepsilon_n}), \label{eq:cor:mt:p:pfin}
\end{align}
follows by combining Equations~\eqref{eq:cor:mt:mt+2:s},~\eqref{eq:cor:mt:1:s},~\eqref{eq:cor:mt:hatu:s},~\eqref{eq:cor:mt:check:s} and the union bound. A more detailed verification of the order term in Equation~\eqref{eq:cor:mt:p:pfin} can be found in Appendix~\ref{app:ot:cor:mt:p:pfin}.

\end{IEEEproof}

\subsection{Order terms}

Before deriving the order terms explicitly, a few inequalities useful inequalities need to be derived for $\alpha \in \left(\frac{\logt n}{n},\frac{1}{8 \ln 2}\right)$, and $n\geq 27$. Letting $\alpha_- = \frac{\logt n}{n}$ and $\alpha_+ = \frac{1}{8 \ln 2}$ these inequalities are as follows:
\begin{align}
2^{-n\alpha}  &\leq 2^{-n \alpha_-} \leq  \frac{1}{n} \label{eq:app:ot:exp}\\
%2 + 5\cdot 2^{-n\alpha} &< 2+ \frac{5}{n} < 2 + \frac{5}{8} < 3\\
\frac{1}{n} &\leq -\sqrt{\alpha} \logt (\alpha ) \cdot \frac{1}{-n\sqrt{\alpha_-} \logt (\alpha_- )} \notag \\
&\leq -\sqrt{\alpha} \logt \alpha \frac{1}{\sqrt{n} \logt^{\frac{3}{2}} n } \label{eq:app:ot:1/n}\\
\alpha%&\leq -\sqrt{\alpha} \logt (\alpha ) \cdot \frac{\sqrt{\alpha}}{-\logt \alpha} \notag\\
&\leq -\sqrt{\alpha} \logt (\alpha )  \frac{\sqrt{\alpha_+}}{-\logt \alpha_+} \notag \\
&\leq -\sqrt{\alpha} \logt (\alpha ) \cdot \frac{1 }{2\sqrt{2 \ln 2} \logt (8 \ln 2)} \label{eq:app:ot:a} \\
\sqrt{\alpha} &\leq -\sqrt{\alpha} \logt (\alpha )  \frac{1}{-\logt \alpha_+} \notag \\
&\leq -\sqrt{\alpha} \logt (\alpha ) \cdot \frac{1 }{\logt (8 \ln 2)} \label{eq:app:ot:sqrta} .
\end{align}

Likewise if $\varepsilon_n = n^{-\frac{1}{|\mcf{X}||\mcf{Y}|+1}}$, $|\mcf{X}|\geq 2$, $|\mcf{Y}| \geq 2$ and $n \geq 27$ then
\begin{align}
\tilde k &= \left( |\mcf{Y}| \left( 1 + \left \lfloor \frac{4 |\mcf{Y}|^2}{\varepsilon_n} \right \rfloor \right) \right)^{|\mcf{X}||\mcf{Y}|} \notag \\
&\leq \left( |\mcf{Y}| + 4 |\mcf{Y}|^3 \right)^{|\mcf{X}||\mcf{Y}|} \varepsilon_n^{-|\mcf{X}||\mcf{Y}|} \notag \\
&= \left( |\mcf{Y}| + 4 |\mcf{Y}|^3 \right)^{|\mcf{X}||\mcf{Y}|} n^{1 - \frac{1}{|\mcf{X}||\mcf{Y}|+1}} \notag \\
&= \left( |\mcf{Y}| + 4 |\mcf{Y}|^3 \right)^{|\mcf{X}||\mcf{Y}|} n \varepsilon_n  \label{eq:app:ot:tk}\\
\varepsilon_n %&= (-\sqrt{\varepsilon_n} \logt \varepsilon_n) \frac{\sqrt{\varepsilon_n}}{-\logt \varepsilon_n} \\
%&= (-\sqrt{\varepsilon_n} \logt \varepsilon_n) \frac{ 2 (|\mcf{X}||\mcf{Y}|+1) }{ n^{\frac{1}{2|\mcf{X}|\mcf{Y}| + 2}} \logt n} \\
&\leq (-\sqrt{\varepsilon_n} \logt \varepsilon_n) \frac{ 2 (|\mcf{X}||\mcf{Y}|+1) }{ \logt n}  \label{eq:app:ot:ve}\\
2^{-n\varepsilon} %&= (-\sqrt{\varepsilon_n} \logt \varepsilon_n) \frac{2^{-n\varepsilon_n}}{-\sqrt{\varepsilon_n} \logt \varepsilon_n}\\
%&=  (-\sqrt{\varepsilon_n} \logt \varepsilon_n)  \frac{ (|\mcf{X}||\mcf{Y}|+1) 2^{-n^{1-\frac{1}{|\mcf{X}||\mcf{Y}|+1}}}}{ n^{\frac{1}{2(|\mcf{X}||\mcf{Y}|+1)}} \logt n}\\
&\leq  (-\sqrt{\varepsilon_n} \logt \varepsilon_n)  \frac{ (|\mcf{X}||\mcf{Y}|+1) }{ 2^{n^{\frac{4}{5}}}  \logt n} \label{eq:app:ot:exp2ve} \\
-\varepsilon_n \logt \varepsilon_n &\leq -\sqrt{\varepsilon_n} \logt \varepsilon_n \label{eq:app:ot:eve} \\
n 2^{-n\varepsilon_n} &\leq (-\sqrt{\varepsilon_n} \logt \varepsilon_n) \frac{n 2^{-n^{1-\frac{1}{|\mcf{X}||\mcf{Y}|+1}}}}{-\sqrt{\varepsilon_n} \logt \varepsilon_n} \notag \\
%&= (-\sqrt{\varepsilon_n} \logt \varepsilon_n)(|\mcf{X}||\mcf{Y}|+1) n^{1 + \frac{1}{2|\mcf{X}||\mcf{Y}|+2}} \notag \\
%&\hspace{50pt} \cdot 2^{-n^{1-\frac{1}{|\mcf{X}||\mcf{Y}|+1}}} \notag \\
&\leq (-\sqrt{\varepsilon_n} \logt \varepsilon_n)(|\mcf{X}||\mcf{Y}|+1) n^{\frac{11}{10}} 2^{-n^{\frac{4}{5}}} \\
&< (-\sqrt{\varepsilon_n} \logt \varepsilon_n)(|\mcf{X}||\mcf{Y}|+1) \label{eq:app:ot:nexpve}
\end{align}

\subsubsection{Equation~\eqref{eq:daco:prop1}}\label{app:ot:daco:prop1p}
In specific the combination of Equations~\eqref{eq:daco:prove1} and~\eqref{eq:daco:prove2} provides
\begin{align}
\Pr \left( |h(\mbf{Y}|U) -  s^* \lambda_n |  < n \tilde \delta + \lambda+ h(U) \middle| u \right) < 3\cdot 2^{-n\alpha } \label{eq:daco:prop1p}.
\end{align}
Hence we need to show that $\tilde\delta + n^{-1}\lambda = O(-\sqrt{\alpha} \logt \alpha)$. Towards this goal, let $B$ be a Bernoulli random variable with parameter $\sqrt{\alpha 2 \ln 2}$, and recall that 
\begin{align}
\tilde \delta + \frac{\lambda}{n}  &=\tau_n(2^{-n\alpha}, 2^{-n\alpha}) + \alpha + 7.19|\mcf{Y}| \frac{\logt n}{n}  + \frac{\lambda}{n}\notag \\
&\leq \tau_n(2^{-n\alpha}, 2^{-n\alpha}) + \alpha + 11.19|\mcf{Y}| \frac{\logt n}{n}, \label{eq:app:ot:daco:prop1p:1}
\end{align}
and
\begin{align}
\tau_n(2^{-n\alpha}, 2^{-n\alpha}) &\leq  \mathbb{H}\left( B \right) + \sqrt{\alpha 2 \ln 2} \logt|\mcf{Y}| \\
&\leq -\sqrt{\alpha 2 \ln 2} \logt \alpha +  \sqrt{\alpha 2 \ln 2} \logt|\mcf{Y}| \label{eq:app:ot:daco:prop1p:2}
%\\&\leq -\sqrt{\alpha} \logt \alpha \left( \frac{\sqrt{2 \ln 2} \logt(\alpha|\mcf{Y}|)}{\logt \alpha} \right)
\end{align}
since the entropy of a Bernoulli random variable with parameter $x$ is less than $ -2 x \logt x$ for all $x < \frac{1}{2}$ (also recall  $\alpha < (8 \ln 2)^{-1}$). 
Combining Equations~\eqref{eq:app:ot:daco:prop1p:1} and~\eqref{eq:app:ot:daco:prop1p:2} with Equations~\eqref{eq:app:ot:1/n}--\eqref{eq:app:ot:sqrta} gives
\begin{align}
\tilde \delta + n^{-1}\lambda &\leq - \mu \sqrt{\alpha} \logt \alpha \label{eq:app:daco:prop1:td}
\end{align}
where 
\begin{align}
\mu &\defn \sqrt{2 \ln 2} + \frac{\sqrt{2 \ln 2} \logt|\mcf{Y}|}{\logt ( 8 \ln 2)} \notag \\
&\hspace{20pt} + \frac{1 }{2\sqrt{2 \ln 2} \logt (8 \ln 2)} +  \frac{11.19 |\mcf{Y}| }{\sqrt{n} \logt^{\frac{1}{2}} n } .
\end{align}
Clearly $\mu$ has a maximum upper bound in terms of $|\mcf{Y}|$ and thus $\tilde \delta + n^{-1}\lambda = O(-\sqrt{\alpha} \logt \alpha)$.

\subsubsection{Equation~\eqref{eq:lem:part2:p:ot}} \label{app:ot:lem:part2:p:ot}

Observe that there exists a positive number $\mu$ such that 
\begin{align}
\delta = -\mu \sqrt{\alpha} \logt \alpha + \alpha + \frac{1}{n}
\end{align}
since $\tilde \delta = O(-\sqrt{\alpha} \logt \alpha)$. It therefore follows that $\delta = O(-\sqrt{\alpha} \logt \alpha)$ by Equations~\eqref{eq:app:ot:1/n} and~\eqref{eq:app:ot:a}.

\subsubsection{Equation~\eqref{eq:thm:mt:p:stab}} \label{app:ot:thm:mt:p:stab}
In particular the combination gives 
\begin{align}
&\Pr \left( \begin{array}{r} |h(\mbf{Y}_i|U,M_j) - \mathbb{H}(\mbf{Y}_i|U,M_j)| \\  > n \delta + (2 + 4\cdot 2^{-n\alpha} ) h(U) \end{array}  \middle| U=  u \right)  <  4\cdot 2^{-n\alpha},
\end{align} 
where 
\begin{align}
\delta = (2 + 4\cdot 2^{-n\alpha}) (\max_{i,j} \delta_{i,j}) + 4 \cdot 2^{-n\alpha} ( \logt|\mcf{Y}| - 4 + 2\alpha).
\end{align} 
In turn then we have 
\begin{align}
&\Pr \left( \begin{array}{r} |h(\mbf{Y}_i|U,M_j) - \mathbb{H}(\mbf{Y}_i|U,M_j)| \\  > n \delta +  3 h(U) \end{array}  \middle| U=  u \right)  <  4\cdot 2^{-n\alpha},
\end{align} 
and
\begin{align}
\delta \leq 3 (\max_{i,j} \delta_{i,j}) + \frac{4}{n} ( \logt|\mcf{Y}|).
\end{align} 
since $\alpha \in ( n^{-1} \logt n, (8 \ln 2)^{-1}) $ and $n\geq 27$ imply $2^{-n\alpha} \leq \frac{1}{n} \leq \frac{1}{27}.$ To finish the proof note that $\delta = O(-\sqrt{\alpha} \logt \alpha)$ since $\delta_{i,j} = O(-\sqrt{\alpha} \logt \alpha)$ for all $i,j$, and $\frac{1}{n} = O(-\sqrt{\alpha} \logt \alpha)$ by Equation~\eqref{eq:app:ot:1/n}.

\subsubsection{Equation~\eqref{eq:mt_aug:1:newstab:fin}}\label{app:mt_aug:1:newstab:fin}
%Begin by noting that $2^{-\beta} \leq \frac{1}{2}$ since $\beta \geq 1$, and that $\logt|\mcf{M}_{j,u}| < \psi$. Therefore, 
Given Equation~\eqref{eq:mt_aug:1:post_fix2} ,
\begin{align}
\Pr  \left( |\mathbb{H}(M_j|U) -  h(M_j|U) | >  \tilde \beta   | U = u \right) < 2^{-\tau}
\end{align}
where
\begin{align}
\tilde \beta = (2+2^{-\rho}) (2 \rho + \logt(|\mcf{U}|+1) +  1) + 2^{-\rho}\psi + 2^{-\rho} 2 \rho ,
\end{align}
follows by Corollary~\ref{cor:daco entropy}. But $\rho\geq 1$ and $|\mcf{U}| \geq 1$, and hence
\begin{align}
\tilde \beta < 8  \rho + 3\logt|\mcf{U}| + 6 + 2^{-\rho} \psi &\leq 14 \rho + 2^{-\rho} \psi + 3 \logt|\mcf{U}|.
\end{align}
Thus 
\begin{align}
\Pr  \left( |\mathbb{H}(M_j|U) -  h(M_j|U) | >  \beta + 3\logt|\mcf{U}|   | U = u \right) < 2^{-\rho}
\end{align}
where $\beta = 14 \rho + 2^{-\rho} \psi= O(\rho + 2^{-\rho}\psi)$.

\subsubsection{Equation~\eqref{eq:cor:mt:mt+2:p}}\label{app:cor:mt:mt+2:p}
The specific bound guaranteed is 
\begin{align}
\Pr \left( |h(\mbf{Y}_{w}|M_j,U) - \mathbb{H}_{U}(\mbf{Y}_{w}|M_j)| > n \breve \delta    \middle| U = u\right) &\notag \\
&\hspace{-80pt} \leq 5\cdot 2^{-n\varepsilon_n},
\end{align}
where
\begin{align}
\breve \delta &=\left( 2 + 5\cdot 2^{-n\varepsilon_n} \right) \left( \tilde \delta + \frac{3 h(U)}{n}  + \varepsilon_n\right) \notag \\
&\hspace{5pt} + 5 \cdot 2^{-n \varepsilon_n} \left( \logt|\mcf{Y}| - \frac{2\logt 5}{n} + 2 \varepsilon_n \right).
\end{align}
%and $\tilde \delta \leq -\mu \sqrt{2\varepsilon_n} \logt 2\varepsilon_n  < -( \mu\sqrt{2}) \sqrt{\varepsilon_n} \logt \varepsilon_n $ for some positive real number $\mu$, and $\varepsilon_n = n^{\frac{-1}{|\mcf{X}||\mcf{Y}|+1}}$. 
Notice that 
\begin{align}
5 \cdot 2^{-n\varepsilon_n} &= 5 \cdot 2^{-n^{1 - \frac{1}{|\mcf{X}||\mcf{Y}|+1} }} \leq 5\cdot 2^{-n^{\frac{4}{5}}} < \frac{1}{3},
\end{align}
and hence
\begin{align}
 \breve \delta \leq \frac{7}{3}\left( \tilde \delta + \varepsilon_n + \frac{3 h(U)}{n} \right) + \frac{ 2 \varepsilon_n}{3}  + 5\cdot 2^{-n\varepsilon_n} \logt|\mcf{Y}|,
\end{align}
since $n \geq 27$ and $|\mcf{Y}|, |\mcf{X}| \geq 2$. Thus 
\begin{align}
\Pr \left( |h(\mbf{Y}_{w}|M_j,U) - \mathbb{H}_{U}(\mbf{Y}_{w}|M_j)| > n \delta + 7 \logt|\mcf{U}|    \middle| U =u\right) &\notag\\
&\hspace{-80pt} \leq 5\cdot 2^{-n\varepsilon_n},
\end{align}
for $$\delta =  \frac{7}{3} \tilde \delta + 3 \varepsilon_n + 5 \cdot 2^{-n\varepsilon_n} \logt|\mcf{Y}|.$$ Furthermore $\delta = O(-\sqrt{\varepsilon_n} \logt \varepsilon_n)$ because of Equations~\eqref{eq:app:ot:ve} and~\eqref{eq:app:ot:exp2ve} and because $\tilde \delta = O(-\sqrt{\varepsilon_n} \logt \varepsilon_n)$.

\subsubsection{Equation~\eqref{eq:cor:mt:p:uot}} \label{app:ot:cor:mt:p:uot}
Begin by observing there exists a positive real number $\mu$ such that
\begin{align}
\logt|\mcf{U}| &\leq \logt|\mcf{T}| +  l(3 \mu n \varepsilon_n + 4)  \logt n \notag \\
&\hspace{10pt} + l\mu n \varepsilon_n \logt (2 \logt |\mcf{Y}|) ,
\end{align}
since $\tilde k = O(n\varepsilon_n)$. 
Factoring out $-l n \varepsilon_n \logt \varepsilon_n$ results in 
\begin{align}
\logt|\mcf{U}| \leq \logt|\mcf{T}| -  \hat \mu l n \varepsilon_n \logt \varepsilon_n,
\end{align}
where
\begin{align} 
\hat \mu &= 3 \mu \frac{\logt n}{-\logt \varepsilon_n} + 4 \frac{\logt n}{-n \varepsilon_n \logt \varepsilon_n} + \mu \frac{\logt (2\logt |\mcf{Y}|)}{-\logt \varepsilon_n} \notag \\
&=(|\mcf{X}||\mcf{Y}|\!+\!1) \left(\!  3 \mu \! +\! 4 n^{-1 + \frac{1}{|\mcf{X}||\mcf{Y}|+1}}\! +\! \mu  \frac{\logt (2\logt |\mcf{Y}|)}{\logt n} \right).
\end{align}
Clearly then
\begin{align}
\logt|\mcf{U}| &= O(\logt|\mcf{T}| - l n \varepsilon_n \logt \varepsilon_n).
\end{align}

\subsubsection{Equation~\eqref{eq:cor:mt:p:nuot}} \label{app:ot:cor:mt:p:nuot}

First note that $\nu_n $ is the maximum of three different terms. If we show that each of these terms is $O(n^{-1} \logt|\mcf{T}| -l \sqrt{\varepsilon_n}\logt \varepsilon_n)$ then it must also follow that $\nu_n = O(n^{-1} \logt|\mcf{T}| -l \sqrt{\varepsilon_n}\logt \varepsilon_n)$. First
\begin{align}
&\delta + 7 \varepsilon_n + 7 \frac{\logt|\mcf{U}|}{n} \notag \\
&= O(-\sqrt{\varepsilon_n} \logt \varepsilon_n) + O( n^{-1} \logt |\mcf{T}| - l \varepsilon_n \logt \varepsilon_n) \notag \\
&\leq O(n^{-1} \logt |\mcf{T}| - l \sqrt{\varepsilon_n} \logt \varepsilon_n)
\end{align}
by Equations~\eqref{eq:app:ot:ve} and~\eqref{eq:app:ot:eve}, and because $\delta = O(-\sqrt{\varepsilon_n} \logt \varepsilon_n)$ and $\logt|\mcf{U}| = O(\logt|\mcf{T}| - ln \varepsilon_n \logt \varepsilon_n)$.

Next 
\begin{align}
&\frac{\beta+3 \logt|\mcf{U}| }{n} \notag \\
&= O(\varepsilon_n + n 2^{-n\varepsilon_n})  + O( n^{-2} \logt |\mcf{T}| - ln^{-1} \varepsilon_n \logt \varepsilon_n) \notag \\
&\leq O(n^{-1} \logt |\mcf{T}| - l \sqrt{\varepsilon_n} \logt \varepsilon_n)
\end{align}
by Equations~\eqref{eq:app:ot:ve} and~\eqref{eq:app:ot:nexpve} and because $\logt|\mcf{U}| = O(\logt|\mcf{T}| - ln \varepsilon_n \logt \varepsilon_n)$.

Finally
\begin{align}
2 \varepsilon_n + \frac{1}{n} \logt(|\mcf{U}| + 1) &\leq  2 \varepsilon_n + 1+ \frac{1}{n} \logt|\mcf{U}| \notag \\
&\leq O(n^{-1} \logt |\mcf{T}| - l \sqrt{\varepsilon_n} \logt \varepsilon_n)
\end{align}
by Equation~\eqref{eq:app:ot:nexpve} and because $\logt|\mcf{U}| = O(\logt|\mcf{T}| - ln \varepsilon_n \logt \varepsilon_n)$. 

Since all three terms are $O(n^{-1} \logt |\mcf{T}| - l \sqrt{\varepsilon_n} \logt \varepsilon_n)$ it also follows that
\begin{align}
\nu_n = O(n^{-1} \logt |\mcf{T}| - l \sqrt{\varepsilon_n} \logt \varepsilon_n).
\end{align}

\subsubsection{Equation~\eqref{eq:cor:mt:p:pfin}} \label{app:ot:cor:mt:p:pfin}

First, clearly,
\begin{align}
(2l+1)2^{-n \varepsilon_n} = O(l2^{-n\varepsilon_n}),
\end{align}
and on the other hand
\begin{align}
\tilde k 2^{-\frac{n}{2} \logt \frac{n}{8}} &\leq (|\mcf{Y}|+4|\mcf{Y}|^3)^{|\mcf{X}||\mcf{Y}|} n\varepsilon_n 2^{-\frac{n}{2} \logt \frac{n}{8}} \\
&\leq (|\mcf{Y}|+4|\mcf{Y}|^3)^{|\mcf{X}||\mcf{Y}|} n 2^{-\frac{n}{2} \logt \frac{n}{8}}\\
&= 8(|\mcf{Y}|+4|\mcf{Y}|^3)^{|\mcf{X}||\mcf{Y}|} 2^{-\left( \frac{n}{2} -1 \right) \logt \frac{n}{8}} \\
&\leq 8(|\mcf{Y}|+4|\mcf{Y}|^3)^{|\mcf{X}||\mcf{Y}|} 2^{-n \varepsilon_n}.
\end{align}
The summation of the two terms is therefore $O(l 2^{-n \varepsilon_n})$.

\bibliographystyle{ieeetr}

\end{document}